\PassOptionsToPackage{unicode}{hyperref}
\PassOptionsToPackage{hyphens}{url}
\pdfoutput=1
\documentclass[
  openany]{article}
\usepackage{lmodern}
\usepackage{amssymb,amsmath}
\usepackage{ifxetex,ifluatex}
\ifnum 0\ifxetex 1\fi\ifluatex 1\fi=0 
  \usepackage[T1]{fontenc}
  \usepackage[utf8]{inputenc}
  \usepackage{textcomp} 
\else 
  \usepackage{unicode-math}
  \defaultfontfeatures{Scale=MatchLowercase}
  \defaultfontfeatures[\rmfamily]{Ligatures=TeX,Scale=1}
\fi
\IfFileExists{upquote.sty}{\usepackage{upquote}}{}
\IfFileExists{microtype.sty}{
  \usepackage[]{microtype}
  \UseMicrotypeSet[protrusion]{basicmath} 
}{}
\makeatletter
\@ifundefined{KOMAClassName}{
  \IfFileExists{parskip.sty}{%
    \usepackage{parskip}
  }{
    \setlength{\parindent}{0pt}
    \setlength{\parskip}{6pt plus 2pt minus 1pt}}
}{
  \KOMAoptions{parskip=half}}
\makeatother
\usepackage{xcolor}
\IfFileExists{xurl.sty}{\usepackage{xurl}}{} 
\IfFileExists{bookmark.sty}{\usepackage{bookmark}}{\usepackage{hyperref}}
\hypersetup{
  pdftitle={New ideas for method comparison: a Monte Carlo power analysis},
  pdfauthor={Dr Giorgio Pioda},
  hidelinks,
  pdfcreator={LaTeX via pandoc}}
\usepackage[margin=1in]{geometry}
\usepackage{longtable,booktabs}
\usepackage{etoolbox}
\makeatletter
\patchcmd\longtable{\par}{\if@noskipsec\mbox{}\fi\par}{}{}
\makeatother
\IfFileExists{footnotehyper.sty}{\usepackage{footnotehyper}}{\usepackage{footnote}}
\makesavenoteenv{longtable}
\usepackage{graphicx}
\makeatletter
\def\maxwidth{\ifdim\Gin@nat@width>\linewidth\linewidth\else\Gin@nat@width\fi}
\def\maxheight{\ifdim\Gin@nat@height>\textheight\textheight\else\Gin@nat@height\fi}
\makeatother
\setkeys{Gin}{width=\maxwidth,height=\maxheight,keepaspectratio}
\makeatletter
\def\fps@figure{htbp}
\makeatother
\usepackage{float}
\usepackage{hyperref}
\usepackage{booktabs}
\usepackage{longtable}
\usepackage{array}
\usepackage{multirow}
\usepackage{wrapfig}
\usepackage{float}
\usepackage{colortbl}
\usepackage{pdflscape}
\usepackage{tabu}
\usepackage{threeparttable}
\usepackage{threeparttablex}
\usepackage[normalem]{ulem}
\usepackage{makecell}
\usepackage{xcolor}
\usepackage[]{natbib}
\title{New ideas for method comparison: a Monte Carlo power analysis}
\author{Dr Giorgio Pioda\footnote{Scuola Superiore Medico Tecnica - Locarno - Switzerland, \href{mailto:giorgio.pioda@edu.ti.ch}{\nolinkurl{giorgio.pioda@edu.ti.ch}}, \href{mailto:gfwp@ticino.com}{\nolinkurl{gfwp@ticino.com}}}}
\date{10/5/2021}

\begin{document}
\maketitle

{
\setcounter{tocdepth}{2}
\tableofcontents
}
\hypertarget{introduction}{%
\section{Introduction}\label{introduction}}

\hypertarget{joint-ellipse-je-vs-confidence-intervals-ci-based-method-comparison}{%
\subsection{Joint ellipse (JE) vs confidence intervals (CI) based method comparison}\label{joint-ellipse-je-vs-confidence-intervals-ci-based-method-comparison}}

Comparison methods are very important for all analytical laboratories and are largely standardized like in the EU ISO 15189 norms collection \citep{Pum2019rev}. For quantitative assessments there are basically two different approaches. The first one is the so-called Bland-Altman (BA) method which relies on a special \(T\) test and on the visual inspection of the so-called BA plot \citep{BlandAltman1983}. The other approach is based on two regression methods: the Deming (Dem) parametric \citep{Linnet1990} and the Passing-Bablok (PaBa) \citep{Passing1983} non-parametric one.

The analytical confidence intervals (CI) of these regressions are available \citep{Linnet1990, Passing1983} but frequently the jackknife \citep{Linnet1990} or the more modern and computer intensive bootstrap procedure is preferred. Simple percentile CI and BCa bootstrapped CI are the most common choices. In this work all the CI are calculated with the popular BCa approach unless otherwise noted.

With the traditional strategy the method comparison takes place twice, one time for the slope and one time for the intercept, with the same \(\alpha=0.05\). Hence, the global family wise error rate (FWER) is not corrected. From a theoretical point of view it would be advisable to adjust the \(\alpha\) levels of the tests to have a global type I error of \(5\)\%. One strategy could be to widen the CI according to a Bonferroni approach.

Here an alternative approach is presented. The intercept and slope pairs \((\beta_{0};\beta_{1})_{i}\) obtained through the usual bootstrap process are combined into a joint multivariate distribution \citep{Fox2002bootstrap}. The resulting multivariate distribution is Gaussian if the underlying data is homoscedastic and show no anomalies like outliers or ties. Unfortunately in most real cases the ideal multivariate distribution is not met. Thus, the determination of the covariance matrix should be performed safely with a robust method.

A lot of robust covariance matrix determinations are known: the Donoho - Stahel \citep{Todorov2009} and the MCD \citep{rousseeuw1999} algorithm proposed here provide a very good alternative. The results obtained with these robust methods are compared with those obtained with the classical covariance.

The following joint ellipse (JE) procedure is proposed. After combining the \(\beta_{1}\) and \(\beta_{0}\) into a multivariate distribution via robust covariance matrix determination of the bootstrapped \(t^{*}\), the Mahalanobis distance of the \(H_{0}:(0;1)\) point from the center of the bootstrapped samples can be calculated. The Mahalanobis distance follows a \(\chi^2\) distribution with \(2\) degree of freedom. Hence, the final test takes place only once in a combined fashion.

The visual part of the validation, plotting the \(\beta\)'s in a 2D graph and adding the covariance ellipses and a rectangular box representing the traditional CI boundaries, provides a great overview for the comparison with the position of the \(H_{O}\) point. Moreover, the coordinates of the center of the covariance ellipse could be seen as a good and balanced values of the desired estimates.

Since all method comparisons rely on a \(H_{0}\) confirmation and not on a \(H_{0}\) rejection process, such a 2D-plot could get an important place, similarly as the visual inspection with Q-Q plots is the preferred normality assessment method. Thus, the power simulation that is discussed in this research should not be overrated and the reader should keep in mind that the main goal of this research is to provide a new graphical tool to approach method comparison problems; the role of the calculated probabilities for the validations should be reduced to mere auxiliary values. In fact the ``absence of evidence is not evidence of absence'' approach should prevail and the emphasis on p-values should be reduced.

This work is based on the use of the R software. Several packages provide method comparison tools. Among them \{mcr\}, \{deming\}, \{MethComp\}, \{groc\}. This study has been simulated and calculated using mainly the \{mcr\}\footnote{A fork of the \{mcr\} package has been created and is going to be soon installable from GitHub via install\_github(``piodag/mcr''). It provides the graphical box and ellipses plot and also the new M- and MM-Deming regressions.} package. Moreover the \{rrcov\} package has been used for the robust covariance matrix determinations.

\hypertarget{robust-deming-regressions}{%
\subsection{Robust Deming regressions}\label{robust-deming-regressions}}

There is a vast literature about the robustification of linear regressions. Starting from the sixties the germinal work of Huber has grown into a multitude of methods, from the original M \citep{huberronchetti2009} regression to the development of the MM \citep{yohai1987high} methods and to the most recent SMDM evolution \citep{koller2013robust}. For the orthogonal regression a robust variant is also known \citep{fekri2004}. Quite surprisingly, to the knowledge of the author, there has been no robustification attempt for the Deming regression, although a weighted variant is already documented since 1990 \citep{Linnet1990}. Perhaps the availability of the consolidated non-parametric Passing-Bablok \citep{Passing1984} regression has reduced the interest in an more ``classical'' robustified Deming variant.

\hypertarget{the-m-deming-mdem}{%
\subsubsection{The M-Deming (MDem)}\label{the-m-deming-mdem}}

The robustification of the Deming regression with M weights is very straightforward because the only coding work needed is to rewrite the algorithm for the calculation of the weights in the weighted Deming (WDem) regression proposed by Linnet \citep{Linnet1990}. The limit of the non descending part of the Huber function has been kept from the original work \citep{huberronchetti2009}. An optimization of this parameter falls outside the scope of this work, although a deeper insight would be very interesting. In fact in the M-Deming (and also in the MM-Deming) regression the weights are applied twice, once for the \(X_{i}\) and once for the \(Y_{i}\). It would be more appropriated to talk about a \(M^{2}\) (and \(MM^{2}\)) Deming regression and for this reason an optimization could be beneficial. The detailed algorithm is readable in the R package.

\hypertarget{the-mm-deming-mmdem}{%
\subsubsection{The MM-Deming (MMDem)}\label{the-mm-deming-mmdem}}

The recursive part of the MM-Deming regression is easily implemented and consists in a plugin of the Tukey bisquare redescending function on the standardized euclidean residuals. The most challenging part, which is still under development, is the determination of the starting values. An S-estimator borrowed from classical linear robust regression is not appropriate because of the ``regression to the mean'' effect \citep{pmid9261910, blandaltman1994}. Indeed, such a starting value would have a systematically too low slope. To prevent this the choice fell on the use of an S-estimator for the covariance matrix obtained with the Sest algorithm \citep{salibian2006fast} available in the \{rrcov\} R package. By using this method it is possible to calculate a ``principal component like'' slope, which is the average of the two possible slopes obtained keeping once \(X\) and then \(Y\) as independent variable. The center of the multivariate distribution is then used to estimate the intercept. Finally the mean of the absolute deviances from the starting model is used to obtain the scale parameter for the MM iteration.

This procedure is mostly stable. Unfortunately with small samples, especially if a nested bootstrap is applied, some occasional singularities are met. For this reason the proposed algorithm contains also a fallback robust covariance starter which uses the Rocke algorithm \citep{rocke1996identification} (also available in \{rrcov\}) activated when the default Sest algorithm fails to deliver usable starting values. It is worth noting that usually with samples \(n>100\) there would be no need for such an implementation, and problems starts to show up with \(n<50\) samples.

Also in the case of the MMDem regression no optimization of the parameters has been performed and the usual values are used, like for the M case.

\hypertarget{methods}{%
\section{Methods}\label{methods}}

\hypertarget{monte-carlo-simulation-models}{%
\subsection{Monte Carlo simulation models}\label{monte-carlo-simulation-models}}

The Monte Carlo power simulation has been performed on four different simulated data sets, according to the most common practical situations\footnote{Some additional data sets generation mechanisms are discussed later for the study of the method validation with heteroscedastic data.}.

\begin{figure}

{\centering \includegraphics{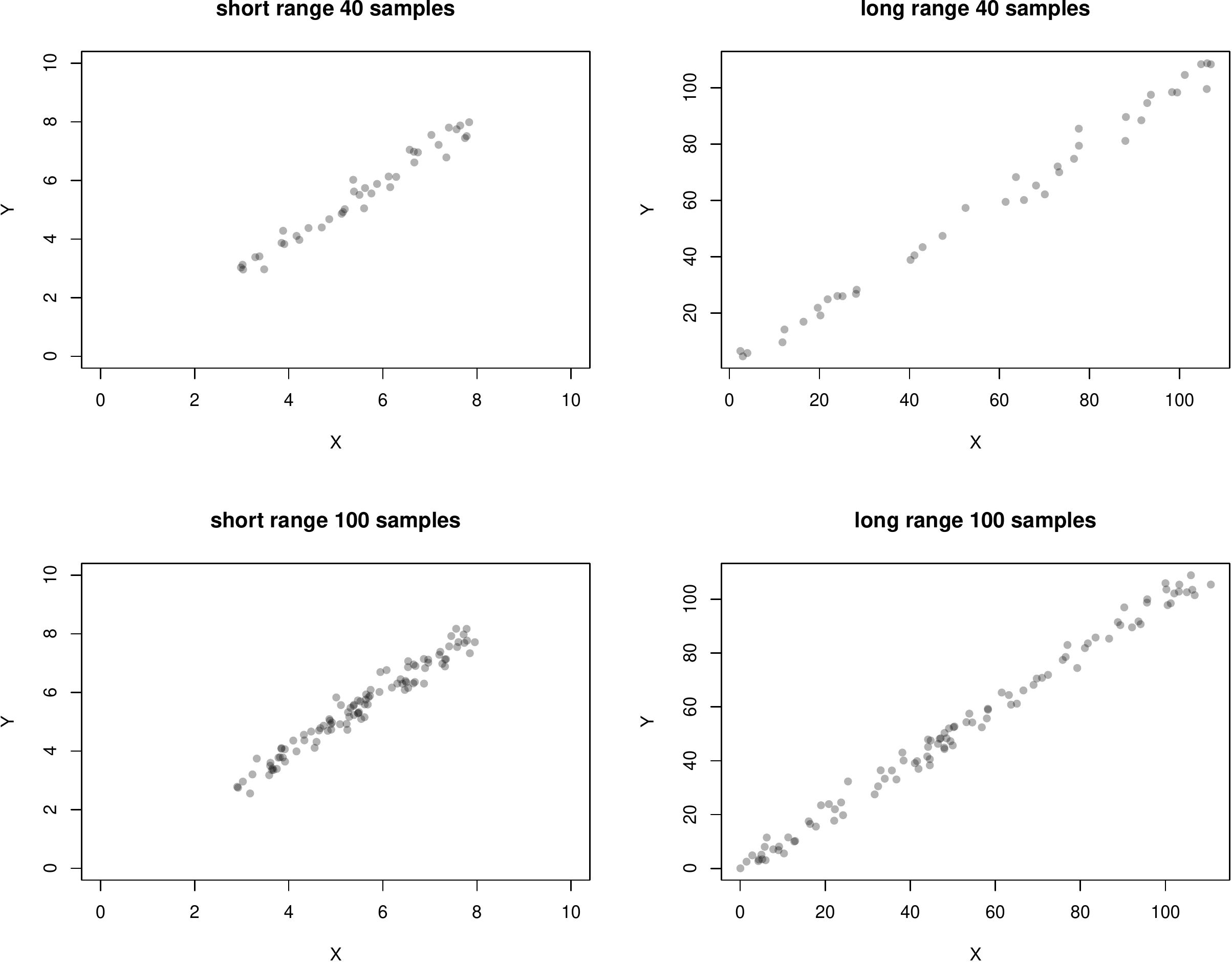} 

}

\caption{Short and long random data set examples.}\label{fig:dsetmods}
\end{figure}

Usually a method validation is done using a 40 bare minimal samples data set or a more ideal 100 samples data set \citep{Pum2019rev}. Moreover, there is a large known power difference that arises from the range of variation of the data. Thus, two fictional ranges has been used. A short one with values between 3 and 8 and a long one with values ranging from 0 to 110. The choice of these limits is totally arbitrary. The four combinations are shown in Fig. \ref{fig:dsetmods}.

In the short range model the data spans across half of a tenth power and the starting point is not too near to the zero. This case is commonly considered a difficult one for method validations; usually a very large data set is recommended because of a known lack of power of the regressions in such a narrow range of values.

In the long range model the data spans over two tenth power and is allowed to be near zero. To better simulate a real case, values lower than 0.1 (a fictitious detection limit of the instrument) has been set to a mean class value of 0.05. This kind of approximation is very frequent in practice. The additive normal error terms for \(X_{i}\) and \(Y_{i}\) (both are random variables) have been kept approximately similar between the long and the short data set, preserving the same \(\frac{\mu_{x_{i}}}{\sigma_{x_{i}}}\) ratio in the center of the data. The R code for the sample generation is available in the appendix \ref{DataGenFunc}. It is worth noting that every generated data set has been analyzed with all the regression methods in parallel so that the different approaches have always been applied on the same data set.

\hypertarget{the-type-i-error}{%
\subsection{The type I error}\label{the-type-i-error}}

The first stage in this study consists in finding an \(\alpha\) type I error for the \(\chi^2\) distribution used in the JE method that gives comparable results with the classical CI based hypothesis testing on samples with slope \(\beta_{1}=1\) and intercept \(\beta_{0}=0\) (ideal samples). A first Monte Carlo comparison of the empirical type I error against the theoretical type I error (the \(\alpha\) values) has been performed.

Simulated samples (10'000) have been generated and for each of them the estimates of the bootstrapped\footnote{For the Deming regression family the \(\lambda = \frac{\sigma_{X}}{\sigma_{Y}} = 1\) has been kept constant.} regressions were recorded. The robust covariance matrix of the slope / intercept pairs of the bootstrapped estimates was subsequently calculated with three methods (SDe, MCD and the classic covariance) and a \(\chi^2\) p-value for the joint \(H_0:(\beta_{0}=0,\beta_{1}=1)\) hypothesis has been determined for each of them.

\begin{figure}

{\centering \includegraphics{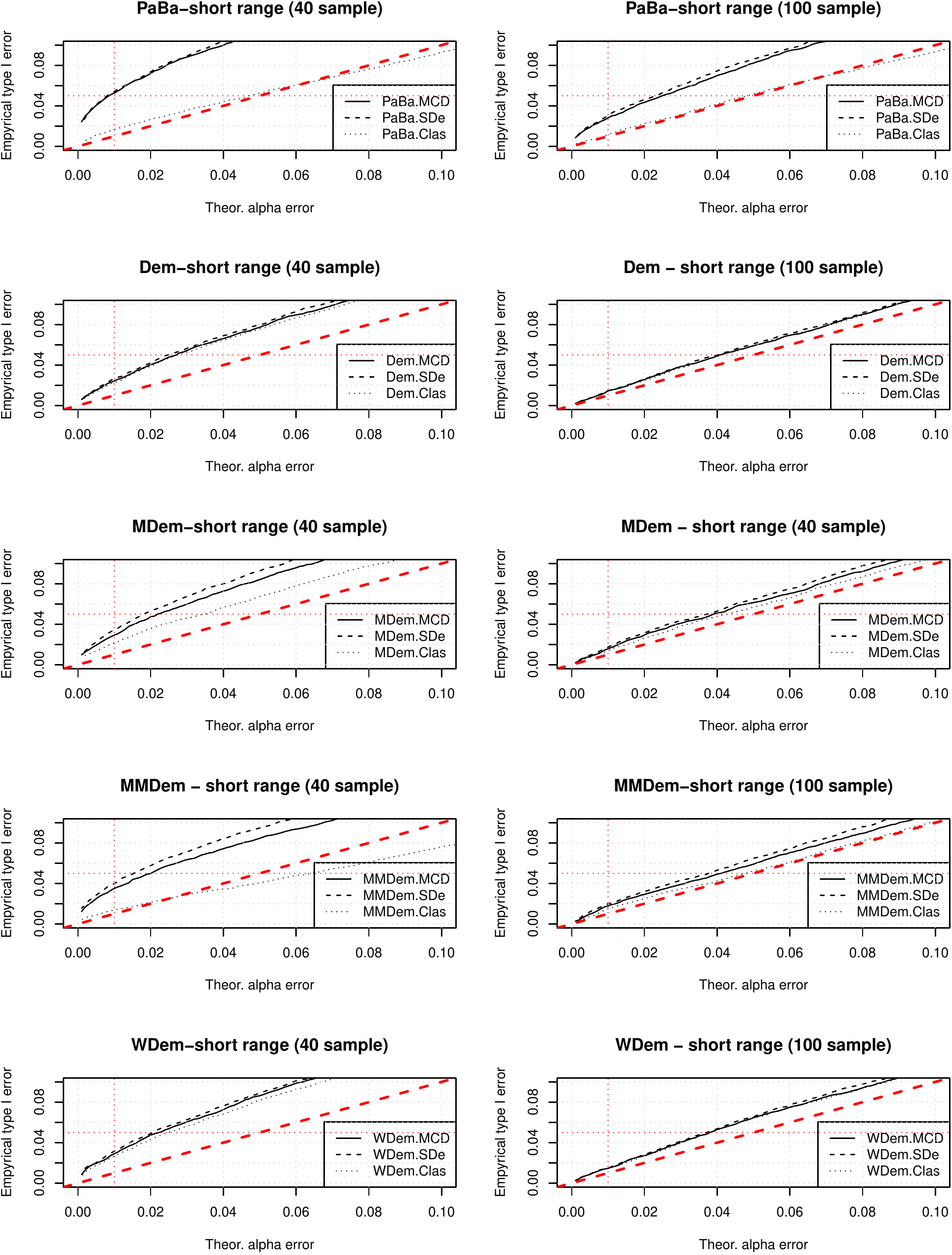} 

}

\caption{P-P plot with short range data comparing 40 and 100 samples data set results.}\label{fig:rejectionsshort}
\end{figure}

The empirical type I error (total rejection ratio) is plotted against the \(\alpha\) value (type I error) of the \(\chi^2\) used for the discrimination in a P-P plot. All methods show decent linear behavior, but none of them perfectly follow the identity line of the P-P plot, remarkably not even using the classical non-robust covariance matrix estimation (see Fig. \ref{fig:rejectionsshort} for the 100 samples experiment).

Indeed, the empirical rejection rate obtained with the classical covariance follows the identity line pretty well only for the PaBa regression and for the MMDem but only with the long range data set. The intuitive idea that rejections calculated with the non-robust Deming regressions combined with the classical covariance should follow the P-P identity line is surprisingly not observed.

The biggest deviation from the identity line is found with the PaBa regression for which a theoretical alpha of 0.01 gives approximately a 0.05 empirical rejection rate when a robust covariance matrix is applied.

The results for the non-robust Deming regression with the three different covariance estimations differs only very slightly, and the same happens for the WDem which is also non-robust. For the robust M- and MM-Deming regression the use of a robust covariance matrix leads to higher rejection rates compared to the classical covariance, but not as high as for the PaBa.

Apparently the MCD covariance method shows a reduced deviation compared to the SDe one, although the difference seems to be very moderate in all cases.

The number of samples plays also a role. With bigger samples the deviation from the identity line is greatly reduced. Only the PaBa regression still shows a notable deviation (see Fig. \ref{fig:rejectionsshort}).

\begin{figure}

{\centering \includegraphics{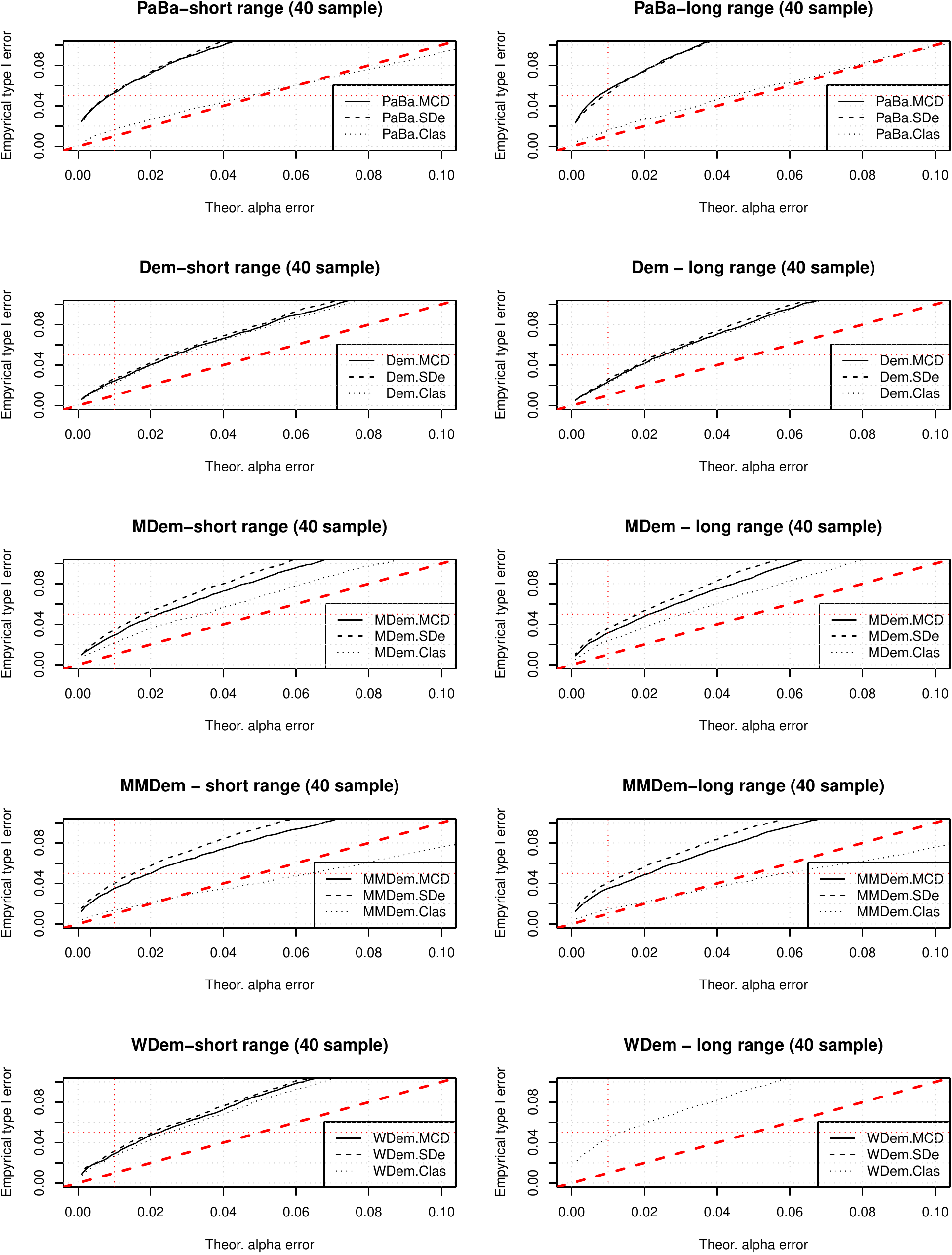} 

}

\caption{P-P plot comparing 40 samples data sets results with short and long range.}\label{fig:rejectionssl}
\end{figure}

The length of the data range has a limited effect as shown in Fig. \ref{fig:rejectionssl}. What is observed is that the WDem deviates very strongly. But in this case the WDem regression suffers the presence of low value outliers generated by the detection cut off that has been introduced in the simulation.

All these deviations from the identity in the P-P plot are a bit counterintuitive because the Donoho Stahel and the MCD procedures in R already contain a correction factor to provide adherence to the \(\chi^{2}\) distribution and the classic covariance should adhere per definition.

In appendix \ref{RejectionLong} the complete rejection plots for the long range (40 and 100 samples) data sets are reported.

Hardin and Rocke \citep{hardin2005distribution} proposed an alternative approach to optimize the p-values of Mahalanobis distances using the \emph{F} distribution with an empirical correction of the \emph{df2}. An extensive attempt to use this method has been performed. Unfortunately it was not possible to find a constant asymptotically corrected \emph{df2} value (also called \(m\) in Hardin's paper). In fact the \(\hat{m}\) (estimated through nonlinear fitting) led to very different values depending on the dimension of the underlying data set (number of points) on the range of the data set and also on the regression method\footnote{This part of the work is not reported but available on request.}.

Anyway, the deep aim of this study is to shift the validation method paradigm more toward the use of a graphical approach. Thus, a strict correction of the \(\chi^{2}\) p-values is not so urgent and the obtained JE probability is only an help to have a better readability of the 2D-plots of the bootstrapped samples.

Analyzing the P-P plots it is manifest that a choice of \(\alpha = 0.01\) for the \(\chi^2\) distribution seems to be safe. As shown in the following paragraphs, the JE method has such an higher power, compared to the traditional CI method, that even testing at \(\alpha=0.01\) gives a huge gain.

The traditional CI based rejection ratios has been also simultaneously recorded in parallel to the JE ones. The global type I empirical error with the traditional CI method is always found higher than \(0.05\) with \(\alpha=0.05\) (see Tab. \ref{tab:type1tab}). The effect of the two steps hypothesis testing without a FWER correction can be reinforced by the fact that the BCa determined CI are commonly considered slightly conservative.

\begin{table}

\caption{\label{tab:type1tab}Type I error table for JE and classical CI methods.}
\centering
\resizebox{\linewidth}{!}{
\begin{tabular}[t]{lcccccccc}
\toprule
  & samples & int.CI5\% & sl.CI5\% & tot.CI5\% & JE.Clas5\% & JE.MCD5\% & JE.SDe5\% & JE.MCD1\%\\
\midrule
PaBa-long & 40 & 0.9474 & 0.9509 & 0.9192 & 0.9445 & 0.8816 & 0.8813 & 0.9442\\
PaBa-short & 40 & 0.9535 & 0.9568 & 0.9409 & 0.9473 & 0.8873 & 0.8830 & 0.9473\\
Dem-long & 40 & 0.9299 & 0.9329 & 0.8963 & 0.9187 & 0.9169 & 0.9130 & 0.9765\\
Dem-short & 40 & 0.9362 & 0.9402 & 0.9232 & 0.9244 & 0.9228 & 0.9201 & 0.9758\\
MDem-long & 40 & 0.9391 & 0.9476 & 0.9130 & 0.9280 & 0.9135 & 0.9040 & 0.9685\\
MDem-short & 40 & 0.9508 & 0.9539 & 0.9404 & 0.9325 & 0.9159 & 0.9073 & 0.9708\\
MMDem-long & 40 & 0.9448 & 0.9564 & 0.9235 & 0.9565 & 0.9140 & 0.9039 & 0.9649\\
MMDem-short & 40 & 0.9566 & 0.9587 & 0.9466 & 0.9594 & 0.9154 & 0.9052 & 0.9649\\
WDem-long & 40 & 0.8753 & 0.8275 & 0.7869 & 0.9062 & 0.4856 & 0.5223 & 0.5418\\
WDem-short & 40 & 0.9164 & 0.9226 & 0.9008 & 0.9180 & 0.9136 & 0.9110 & 0.9710\\
PaBa-long & 100 & 0.9508 & 0.9506 & 0.9231 & 0.9491 & 0.9189 & 0.9121 & 0.9721\\
PaBa-short & 100 & 0.9525 & 0.9544 & 0.9395 & 0.9485 & 0.9177 & 0.9138 & 0.9719\\
Dem-long & 100 & 0.9450 & 0.9445 & 0.9158 & 0.9388 & 0.9390 & 0.9365 & 0.9841\\
Dem-short & 100 & 0.9460 & 0.9463 & 0.9333 & 0.9405 & 0.9410 & 0.9390 & 0.9853\\
MDem-long & 100 & 0.9481 & 0.9484 & 0.9235 & 0.9420 & 0.9385 & 0.9360 & 0.9827\\
MDem-short & 100 & 0.9508 & 0.9505 & 0.9378 & 0.9430 & 0.9386 & 0.9352 & 0.9842\\
MMDem-long & 100 & 0.9535 & 0.9536 & 0.9308 & 0.9479 & 0.9421 & 0.9379 & 0.9826\\
MMDem-short & 100 & 0.9553 & 0.9551 & 0.9430 & 0.9476 & 0.9410 & 0.9351 & 0.9821\\
WDem-long & 100 & 0.9016 & 0.8776 & 0.8479 & 0.9200 & 0.6151 & 0.6411 & 0.6725\\
WDem-short & 100 & 0.9307 & 0.9320 & 0.9158 & 0.9357 & 0.9353 & 0.9345 & 0.9854\\
\bottomrule
\end{tabular}}
\end{table}

Deming's regression CI validation shows higher type I error than the validation performed with the PaBa and MDem regressions. The difference is more important in case of small sample sizes. In general the JE method at 0.01 show empirical type I error smaller than 0.05 with the only exception of the PaBa regression in which case the values are just below 0.95 (and of the distorted WDem on the long range experiment). Thus, a 0.01 level for the JE method is confirmed as a good compromise.

\clearpage

\hypertarget{type-ii-error-and-power-comparison-the-slope}{%
\subsection{Type II error and power comparison: the slope}\label{type-ii-error-and-power-comparison-the-slope}}

Chosen \(\alpha = 0.01\) for the JE method it is possible to compare the statistical power offered by this method in comparison to the power of the traditional CI validation process. For this purpose a Monte Carlo simulation study has been constructed using the same previous data generating functions and recording the empirical rejection rate (which is configured as a Type II error unless the exact \(H_{0}\) value is met) in function of the varied parameter (slope and intercept). For each chosen value of slope (and intercept) 400 randomly generated models were each evaluated through bootstrap for the empirical rejection.

\hypertarget{CovMatrixMeth}{%
\subsubsection{Comparison of the method for covariance matrix determination}\label{CovMatrixMeth}}

A first set of graphs is drawn in order to compare the different robust covariance matrix algorithms with the non-robust method. The model with a short range sample and 40 data pairs is used as reference. It is the model where the strongest differences between classical CI and JE methods are observed. The two main observations are the peak height at slope \(0\) (type I error, can be considered an independent check of the data in Tab. \ref{tab:type1tab} obtained through the curve fitting described in the Paragraph \ref{SimDataFit}) and the level which corresponds to a power of 80\%, thus a 20\% empirical rejection rate. These levels are shown with a dashed green line in the charts. For sake of completeness, the data of the ellipses at \(\alpha=0.05\) are also reported.

\begin{figure}

{\centering \includegraphics{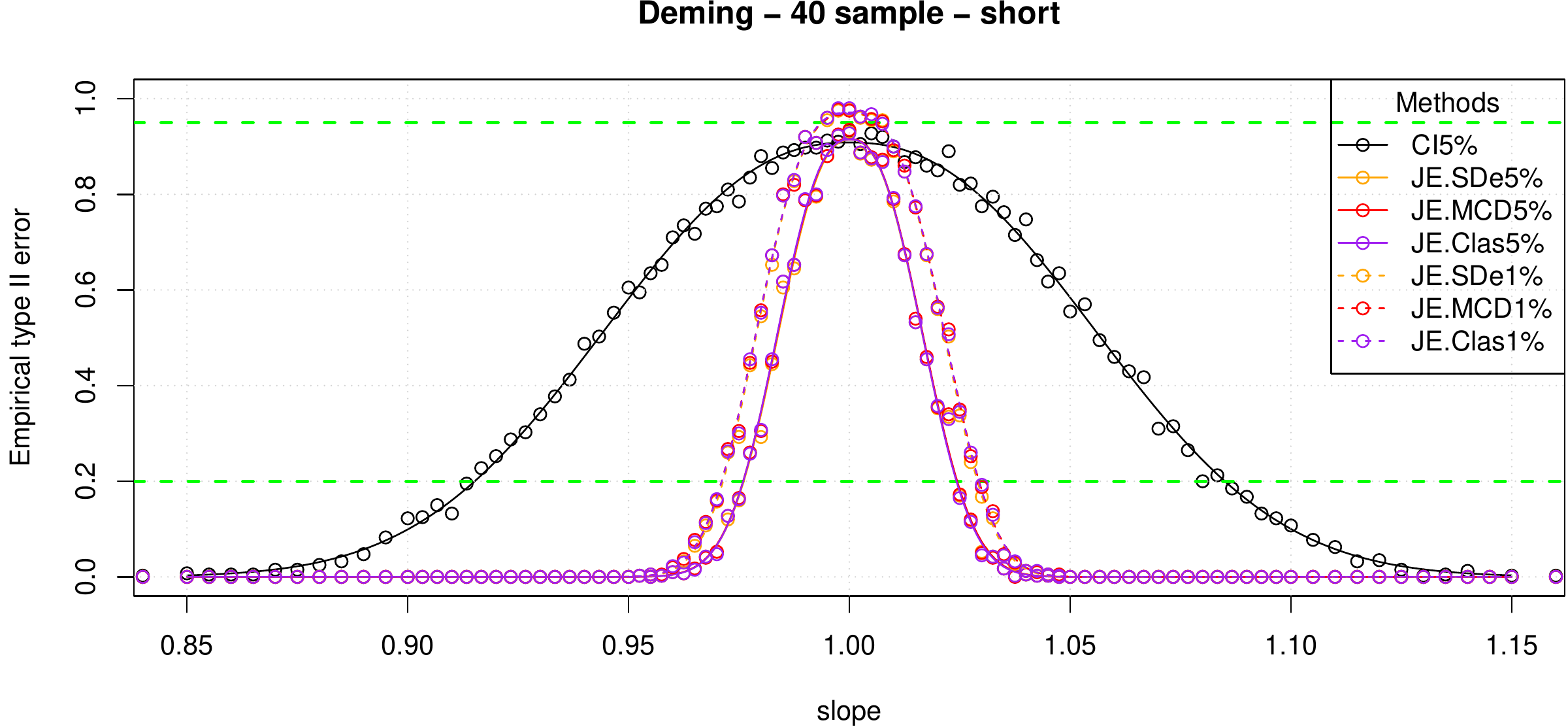} 

}

\caption{Classical, SDe and MCD covariance matrix at 5\% and 1\% levels for the non-robust Deming regression.}\label{fig:dem40short}
\end{figure}

\begin{figure}

{\centering \includegraphics{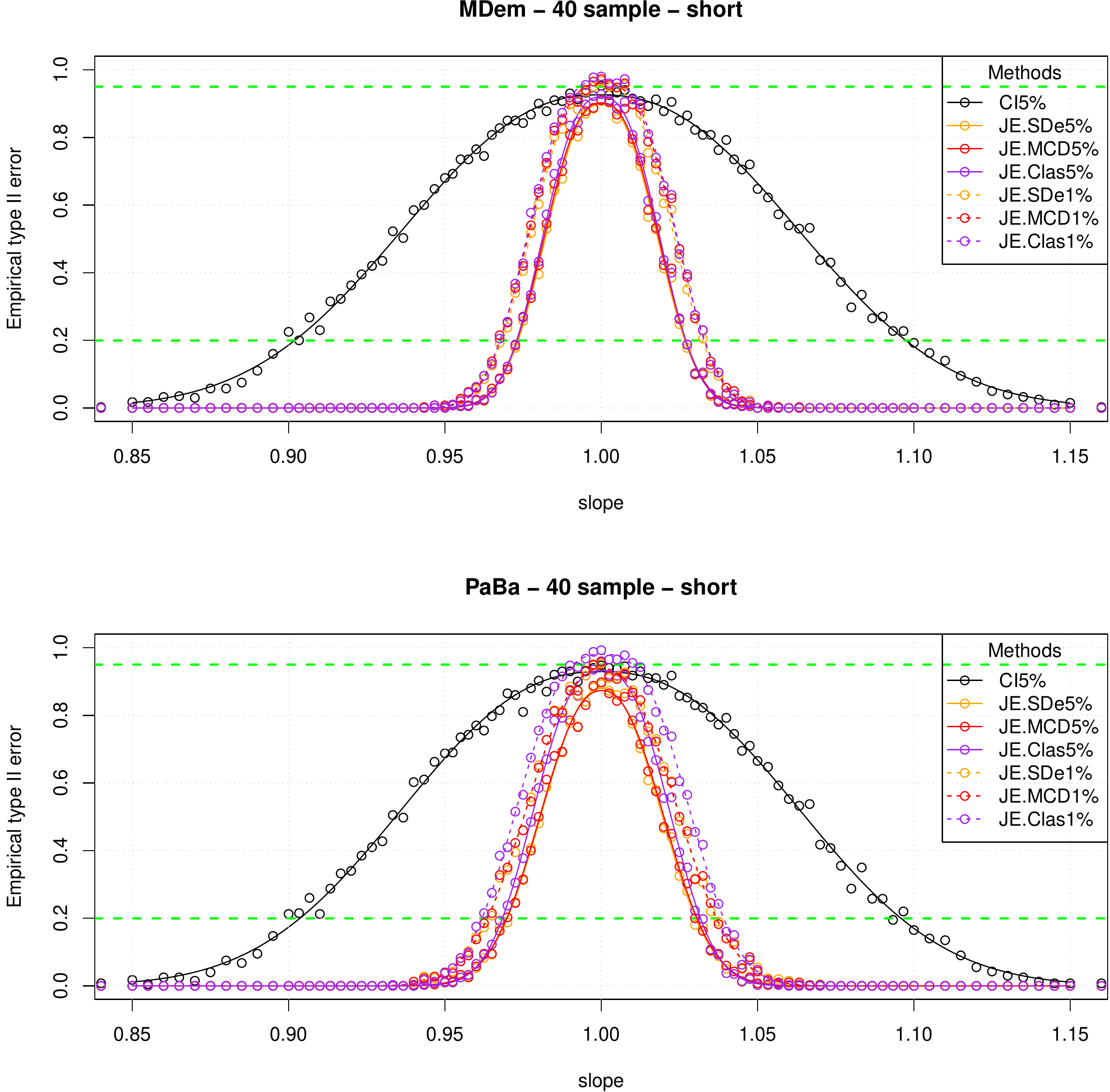} 

}

\caption{Classical, SDe and MCD covariance matrix at 5\% and 1\% levels: MDem and PaBa regressions.}\label{fig:mdempaba40short}
\end{figure}

\begin{figure}

{\centering \includegraphics{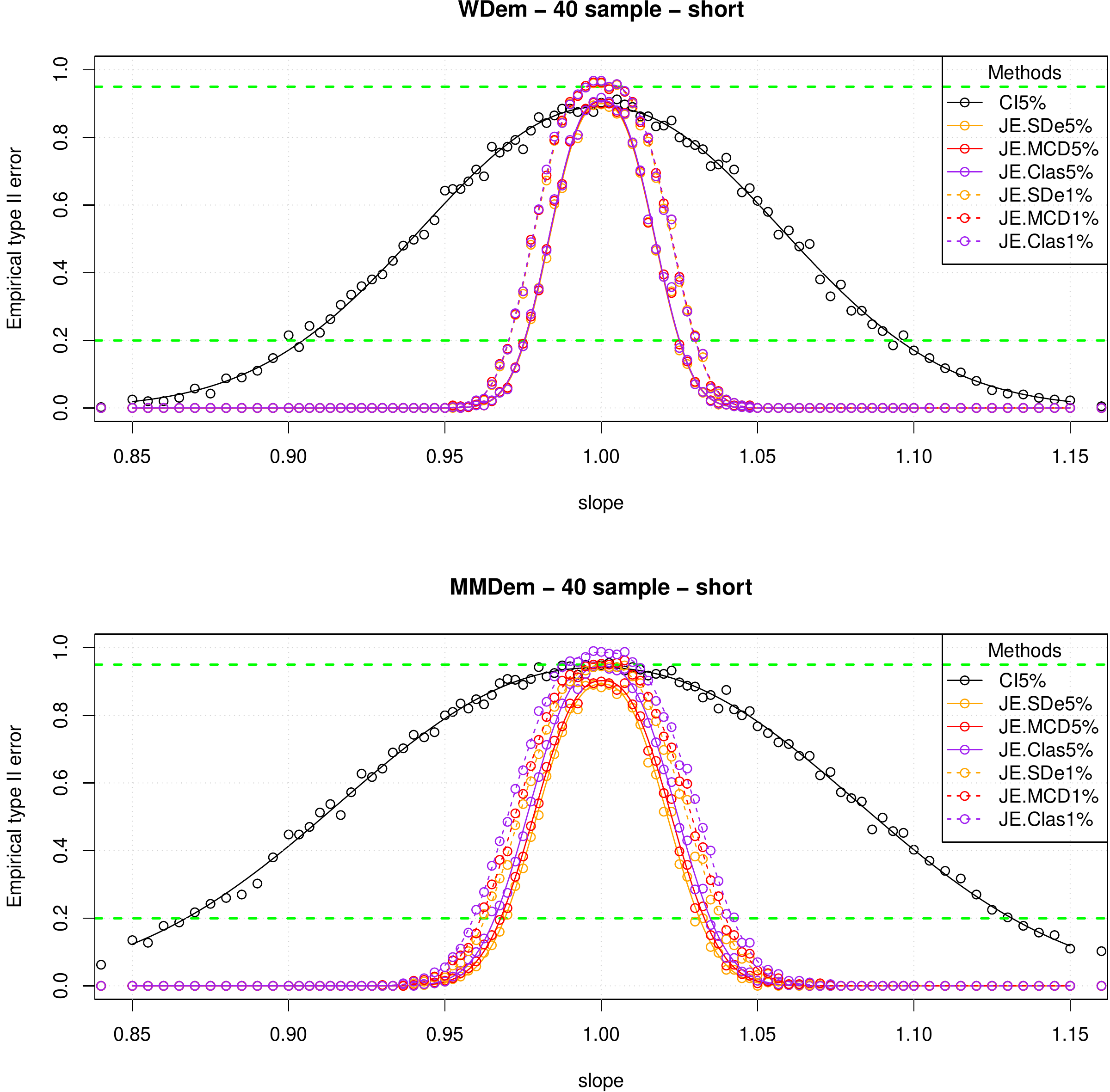} 

}

\caption{Classical, SDe and MCD covariance matrix at 5\% and 1\% levels: WDem and MMDem regressions.}\label{fig:wdemmmdem40short}
\end{figure}

For the non-robust Deming (Fig. \ref{fig:dem40short}) (and also the WDem (Fig. \ref{fig:wdemmmdem40short})) regression there is almost no difference using all three covariance matrix determination methods. For the MDem (Fig. \ref{fig:mdempaba40short}) regression the differences are very weak; they are more marked for the MM and the PaBa regressions (Fig. \ref{fig:wdemmmdem40short} and \ref{fig:mdempaba40short}). In general the results obtained with SDe and the MCD covariances are almost indistinguishable; only in the case of the MMDem regression the MCD method seems to be slightly but consistently more tolerant\footnote{For this reason in the rest of the work, when the readability of the plots will be a concern, only the MCD results will be plotted.}. For a numerical estimation of these differences see Tab. \ref{tab:40shorttable} in Paragraph \ref{SimDataFit}. The main differences are registered between the classical and the robust covariances.

The power of the JE method largely exceed the power of the traditional CI method; the rejection curves are much narrower. Thus, the choice of a more conservative \(\alpha=0.01\) level is fully motivated.

\hypertarget{effect-of-sample-size-and-sample-range}{%
\subsubsection{Effect of sample size and sample range}\label{effect-of-sample-size-and-sample-range}}

\begin{figure}

{\centering \includegraphics{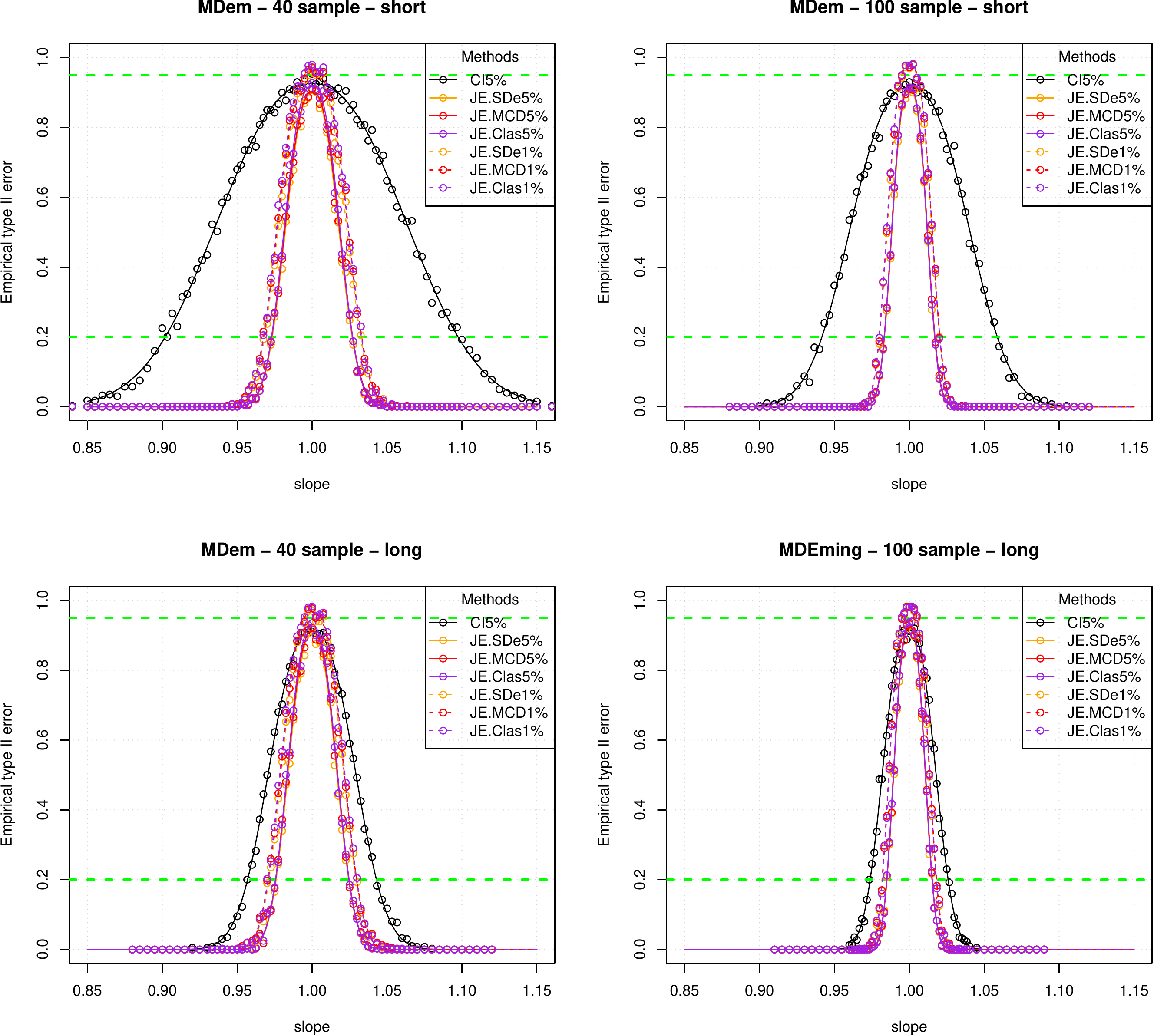} 

}

\caption{Classic CI vs JE type II errors for Passing Bablok regression.}\label{fig:mcompare}
\end{figure}

The plots in Fig. \ref{fig:mcompare} summarizes the simulation results which were performed on long range and on short range data sets each of them for small (40 pairs) and large (100 pairs) sample sizes using the MDem regression method.

\begin{figure}

{\centering \includegraphics{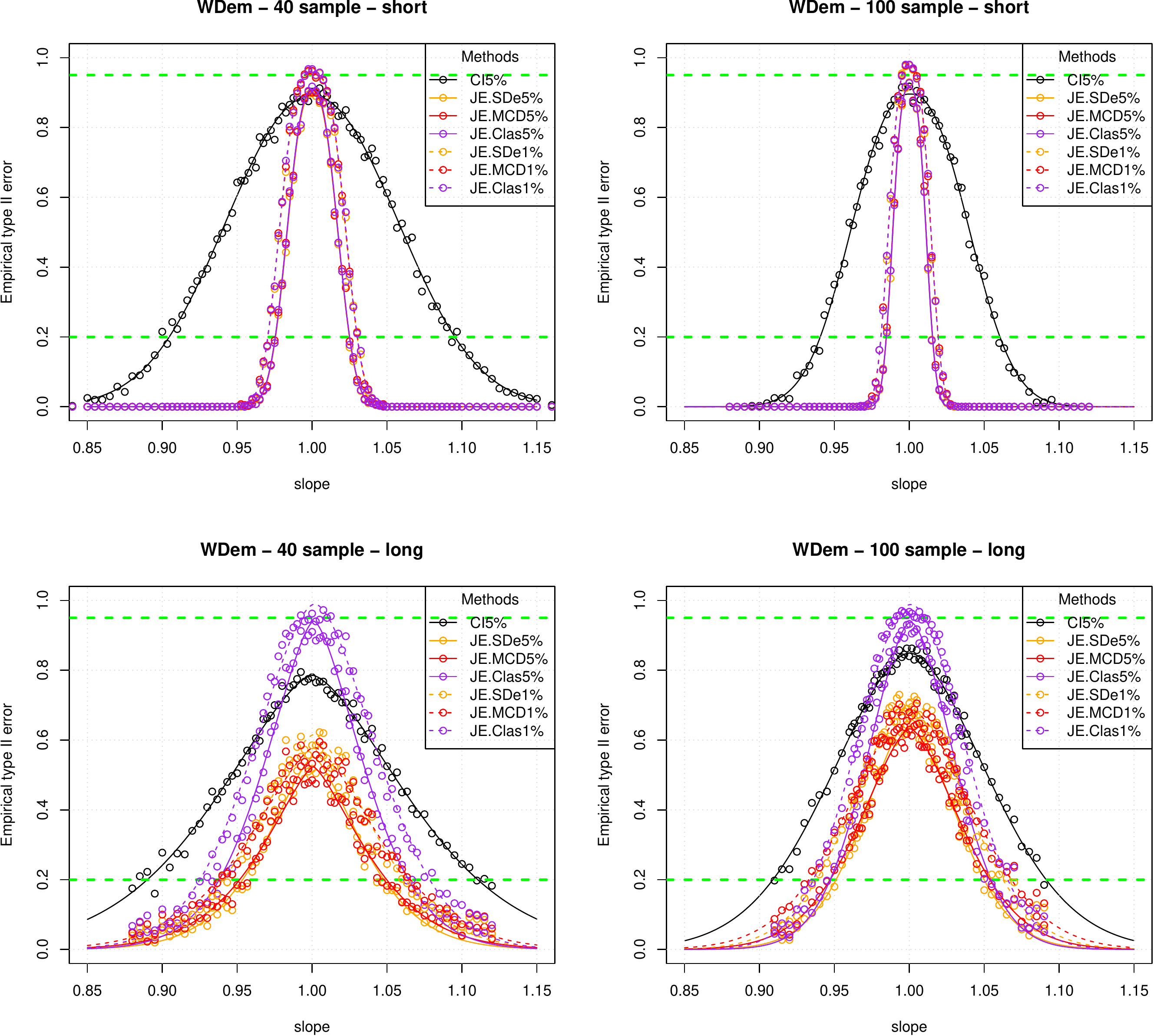} 

}

\caption{Classic CI vs JE type II errors for Passing Bablok regression.}\label{fig:wcompare}
\end{figure}

All the other regressions show the same trends for the sample range and the sample size, except for the WDem on the long range data (40 and 100 data points). In this case the irregularities generated by the simulation detection limits of the methods leads to broad and unreliable curves as shown in the Fig. \ref{fig:wcompare}.

The analogous plots for the other regressions (Deming, Paba and MMDem) are reported in the Appendix \ref{SampleRange}.

The JE method at \(\alpha\) = 0.01 applied on the bootstrap sample of the regression outperforms the classical CI method in every analyzed situation (see Fig. \ref{fig:mcompare}). The JE empirical type II error curves are narrower; the minimal slope difference from \(1\) that can be detected with a power of 80\% is smaller. Moreover, with the classical CI method in most cases even at slope = \(1\) the total empirical rejection is greater than 0.05, confirming once again the results presented in the precedent paragraphs, and anyway greater than the one of the JE method. Thus, the classical CI testing has a higher likelihood to reject legitimate validations and, at the same time, has less power to detect an alternative slope.

The data set with a short range show greater differences than the data set with long range data. In case of a short range method validation the use of the JE method would lead to a major benefit.

In general large data set show increased power for both the CI method and the JE method. With the available simulations is not possible to understand if the power of the two methods asymptotically converge at very large sample sizes.

\hypertarget{SimDataFit}{%
\subsubsection{Simulation data fit with the exponential power distribution}\label{SimDataFit}}

All the previous figures report the empirical type II errors against the varied parameter (the slope and, later on, the intercept) with a curve fitting the points. Attempts to fit the data with the most common non-parametric procedures were unsatisfactory because the bottom line was not stable, leading to negative fitted values. For this reason a parametric approach was chosen.

The empirical distributions for all the type II errors do not follow a pure Gaussian model and are much better modeled by an exponential power distribution \citep{Nadarajah2005} (also called Subbotin \citep{Subbotin1923} distribution). An additional \(\alpha\) parameter has been introduced for the density \(d\) scaling since the empirical curves are not normalized. Hence, the four parameters function has the following form:

\[ d\langle x,\alpha, \beta, \sigma, \mu \rangle = \alpha \cdot \frac{\beta}{2 \cdot \sigma \cdot \Gamma \langle \frac{1}{\beta} \rangle } \cdot exp \langle - (\frac{\big| x-\mu \big|}{\sigma})^{\beta} \rangle\]

The fitting was performed with the R nls() function. For the starting values, only the choice of \(\sigma\) was tricky. Too low or too high \(\sigma\) values leaded to a failure of the convergence. All the parameters were roughly estimated observing the plots; the \(\beta=2\) was a good starting value although almost all the fits showed a clear platykurtic character.

The profile traces (see Appendix \ref{NonLinOpt}) of the non-linear regressions showed a good independence of the parameters and the plots of the residuals did not show any additional structure. Here below in Fig. \ref{fig:residuals} an example of the residual analysis of the MDem total rejection ratio fitting for the classical CI testing at \(\alpha=0.05\) against the varying slope (short range 40 sample).

\begin{figure}

{\centering \includegraphics{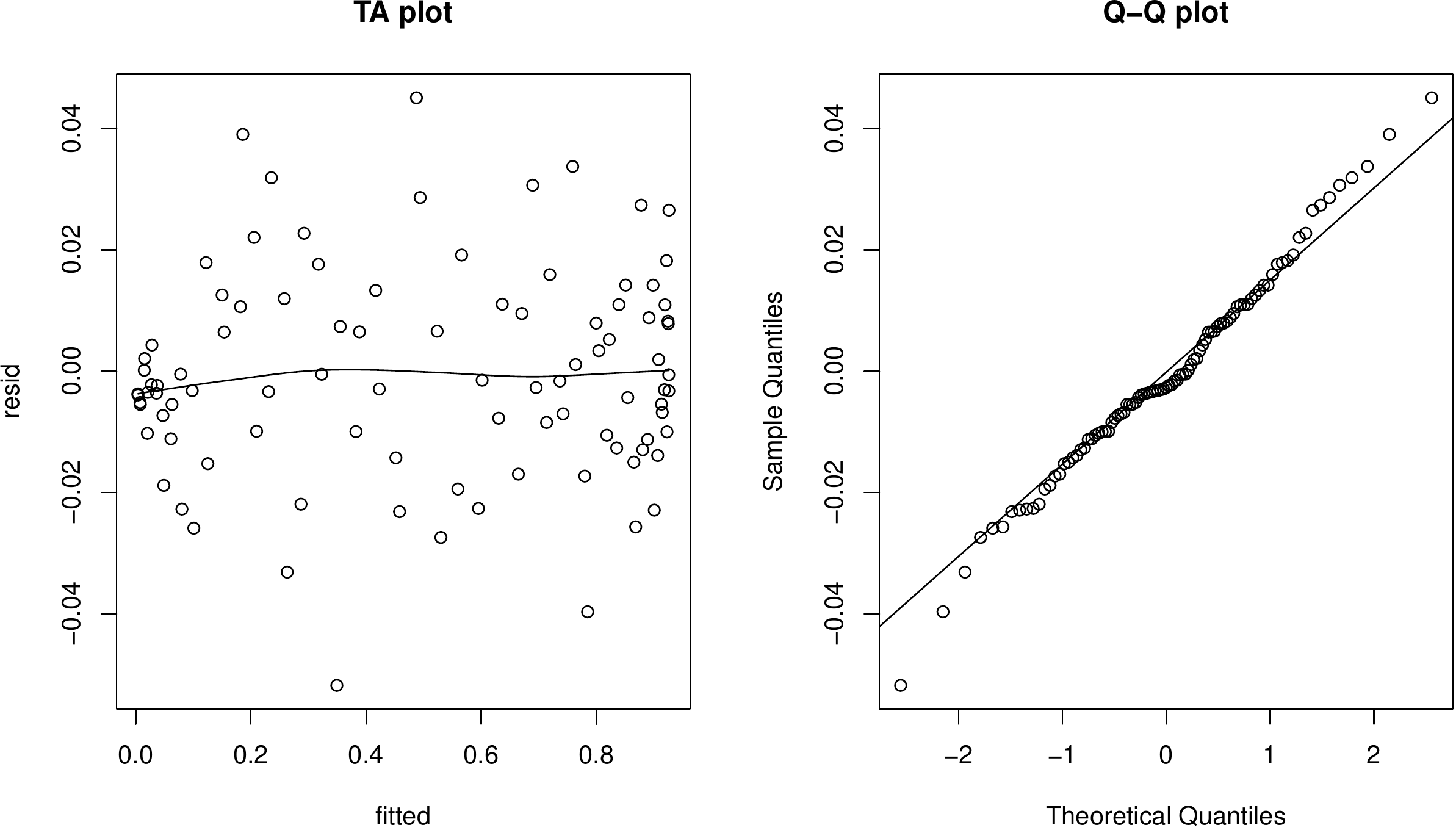} 

}

\caption{Residuals analysis example for the exponential power fit: MDem regression, classical CI type II error vs the slope in the 40 samples short range experiment.}\label{fig:residuals}
\end{figure}

The theoretical nature of these type II error distributions is unknown. Hence, the CI for the type I and II errors are just calculated for comparative reasons.

For the evaluation of the standard errors (not for the fitting) the gradient matrix has been calculated on a modified \(\tilde{d}\) removing the absolute value function from the original \(d\) and reducing the domain to \(D_{\tilde{d}} = (\mu ; +\infty)\).

\[ \nabla d = \nabla \tilde {d} \]
The gradient at \(x=\mu\) is also estimated as limit\footnote{In general for the slope \(\mu = 1\)} as shown here:

\[\nabla \tilde{d}\langle x= \mu \rangle = lim_{x \to \mu^{+}} \nabla \tilde {d}\langle x \rangle\]
with

\[ \tilde{d}\langle x | x>\mu,\alpha, \beta, \sigma, \mu \rangle = \alpha \cdot \frac{\beta}{2 \cdot \sigma \cdot \Gamma \langle \frac{1}{\beta} \rangle } \cdot exp \langle - (\frac{ x-\mu}{\sigma})^{\beta} \rangle \]

Using the nonlinear fitting with the above calculation tricks it is possible to compare the slope (and later the intercept) level at which a rejection power of 80\% is reached and, at the same time, it is also possible to estimate the type I error at slope \(x=1\), both of them with confidence intervals. For the determination of the power level a calibration strategy has been used, numerically inverting the fitting function and its CI levels\footnote{The numerical method for the determination of the CI of the calibration curve was borrowed from the lecture notes and exercises of Prof.~A. Ruckstuhl - ETH - Zürich.}.

The data in Tab. \ref{tab:mdemtable} helps with the comparison of the methods in function of data the range and sample size. The table confirms the visual inspection of the plots in Fig. \ref{fig:mcompare}.

\pagebreak

\begin{longtable}[]{@{}ccccccccc@{}}
\caption{\label{tab:mdemtable}Power and type I error comparison for the slope with the MDem regression.}\tabularnewline
\toprule
Range & sample & Method & p80\% LCI & p80\% est & p80\% UCI & T-I LCI & T-I est & T-I UCI\tabularnewline
\midrule
\endfirsthead
\toprule
Range & sample & Method & p80\% LCI & p80\% est & p80\% UCI & T-I LCI & T-I est & T-I UCI\tabularnewline
\midrule
\endhead
short & 40 & CI5\% & 1.0928 & 1.0974 & 1.1026 & 0.89057 & 0.92598 & 0.96140\tabularnewline
short & 40 & JE.SDe1\% & 1.0315 & 1.0323 & 1.0332 & 0.92968 & 0.95310 & 0.97651\tabularnewline
short & 40 & JE.MCD1\% & 1.0323 & 1.0331 & 1.0339 & 0.93752 & 0.95983 & 0.98215\tabularnewline
short & 40 & JE.Clas1\% & 1.0321 & 1.0329 & 1.0337 & 0.94615 & 0.96859 & 0.99103\tabularnewline
short & 100 & CI5\% & 1.0560 & 1.0588 & 1.0618 & 0.87659 & 0.91411 & 0.95163\tabularnewline
short & 100 & JE.SDe1\% & 1.0200 & 1.0203 & 1.0207 & 0.95780 & 0.97465 & 0.99150\tabularnewline
short & 100 & JE.MCD1\% & 1.0200 & 1.0204 & 1.0208 & 0.95745 & 0.97586 & 0.99427\tabularnewline
short & 100 & JE.Clas1\% & 1.0200 & 1.0204 & 1.0207 & 0.96287 & 0.97910 & 0.99533\tabularnewline
long & 40 & CI5\% & 1.0416 & 1.0429 & 1.0444 & 0.88799 & 0.91237 & 0.93676\tabularnewline
long & 40 & JE.SDe1\% & 1.0285 & 1.0293 & 1.0302 & 0.93556 & 0.96050 & 0.98543\tabularnewline
long & 40 & JE.MCD1\% & 1.0289 & 1.0297 & 1.0306 & 0.94262 & 0.96676 & 0.99090\tabularnewline
long & 40 & JE.Clas1\% & 1.0291 & 1.0298 & 1.0305 & 0.95086 & 0.97261 & 0.99436\tabularnewline
long & 100 & CI5\% & 1.0249 & 1.0259 & 1.0271 & 0.88127 & 0.91408 & 0.94689\tabularnewline
long & 100 & JE.SDe1\% & 1.0177 & 1.0180 & 1.0183 & 0.96414 & 0.98090 & 0.99767\tabularnewline
long & 100 & JE.MCD1\% & 1.0178 & 1.0182 & 1.0185 & 0.96334 & 0.98168 & 1.00002\tabularnewline
long & 100 & JE.Clas1\% & 1.0177 & 1.0181 & 1.0185 & 0.96279 & 0.98433 & 1.00587\tabularnewline
\bottomrule
\end{longtable}

\clearpage

The range of the tested values is important and the power increases with the longer range. The lack of power expected for short range data is manifest, especially in the small sample experiment. Once again, the classical CI method shows in general higher than desired Type I errors (more than 5\%). It is worth noting that some UCI for the type I error go beyond \(1\) and this has no practical sense.

In the Tab. \ref{tab:40shorttable} the focus is set on the different regression methods using the short range 40 samples experiment.

\begin{longtable}[]{@{}cccccccc@{}}
\caption{\label{tab:40shorttable}Power and type I error comparison for the slope with short range 40 samples simulations.}\tabularnewline
\toprule
Regression & Method & p80\% LCI & p80\% est & p80\% UCI & T-I LCI & T-I est & T-I UCI\tabularnewline
\midrule
\endfirsthead
\toprule
Regression & Method & p80\% LCI & p80\% est & p80\% UCI & T-I LCI & T-I est & T-I UCI\tabularnewline
\midrule
\endhead
Deming & CI5\% & 1.0811 & 1.0850 & 1.0894 & 0.87433 & 0.90834 & 0.94236\tabularnewline
Deming & JE.SDe1\% & 1.0284 & 1.0290 & 1.0297 & 0.94656 & 0.96790 & 0.98925\tabularnewline
Deming & JE.MCD1\% & 1.0287 & 1.0294 & 1.0301 & 0.94639 & 0.96875 & 0.99112\tabularnewline
Deming & JE.Clas1\% & 1.0286 & 1.0293 & 1.0300 & 0.95034 & 0.97118 & 0.99202\tabularnewline
MDem & CI5\% & 1.0928 & 1.0974 & 1.1026 & 0.89057 & 0.92598 & 0.96140\tabularnewline
MDem & JE.SDe1\% & 1.0315 & 1.0323 & 1.0332 & 0.92968 & 0.95310 & 0.97651\tabularnewline
MDem & JE.MCD1\% & 1.0323 & 1.0331 & 1.0339 & 0.93752 & 0.95983 & 0.98215\tabularnewline
MDem & JE.Clas1\% & 1.0321 & 1.0329 & 1.0337 & 0.94615 & 0.96859 & 0.99103\tabularnewline
PaBa & CI5\% & 1.0908 & 1.0959 & 1.1016 & 0.88803 & 0.92895 & 0.96988\tabularnewline
PaBa & JE.SDe1\% & 1.0353 & 1.0367 & 1.0382 & 0.90850 & 0.93751 & 0.96652\tabularnewline
PaBa & JE.MCD1\% & 1.0355 & 1.0369 & 1.0383 & 0.91024 & 0.93845 & 0.96666\tabularnewline
PaBa & JE.Clas1\% & 1.0381 & 1.0391 & 1.0402 & 0.95651 & 0.98045 & 1.00440\tabularnewline
MMDem & CI5\% & 1.1253 & 1.1313 & 1.1381 & 0.90772 & 0.94090 & 0.97407\tabularnewline
MMDem & JE.SDe1\% & 1.0372 & 1.0383 & 1.0394 & 0.92305 & 0.94758 & 0.97211\tabularnewline
MMDem & JE.MCD1\% & 1.0388 & 1.0398 & 1.0409 & 0.93230 & 0.95544 & 0.97859\tabularnewline
MMDem & JE.Clas1\% & 1.0411 & 1.0420 & 1.0430 & 0.96489 & 0.98605 & 1.00721\tabularnewline
WDem & CI5\% & 1.0900 & 1.0953 & 1.1013 & 0.85356 & 0.89079 & 0.92803\tabularnewline
WDem & JE.SDe1\% & 1.0295 & 1.0300 & 1.0307 & 0.93664 & 0.95482 & 0.97300\tabularnewline
WDem & JE.MCD1\% & 1.0297 & 1.0303 & 1.0309 & 0.93734 & 0.95569 & 0.97405\tabularnewline
WDem & JE.Clas1\% & 1.0295 & 1.0301 & 1.0308 & 0.93739 & 0.95793 & 0.97848\tabularnewline
\bottomrule
\end{longtable}

For the long range 100 sample data set the analogous results are in Tab. \ref{tab:100longtable}. The other tables are in Appendix \ref{ComparisonTables}.

The highest power is attained with the non-robust Deming regression both comparing the CI method and also the JE methods among them. The WDem shows similar results. The PaBa and the MDem regressions show no significant difference for the classical CI method; using the JE method gives instead a small but significant power advantage to the MDem regression. The MMDem shows the lowest power in all the situations.

The role of the covariance matrix determination seems to be pretty marginal as already observed in Paragraph \ref{CovMatrixMeth}, although some of them are significant. In particular for Deming and MDem there is no significant difference and for MMDem and PaBa there is a significant difference between the robust and the classical covariance. Apparently the more ``redescending'' the regression is, the bigger the divergence caused by the method of the covariance matrix determination is.

Going from 40 to 100 samples the power for all methods and regressions increases as expected. The gain in power of the JE method remains remarkable for the experiments with larger data sets, especially for the short range one. Taking the MDem regression as reference, a 100 sample analyzed with traditional CI testing (1.056) has less power than a 40 sample probe analyzed with the ellipse method (1.03228 for the MCD).

In an additional trial a simulation with 250 samples probes was run for the short range as shown in Tab. \ref{tab:m250shorttable} and \ref{tab:p250shorttable}.

\begin{longtable}[]{@{}cccccccc@{}}
\caption{\label{tab:m250shorttable}MDem regression: classical CI method with 250 samples vs JE method with 40 samples, power and Type I error comparison.}\tabularnewline
\toprule
Samples & Method & p80\% LCI & p80\% est & p80\% UCI & T-I LCI & T-I est & T-I UCI\tabularnewline
\midrule
\endfirsthead
\toprule
Samples & Method & p80\% LCI & p80\% est & p80\% UCI & T-I LCI & T-I est & T-I UCI\tabularnewline
\midrule
\endhead
250 & CI5\% & 1.0347 & 1.0365 & 1.0384 & 0.88217 & 0.92058 & 0.95900\tabularnewline
250 & JE.SDe1\% & 1.0125 & 1.0127 & 1.0130 & 0.96884 & 0.98636 & 1.00388\tabularnewline
250 & JE.MCD1\% & 1.0126 & 1.0128 & 1.0130 & 0.96944 & 0.98589 & 1.00234\tabularnewline
250 & JE.Clas1\% & 1.0126 & 1.0128 & 1.0130 & 0.97202 & 0.98929 & 1.00656\tabularnewline
40 & CI5\% & 1.0928 & 1.0974 & 1.1026 & 0.89057 & 0.92598 & 0.96140\tabularnewline
40 & JE.SDe1\% & 1.0315 & 1.0323 & 1.0332 & 0.92968 & 0.95310 & 0.97651\tabularnewline
40 & JE.MCD1\% & 1.0323 & 1.0331 & 1.0339 & 0.93752 & 0.95983 & 0.98215\tabularnewline
40 & JE.Clas1\% & 1.0321 & 1.0329 & 1.0337 & 0.94615 & 0.96859 & 0.99103\tabularnewline
\bottomrule
\end{longtable}

Using the MDem regression a 40 samples probe tested with the JE method showed a slight but significant higher power 1.03308 than a 250 samples probe tested using the classical CI approach 1.03646.

\begin{longtable}[]{@{}cccccccc@{}}
\caption{\label{tab:p250shorttable}PaBa regression: classical CI method with 250 samples vs JE method with 40 samples, power and Type I error comparison.}\tabularnewline
\toprule
Samples & Method & p80\% LCI & p80\% est & p80\% UCI & T-I LCI & T-I est & T-I UCI\tabularnewline
\midrule
\endfirsthead
\toprule
Samples & Method & p80\% LCI & p80\% est & p80\% UCI & T-I LCI & T-I est & T-I UCI\tabularnewline
\midrule
\endhead
250 & CI5\% & 1.0334 & 1.0350 & 1.0368 & 0.88280 & 0.91905 & 0.95529\tabularnewline
250 & JE.SDe1\% & 1.0144 & 1.0148 & 1.0152 & 0.95873 & 0.98154 & 1.00435\tabularnewline
250 & JE.MCD1\% & 1.0144 & 1.0148 & 1.0152 & 0.95798 & 0.98244 & 1.00690\tabularnewline
250 & JE.Clas1\% & 1.0146 & 1.0149 & 1.0153 & 0.97236 & 0.99723 & 1.02209\tabularnewline
40 & CI5\% & 1.0908 & 1.0959 & 1.1016 & 0.88803 & 0.92895 & 0.96988\tabularnewline
40 & JE.SDe1\% & 1.0353 & 1.0367 & 1.0382 & 0.90850 & 0.93751 & 0.96652\tabularnewline
40 & JE.MCD1\% & 1.0355 & 1.0369 & 1.0383 & 0.91024 & 0.93845 & 0.96666\tabularnewline
40 & JE.Clas1\% & 1.0381 & 1.0391 & 1.0402 & 0.95651 & 0.98045 & 1.00440\tabularnewline
\bottomrule
\end{longtable}

The same comparison with the PaBa regression shows a weak but significant advantage of the traditional CI (1.03497 vs 1.03688). This result is pretty surprising and could turn out to be a major finding, allowing a substantial reduction of the minimal number of needed samples in many biomedical method comparisons; for example for the validation of the devices measuring the potassium concentration in blood. In Appendix \ref{Comparison250Samples} the analogous tables for Deming and MMDem regressions.

\hypertarget{comparing-the-regressions}{%
\subsubsection{Comparing the regressions}\label{comparing-the-regressions}}

\begin{figure}

{\centering \includegraphics{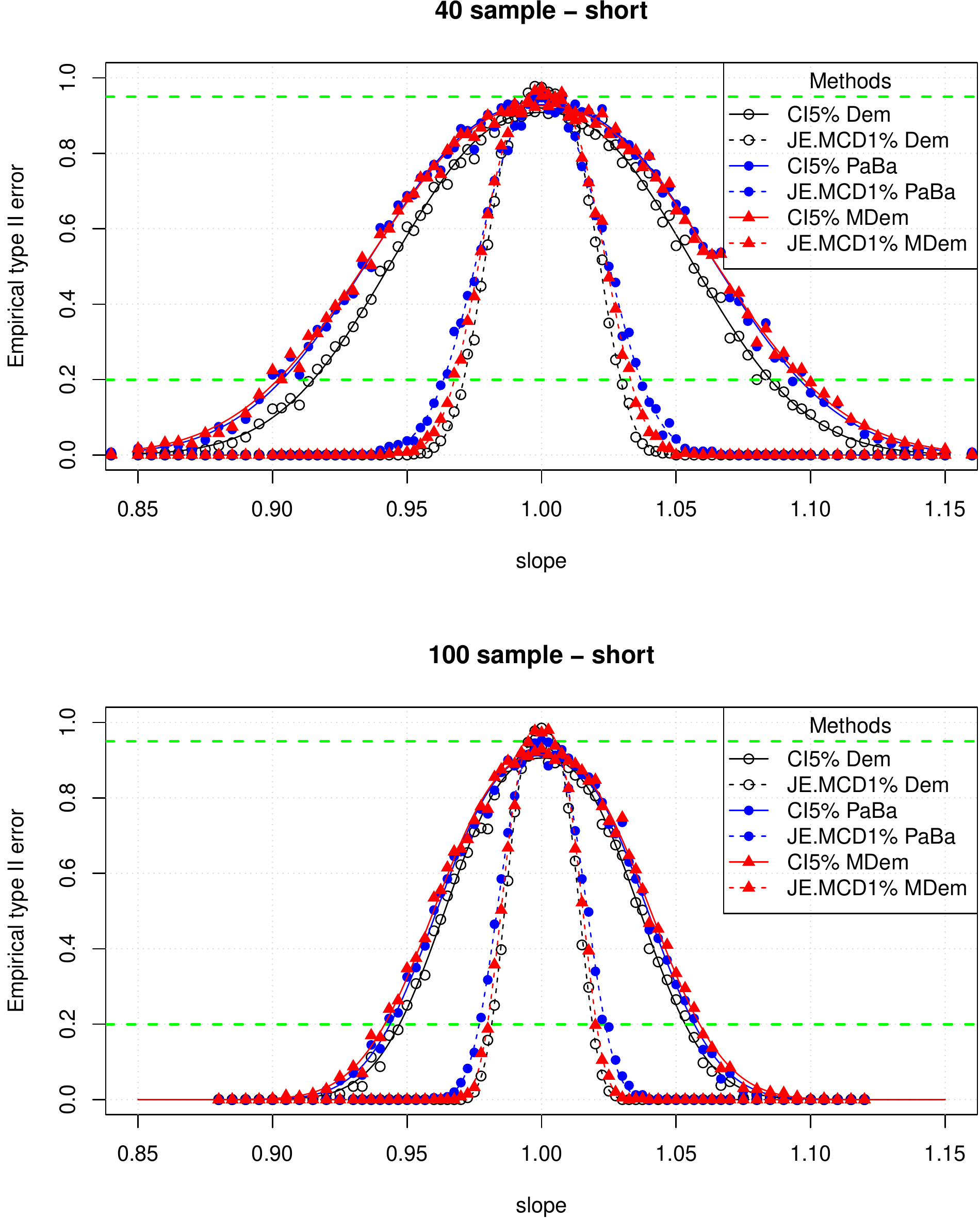} 

}

\caption{Short range regressions: slope power comparison.}\label{fig:slope40short100}
\end{figure}

\begin{figure}

{\centering \includegraphics{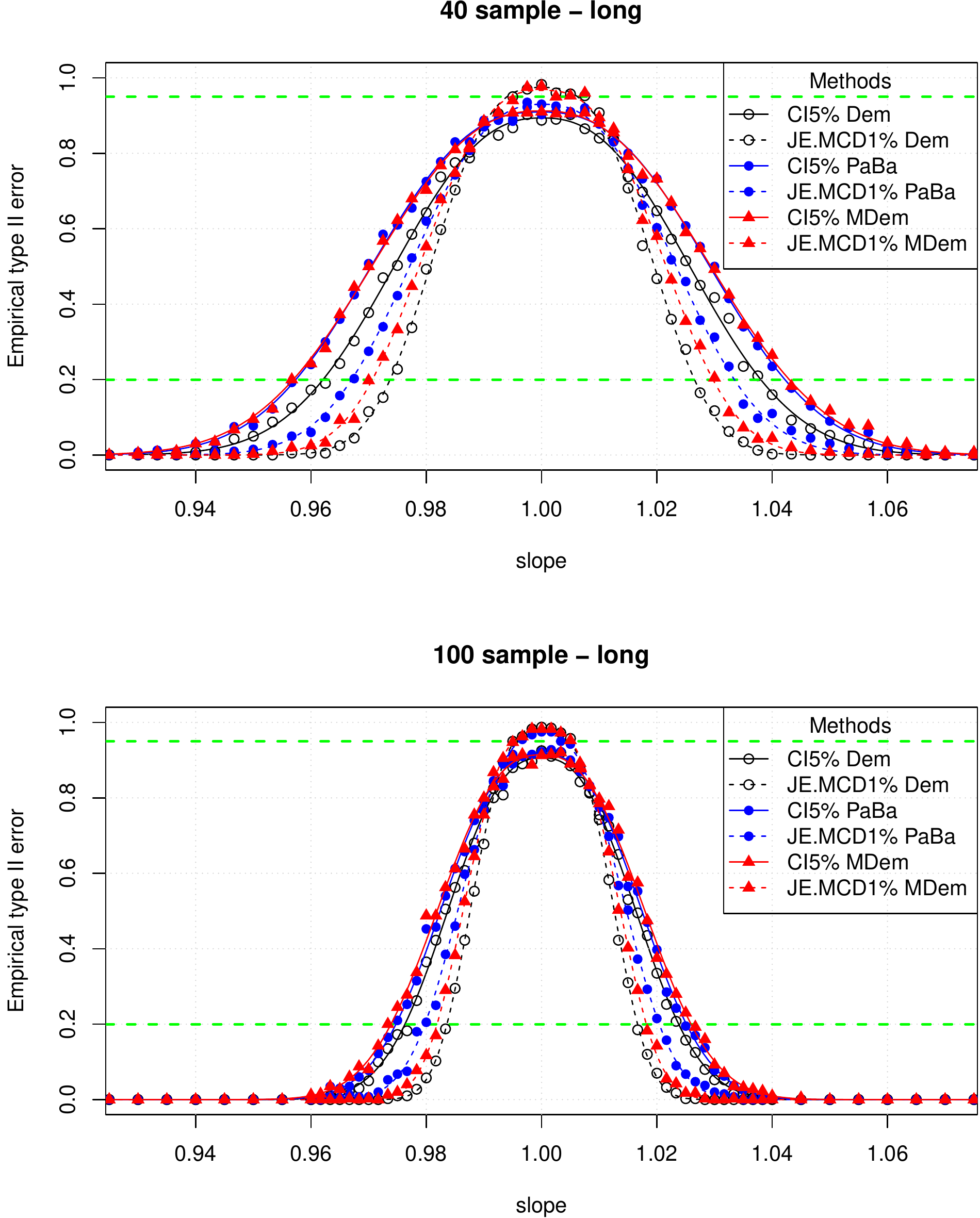} 

}

\caption{Long range regressions: slope power comparison.}\label{fig:slope40long100}
\end{figure}

It is important once again to highlight the power differences between Deming, Paba and MDem regressions in all the possible combinations of range and sample sizes. In Fig. \ref{fig:slope40short100} and \ref{fig:slope40long100} for each experimental situation a separate summary plot is drawn. Long and short ranges are paired separately to allow different ranges in the slopes. To ease the readability, the MMDem and the WDem regressions are not drawn and for the JE method only the results obtained with the MCD covariances are reported.

The two robust methods shows as expected a slightly lower power as the non-robust one. The PaBa regression seems to be slightly more powerful than the MDem if used with the traditional CI testing. In the JE testing the difference is inverted: the MDem shows better power.

Hence, it is possible to conclude that, if a robust method is needed, the best power is reached combining the MDem regression with the JE method. Since the presence of outliers is often hard to be excluded and since the power difference between the classical Deming and the MDem regression is moderate when used in combination with the JE method, the MDem JE method could be potentially considered as a first choice method for validation experiments in general.

\clearpage

\hypertarget{type-ii-error-and-power-comparison-the-intercept}{%
\subsection{Type II error and power comparison: the intercept}\label{type-ii-error-and-power-comparison-the-intercept}}

\begin{figure}

{\centering \includegraphics{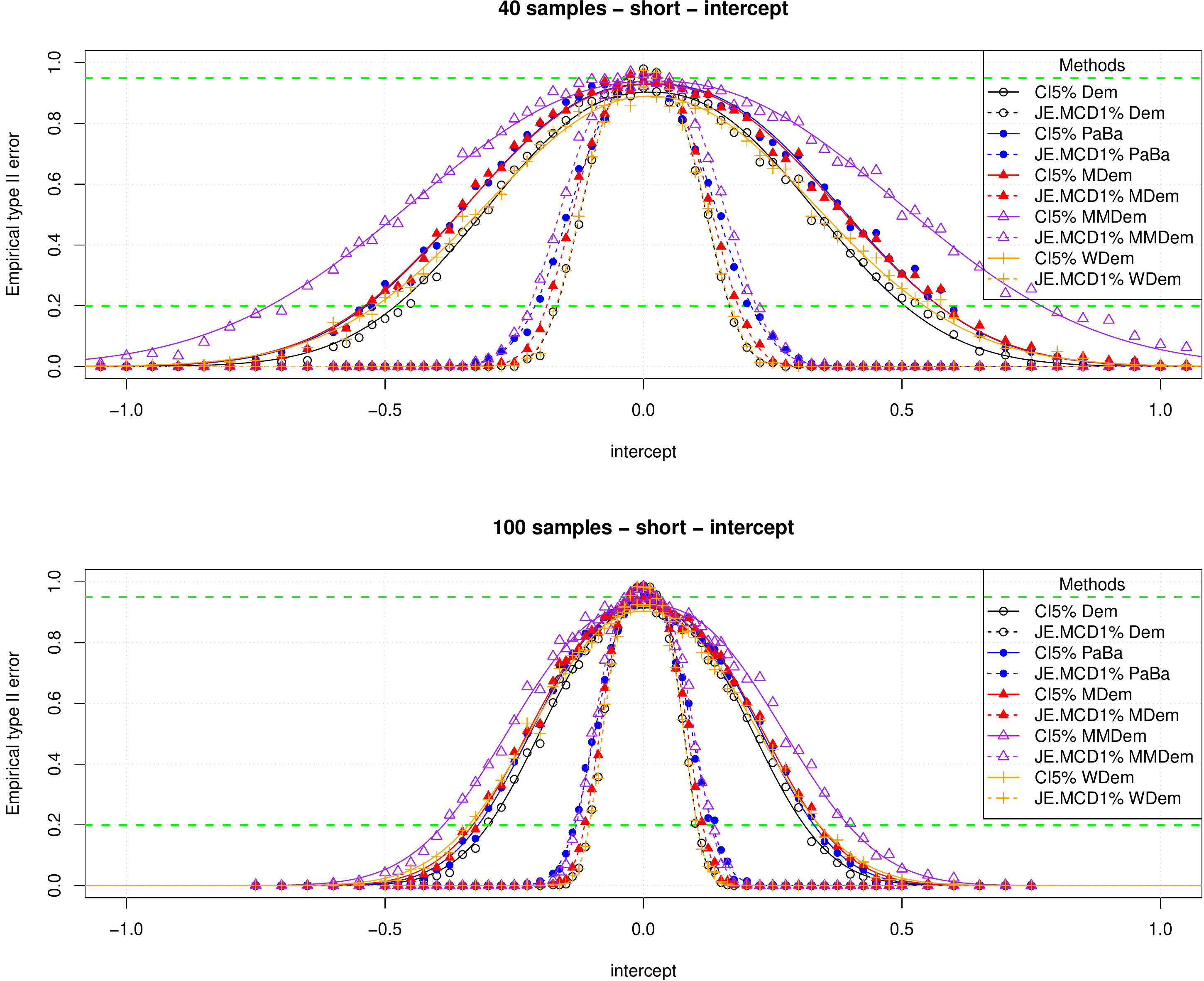} 

}

\caption{Short range regressions: intercept power comparison.}\label{fig:int40short100}
\end{figure}

\begin{figure}

{\centering \includegraphics{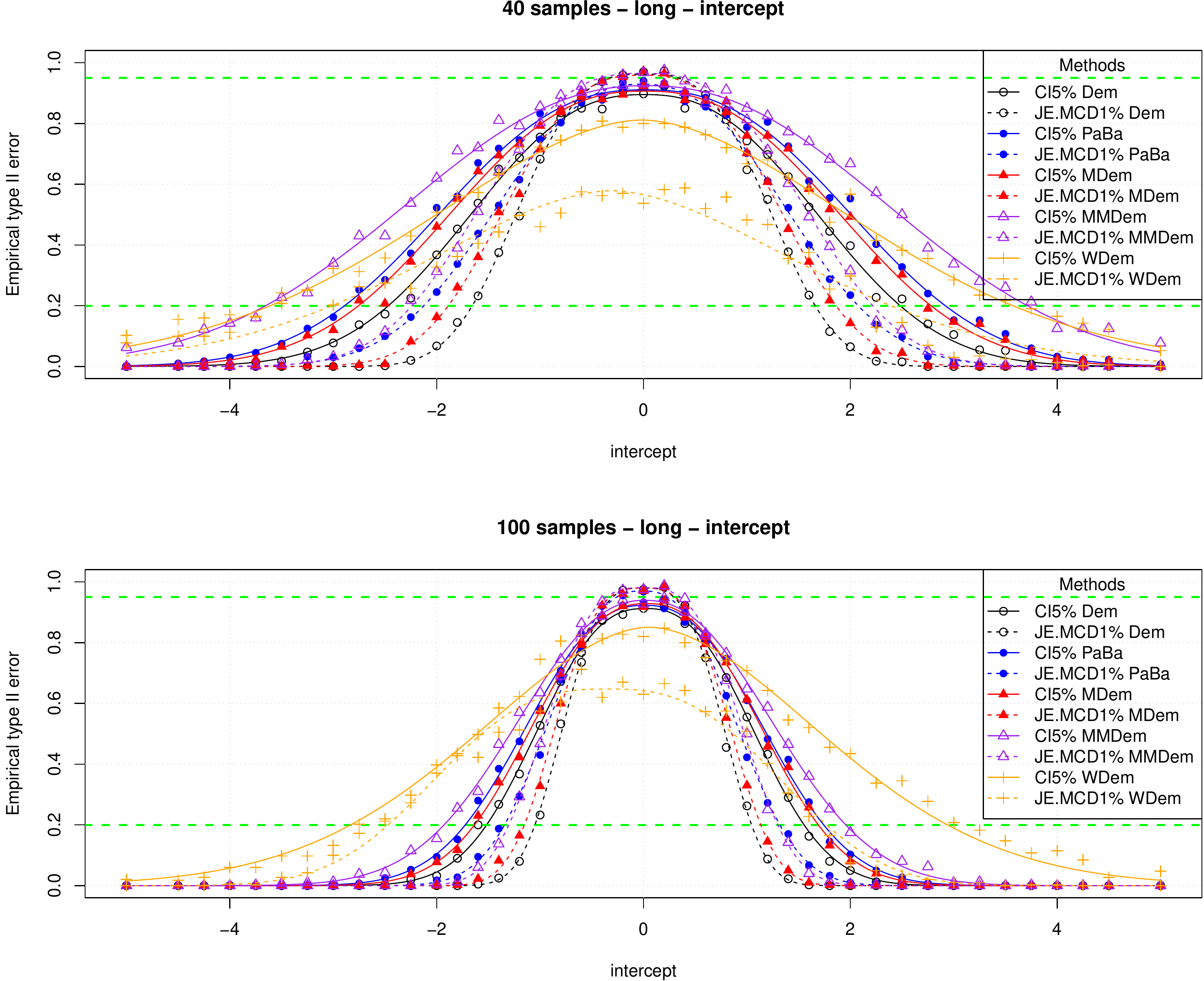} 

}

\caption{Long range regressions: intercept power comparison.}\label{fig:int40long100}
\end{figure}

The variation of the intercept, keeping the slope \(b=1\), gives analogous results. We replicate here a similar plot as in Fig. \ref{fig:slope40short100} and \ref{fig:slope40long100}. For sake of completeness also the less efficient MMDem and the problematic WDem are added (see Fig. \ref{fig:int40short100} and \ref{fig:int40long100}).

Once again, the bad distortion of the WDem regression applied on the long data set is visible (see Fig. \ref{fig:int40long100}). In this case a bias is registered. This bias it's not a surprise since negative values are rounded to the fictive detection limits. A little quantity of low level outliers are generated simulating negative intercepts and the WDem regression is very sensitive to them.

The MMDem shows in general less power and is not recommended.

The short and the long data set cases have very different intercept rejection ranges but for the rest all the previous observations are valid here, too. The JE testing method is largely more powerful and could permit a drastic reduction of the sample sizes.

In Tab. \ref{tab:intcomptab} the type II errors for the variation of the intercept (using the MDem regression) are reported similarly as in Tab. \ref{tab:mdemtable} for the slope.

\begin{longtable}[]{@{}ccccccccc@{}}
\caption{\label{tab:intcomptab}Power and type I error comparison for the intercept with the MDem regression.}\tabularnewline
\toprule
Range & Sample & Method & p80\% LCI & p80\% est & p80\% UCI & T-I LCI & T-I est. & T-I UCI\tabularnewline
\midrule
\endfirsthead
\toprule
Range & Sample & Method & p80\% LCI & p80\% est & p80\% UCI & T-I LCI & T-I est. & T-I UCI\tabularnewline
\midrule
\endhead
short & 40 & CI5\% & 0.5388 & 0.5709 & 0.6078 & 0.8841 & 0.9285 & 0.9728\tabularnewline
short & 40 & JE.SDe1\% & 0.1758 & 0.1790 & 0.1823 & 0.9356 & 0.9518 & 0.9679\tabularnewline
short & 40 & JE.MCD1\% & 0.1758 & 0.1790 & 0.1823 & 0.9356 & 0.9518 & 0.9679\tabularnewline
short & 40 & JE.Clas1\% & 0.1758 & 0.1790 & 0.1823 & 0.9356 & 0.9518 & 0.9679\tabularnewline
short & 100 & CI5\% & 0.3231 & 0.3389 & 0.3564 & 0.8834 & 0.9199 & 0.9564\tabularnewline
short & 100 & JE.SDe1\% & 0.1105 & 0.1136 & 0.1170 & 0.9467 & 0.9739 & 1.0010\tabularnewline
short & 100 & JE.MCD1\% & 0.1105 & 0.1136 & 0.1170 & 0.9467 & 0.9739 & 1.0010\tabularnewline
short & 100 & JE.Clas1\% & 0.1105 & 0.1136 & 0.1170 & 0.9467 & 0.9739 & 1.0010\tabularnewline
long & 40 & CI5\% & 2.6351 & 2.7799 & 2.9434 & 0.8650 & 0.9065 & 0.9479\tabularnewline
long & 40 & JE.SDe1\% & 1.7653 & 1.8364 & 1.9148 & 0.9265 & 0.9603 & 0.9941\tabularnewline
long & 40 & JE.MCD1\% & 1.7653 & 1.8364 & 1.9148 & 0.9265 & 0.9603 & 0.9941\tabularnewline
long & 40 & JE.Clas1\% & 1.7653 & 1.8364 & 1.9148 & 0.9265 & 0.9603 & 0.9941\tabularnewline
long & 100 & CI5\% & 1.6450 & 1.6904 & 1.7387 & 0.9067 & 0.9284 & 0.9502\tabularnewline
long & 100 & JE.SDe1\% & 1.1147 & 1.1295 & 1.1448 & 0.9650 & 0.9777 & 0.9904\tabularnewline
long & 100 & JE.MCD1\% & 1.1147 & 1.1295 & 1.1448 & 0.9650 & 0.9777 & 0.9904\tabularnewline
long & 100 & JE.Clas1\% & 1.1147 & 1.1295 & 1.1448 & 0.9650 & 0.9777 & 0.9904\tabularnewline
\bottomrule
\end{longtable}

For sake of completeness, it is clear that a grid slope/intercept power investigation would be interesting, but it would also be very computational intensive. For this reason this work was concentrated only on the variation of one single parameter at time.

In Tab. \ref{tab:int40shorttable} a comparison between regressions for the 40 samples short range experiments on the intercept. The analogous tables for the other conditions are in Appendix \ref{InterceptTables}

\begin{longtable}[]{@{}cccccccc@{}}
\caption{\label{tab:int40shorttable}Power and type I error comparison for the intercept with the short range 40 samples simulations.}\tabularnewline
\toprule
Regression & Method & p80\% LCI & p80\% est & p80\% UCI & T-I LCI & T-I est. & T-I UCI\tabularnewline
\midrule
\endfirsthead
\toprule
Regression & Method & p80\% LCI & p80\% est & p80\% UCI & T-I LCI & T-I est. & T-I UCI\tabularnewline
\midrule
\endhead
Deming & CI5\% & 0.47803 & 0.50479 & 0.53512 & 0.86243 & 0.90292 & 0.94342\tabularnewline
Deming & JE.SDe1\% & 0.16254 & 0.16581 & 0.16926 & 0.94346 & 0.96154 & 0.97962\tabularnewline
Deming & JE.MCD1\% & 0.16254 & 0.16581 & 0.16926 & 0.94346 & 0.96154 & 0.97962\tabularnewline
Deming & JE.Clas1\% & 0.16254 & 0.16581 & 0.16926 & 0.94346 & 0.96154 & 0.97962\tabularnewline
MDem & CI5\% & 0.53878 & 0.57091 & 0.60782 & 0.88410 & 0.92846 & 0.97282\tabularnewline
MDem & JE.SDe1\% & 0.17581 & 0.17903 & 0.18233 & 0.93562 & 0.95177 & 0.96792\tabularnewline
MDem & JE.MCD1\% & 0.17581 & 0.17903 & 0.18233 & 0.93562 & 0.95177 & 0.96792\tabularnewline
MDem & JE.Clas1\% & 0.17581 & 0.17903 & 0.18233 & 0.93562 & 0.95177 & 0.96792\tabularnewline
PaBa & CI5\% & 0.54044 & 0.57092 & 0.60574 & 0.88751 & 0.93037 & 0.97323\tabularnewline
PaBa & JE.SDe1\% & 0.20277 & 0.20862 & 0.21485 & 0.92249 & 0.94457 & 0.96665\tabularnewline
PaBa & JE.MCD1\% & 0.20277 & 0.20862 & 0.21485 & 0.92249 & 0.94457 & 0.96665\tabularnewline
PaBa & JE.Clas1\% & 0.20277 & 0.20862 & 0.21485 & 0.92249 & 0.94457 & 0.96665\tabularnewline
MMDem & CI5\% & 0.72370 & 0.76861 & 0.82049 & 0.89493 & 0.93953 & 0.98413\tabularnewline
MMDem & JE.SDe1\% & 0.20617 & 0.21082 & 0.21574 & 0.93829 & 0.95836 & 0.97844\tabularnewline
MMDem & JE.MCD1\% & 0.20617 & 0.21082 & 0.21574 & 0.93829 & 0.95836 & 0.97844\tabularnewline
MMDem & JE.Clas1\% & 0.20617 & 0.21082 & 0.21574 & 0.93829 & 0.95836 & 0.97844\tabularnewline
WDem & CI5\% & 0.51056 & 0.54283 & 0.57963 & 0.84829 & 0.88965 & 0.93102\tabularnewline
WDem & JE.SDe1\% & 0.16267 & 0.16577 & 0.16905 & 0.93319 & 0.95053 & 0.96786\tabularnewline
WDem & JE.MCD1\% & 0.16267 & 0.16577 & 0.16905 & 0.93319 & 0.95053 & 0.96786\tabularnewline
WDem & JE.Clas1\% & 0.16267 & 0.16577 & 0.16905 & 0.93319 & 0.95053 & 0.96786\tabularnewline
\bottomrule
\end{longtable}

\clearpage

\hypertarget{the-heteroscedastic-case}{%
\subsection{The heteroscedastic case}\label{the-heteroscedastic-case}}

Method validators frequently have to deal with heteroscedastic data sets or with data sets in which the normality of the residuals is at least questionable. A power analysis investigation with heteroscedastic data (multiplicative error) and with mixed (50\% multiplicative and 50\% additive) errors has been performed and compared to the previous results. The research has been performed only on the long range data set since on the short range the multiplicative effect can be considered negligible. The details of the data generating functions are available in the Appendix \ref{DataGenFunc}. The random generated data sets have the same standard error ratio as in the homoscedastic case in the center (at \(X=55\)), but in the extreme positions the multiplicative nature shows its pattern.

\begin{figure}

{\centering \includegraphics{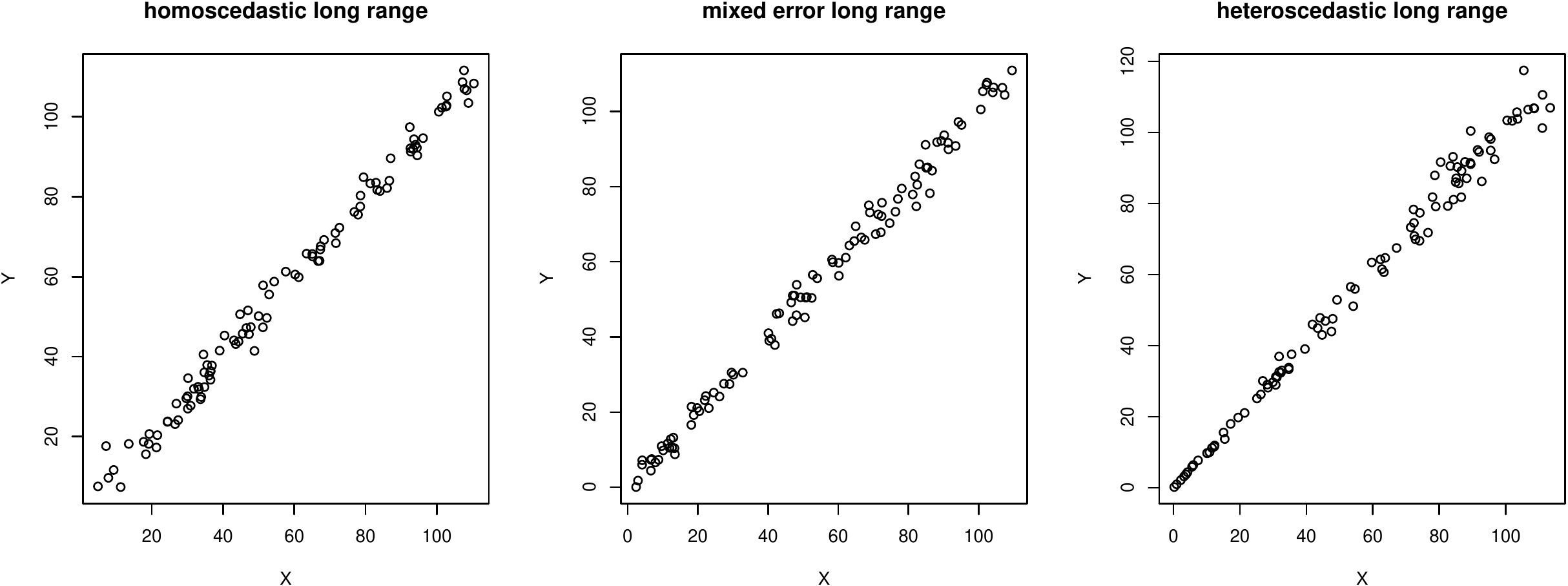} 

}

\caption{Homoscedastic, mixed and pure heteroscedastic data sets.}\label{fig:homomixhetdatagen}
\end{figure}

Since the total error (variance) of a single measurement method is always a sum of errors with different sources, it is reasonable to suppose that the mixed error situation is the most realistic one.

\begin{figure}

{\centering \includegraphics{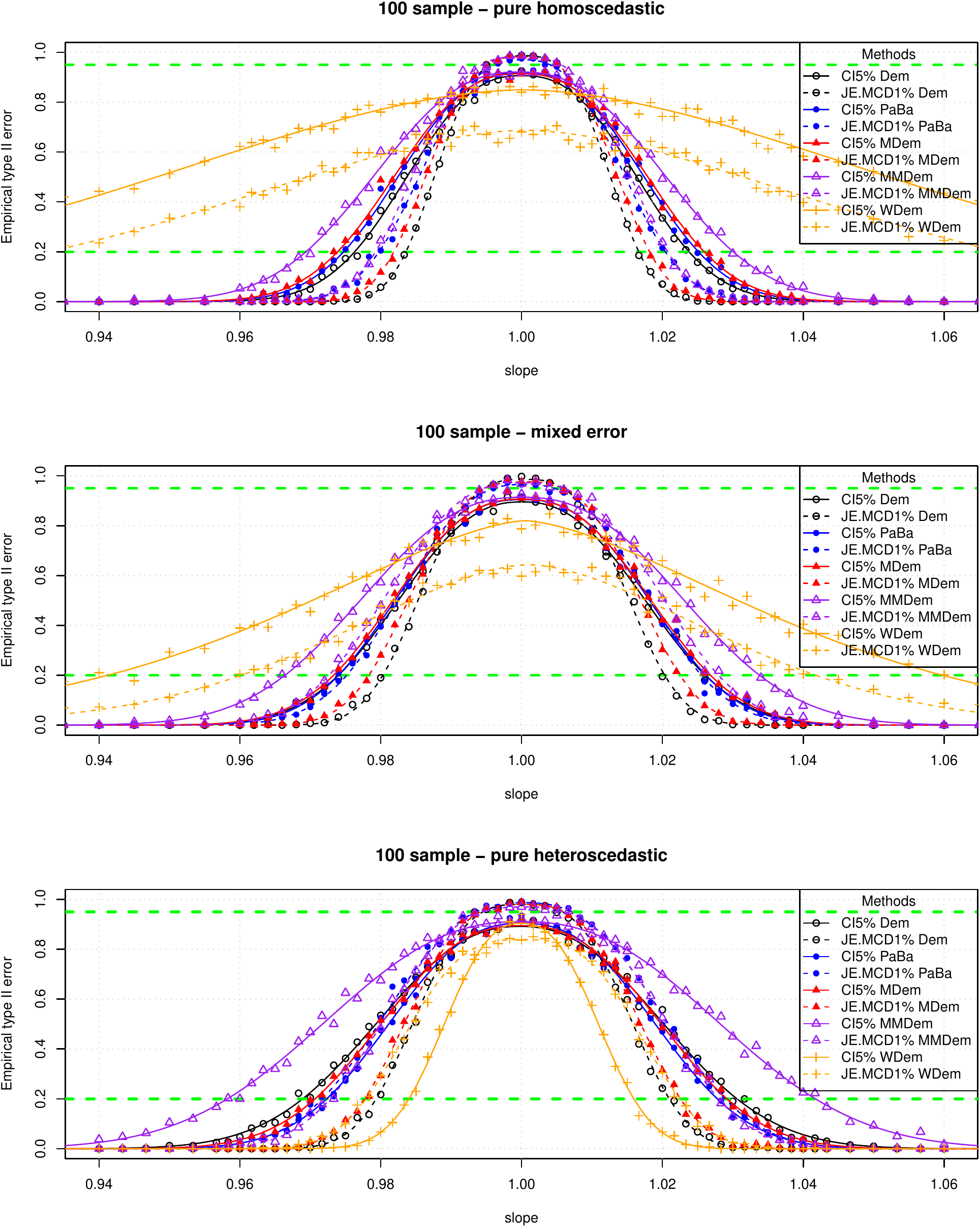} 

}

\caption{Effect of heteroscedasticity on long range 100 samples data sets.}\label{fig:homomixhetplot100}
\end{figure}

The Fig. \ref{fig:homomixhetplot100} summarize the results of the 3 cases (additive, mixed and pure multiplicative) power simulations for the 100 samples experiments.

\begin{figure}

{\centering \includegraphics{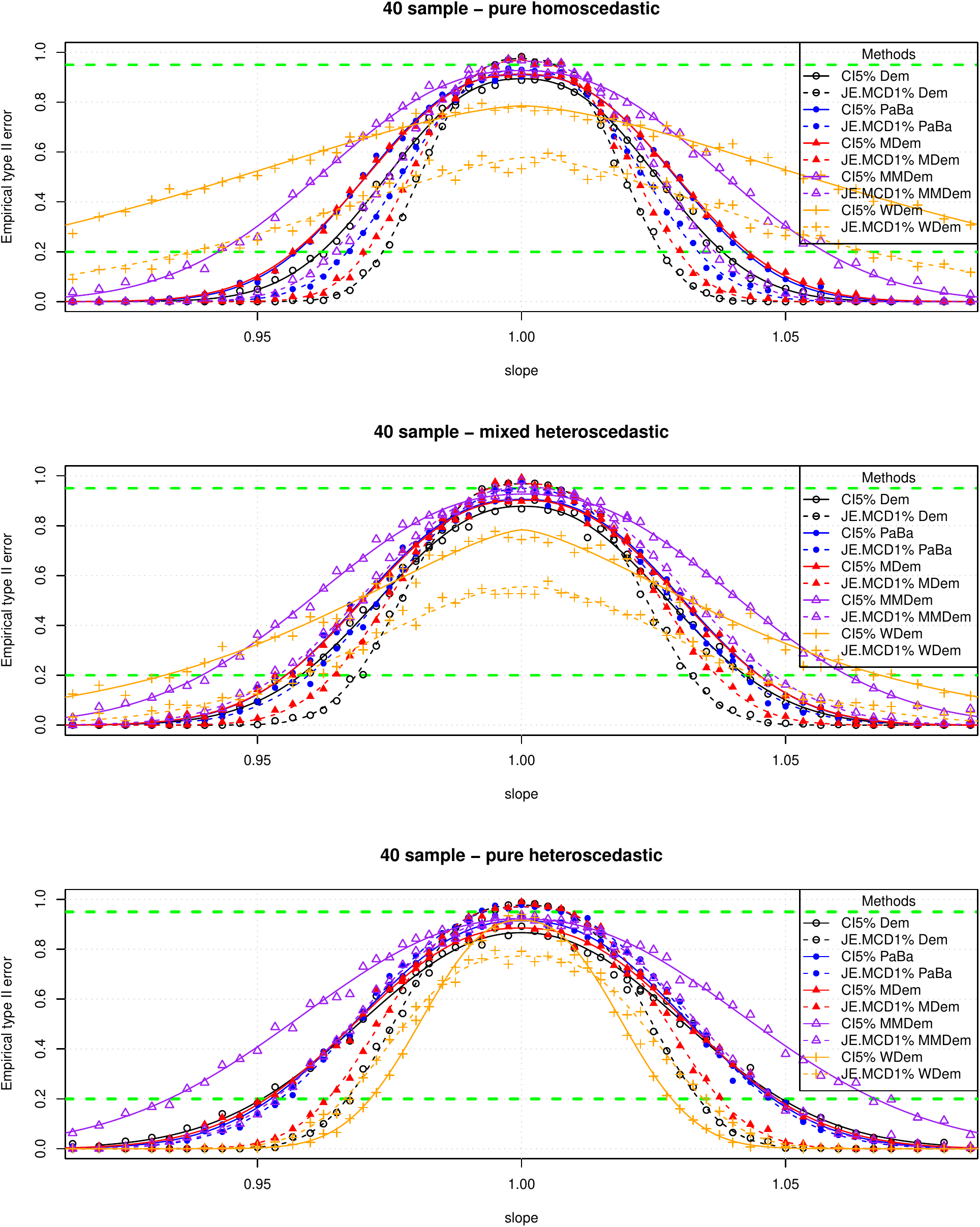} 

}

\caption{Effect of heteroscedasticity on long range 40 samples data sets.}\label{fig:homomixhetplot40}
\end{figure}

The 40 sample simulations in Fig. \ref{fig:homomixhetplot40} provide similar results with wider curves.

It is manifest that in general heteroscedasticity reduces the power. In a pure and ideal heteroscedastic model the WDem regression performs with the best power, and it is the only case in which the CI method outperforms the JE one. In pure heteroscedastic model the nature of the multiplicative error is such that the simulation of the detection cut off is not relevant. But with mixed errors some detection limits are reached and the WDem regression loses completely its power.

For comparative purposes it is worth plotting the homoscedastic, mixed and heteroscedastic cases on the same plot (see Fig. \ref{fig:homohetmixcomp}). An analogous plot for WDem and MMDem is in Appendix \ref{HetCompPlot}

\begin{figure}

{\centering \includegraphics{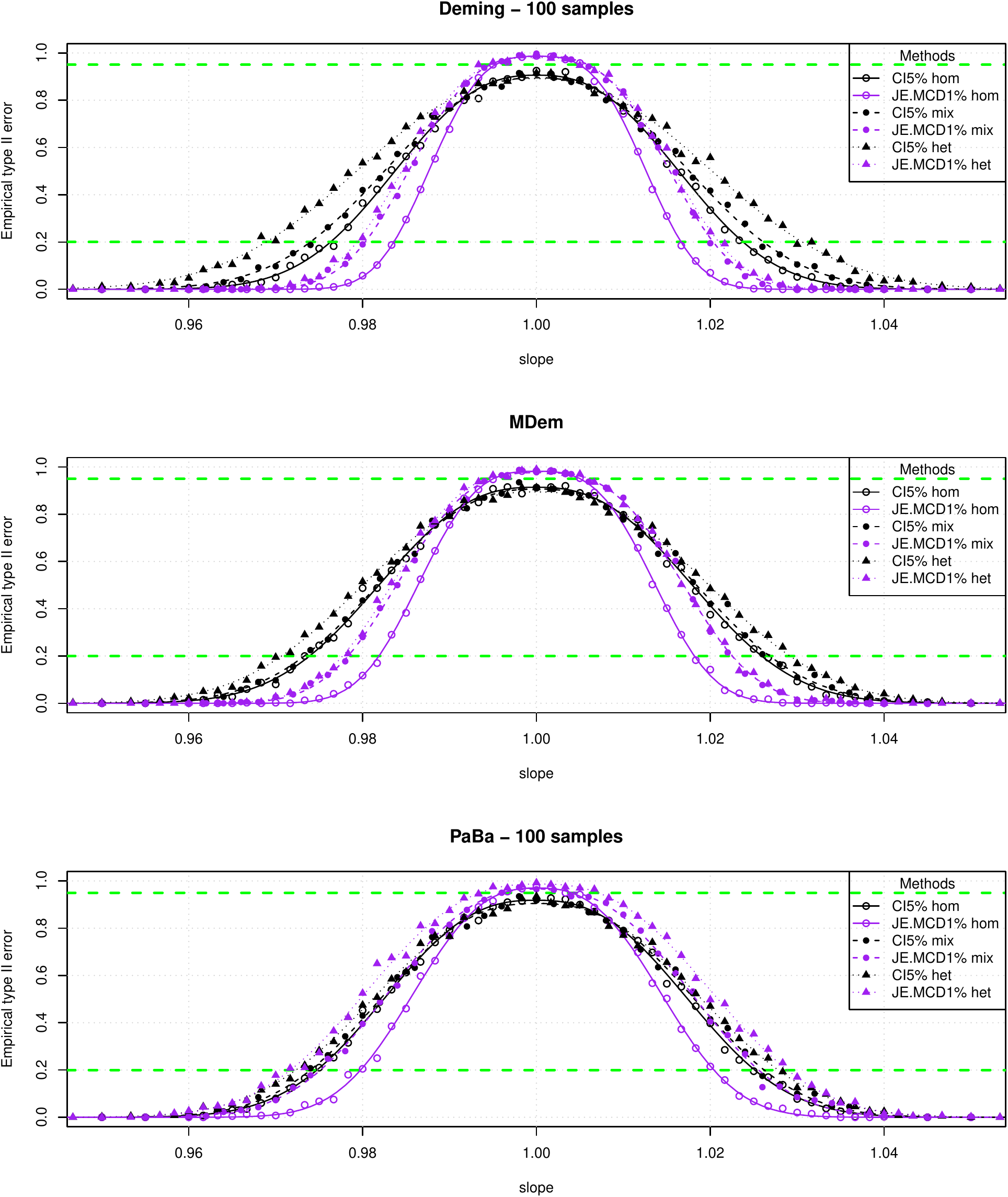} 

}

\caption{Role of the heteroscedasticity: one plot per regression method with the 100 sample data sets.}\label{fig:homohetmixcomp}
\end{figure}

For the classic CI method Deming regression has the biggest power loss moving from homoscedasticity to heteroscedasticity. MDem and PaBa show similar power losses among them, less marked than for the Deming method.

With the JE method the behavior of Deming, MDem and Paba regressions diverges. The JE MDem and Deming combinations partially preserve their advantages compared to the CI method. The JE PaBa method instead loses completely its advantage compared to its classical CI approach.

In Tab. \ref{tab:100longtable}, \ref{tab:100mixtable} and \ref{tab:100hettable} the estimated type I errors and the powers of the three situations are reported for the 100 samples simulations.

\begin{longtable}[]{@{}cccccccc@{}}
\caption{\label{tab:100longtable}Power and type I error comparison for the long range 100 samples pure homoscedastic experiments.}\tabularnewline
\toprule
Regression & Method & p80\% LCI & p80\% est & p80\% UCI & T-I LCI & T-I est & T-I UCI\tabularnewline
\midrule
\endfirsthead
\toprule
Regression & Method & p80\% LCI & p80\% est & p80\% UCI & T-I LCI & T-I est & T-I UCI\tabularnewline
\midrule
\endhead
Deming & CI5\% & 1.0230 & 1.0238 & 1.0247 & 0.87859 & 0.90603 & 0.93346\tabularnewline
Deming & JE.SDe1\% & 1.0163 & 1.0166 & 1.0169 & 0.96812 & 0.98409 & 1.00007\tabularnewline
Deming & JE.MCD1\% & 1.0164 & 1.0166 & 1.0169 & 0.97061 & 0.98602 & 1.00143\tabularnewline
Deming & JE.Clas1\% & 1.0163 & 1.0166 & 1.0168 & 0.97090 & 0.98660 & 1.00230\tabularnewline
MDem & CI5\% & 1.0249 & 1.0259 & 1.0271 & 0.88127 & 0.91408 & 0.94689\tabularnewline
MDem & JE.SDe1\% & 1.0177 & 1.0180 & 1.0183 & 0.96414 & 0.98090 & 0.99767\tabularnewline
MDem & JE.MCD1\% & 1.0178 & 1.0182 & 1.0185 & 0.96334 & 0.98168 & 1.00002\tabularnewline
MDem & JE.Clas1\% & 1.0177 & 1.0181 & 1.0185 & 0.96279 & 0.98433 & 1.00587\tabularnewline
PaBa & CI5\% & 1.0242 & 1.0251 & 1.0261 & 0.88927 & 0.91827 & 0.94727\tabularnewline
PaBa & JE.SDe1\% & 1.0195 & 1.0200 & 1.0205 & 0.94536 & 0.96582 & 0.98627\tabularnewline
PaBa & JE.MCD1\% & 1.0199 & 1.0204 & 1.0210 & 0.94745 & 0.96988 & 0.99232\tabularnewline
PaBa & JE.Clas1\% & 1.0199 & 1.0203 & 1.0208 & 0.96443 & 0.98706 & 1.00969\tabularnewline
MMDem & CI5\% & 1.0288 & 1.0300 & 1.0314 & 0.88866 & 0.92104 & 0.95342\tabularnewline
MMDem & JE.SDe1\% & 1.0200 & 1.0205 & 1.0211 & 0.95578 & 0.98100 & 1.00621\tabularnewline
MMDem & JE.MCD1\% & 1.0203 & 1.0208 & 1.0213 & 0.95887 & 0.98240 & 1.00593\tabularnewline
MMDem & JE.Clas1\% & 1.0204 & 1.0209 & 1.0213 & 0.96574 & 0.98709 & 1.00844\tabularnewline
WDem & CI5\% & 1.0844 & 1.0914 & 1.0995 & 0.80774 & 0.84926 & 0.89079\tabularnewline
WDem & JE.SDe1\% & 1.0559 & 1.0625 & 1.0704 & 0.66574 & 0.71760 & 0.76946\tabularnewline
WDem & JE.MCD1\% & 1.0608 & 1.0681 & 1.0769 & 0.63636 & 0.68714 & 0.73792\tabularnewline
WDem & JE.Clas1\% & 1.0607 & 1.0658 & 1.0719 & 0.93547 & 0.98826 & 1.04106\tabularnewline
\bottomrule
\end{longtable}

\begin{longtable}[]{@{}cccccccc@{}}
\caption{\label{tab:100mixtable}Power and type I error comparison for the long range 100 samples mixed error experiments.}\tabularnewline
\toprule
Regression & Method & p80\% LCI & p80\% est & p80\% UCI & T-I LCI & T-I est & T-I UCI\tabularnewline
\midrule
\endfirsthead
\toprule
Regression & Method & p80\% LCI & p80\% est & p80\% UCI & T-I LCI & T-I est & T-I UCI\tabularnewline
\midrule
\endhead
Deming & CI5\% & 1.0252 & 1.0261 & 1.0271 & 0.86808 & 0.89539 & 0.92270\tabularnewline
Deming & JE.SDe1\% & 1.0198 & 1.0202 & 1.0206 & 0.96357 & 0.98391 & 1.00424\tabularnewline
Deming & JE.MCD1\% & 1.0199 & 1.0203 & 1.0208 & 0.96340 & 0.98573 & 1.00806\tabularnewline
Deming & JE.Clas1\% & 1.0198 & 1.0202 & 1.0208 & 0.95999 & 0.98394 & 1.00790\tabularnewline
MDem & CI5\% & 1.0256 & 1.0269 & 1.0282 & 0.87067 & 0.90664 & 0.94262\tabularnewline
MDem & JE.SDe1\% & 1.0218 & 1.0224 & 1.0230 & 0.94780 & 0.97238 & 0.99697\tabularnewline
MDem & JE.MCD1\% & 1.0219 & 1.0225 & 1.0231 & 0.95131 & 0.97591 & 1.00050\tabularnewline
MDem & JE.Clas1\% & 1.0217 & 1.0223 & 1.0230 & 0.95209 & 0.97818 & 1.00427\tabularnewline
PaBa & CI5\% & 1.0251 & 1.0262 & 1.0274 & 0.87099 & 0.90470 & 0.93842\tabularnewline
PaBa & JE.SDe1\% & 1.0240 & 1.0248 & 1.0257 & 0.93364 & 0.96231 & 0.99098\tabularnewline
PaBa & JE.MCD1\% & 1.0247 & 1.0255 & 1.0264 & 0.93728 & 0.96566 & 0.99404\tabularnewline
PaBa & JE.Clas1\% & 1.0245 & 1.0252 & 1.0260 & 0.95502 & 0.98333 & 1.01164\tabularnewline
MMDem & CI5\% & 1.0324 & 1.0338 & 1.0354 & 0.88078 & 0.91399 & 0.94720\tabularnewline
MMDem & JE.SDe1\% & 1.0268 & 1.0277 & 1.0286 & 0.94485 & 0.97135 & 0.99785\tabularnewline
MMDem & JE.MCD1\% & 1.0271 & 1.0280 & 1.0289 & 0.94462 & 0.97261 & 1.00060\tabularnewline
MMDem & JE.Clas1\% & 1.0272 & 1.0281 & 1.0292 & 0.94565 & 0.97790 & 1.01015\tabularnewline
WDem & CI5\% & 1.0536 & 1.0598 & 1.0673 & 0.76832 & 0.82001 & 0.87171\tabularnewline
WDem & JE.SDe1\% & 1.0346 & 1.0384 & 1.0428 & 0.62640 & 0.67499 & 0.72358\tabularnewline
WDem & JE.MCD1\% & 1.0372 & 1.0418 & 1.0473 & 0.59291 & 0.64358 & 0.69424\tabularnewline
WDem & JE.Clas1\% & 1.0380 & 1.0414 & 1.0456 & 0.92845 & 0.98869 & 1.04892\tabularnewline
\bottomrule
\end{longtable}

\begin{longtable}[]{@{}cccccccc@{}}
\caption{\label{tab:100hettable}Power and type I error comparison for the long range 100 samples pure heteroscedastic experiments.}\tabularnewline
\toprule
Regression & Method & p80\% LCI & p80\% est & p80\% UCI & T-I LCI & T-I est & T-I UCI\tabularnewline
\midrule
\endfirsthead
\toprule
Regression & Method & p80\% LCI & p80\% est & p80\% UCI & T-I LCI & T-I est & T-I UCI\tabularnewline
\midrule
\endhead
Deming & CI5\% & 1.0290 & 1.0306 & 1.0324 & 0.85300 & 0.89169 & 0.93038\tabularnewline
Deming & JE.SDe1\% & 1.0203 & 1.0208 & 1.0214 & 0.95173 & 0.97885 & 1.00596\tabularnewline
Deming & JE.MCD1\% & 1.0204 & 1.0210 & 1.0216 & 0.95156 & 0.98059 & 1.00962\tabularnewline
Deming & JE.Clas1\% & 1.0202 & 1.0208 & 1.0214 & 0.95141 & 0.97883 & 1.00624\tabularnewline
MDem & CI5\% & 1.0281 & 1.0292 & 1.0303 & 0.86666 & 0.89447 & 0.92227\tabularnewline
MDem & JE.SDe1\% & 1.0218 & 1.0224 & 1.0230 & 0.94825 & 0.97444 & 1.00064\tabularnewline
MDem & JE.MCD1\% & 1.0220 & 1.0225 & 1.0232 & 0.95463 & 0.98008 & 1.00554\tabularnewline
MDem & JE.Clas1\% & 1.0219 & 1.0224 & 1.0230 & 0.95463 & 0.97942 & 1.00421\tabularnewline
PaBa & CI5\% & 1.0267 & 1.0278 & 1.0291 & 0.87197 & 0.90448 & 0.93699\tabularnewline
PaBa & JE.SDe1\% & 1.0259 & 1.0269 & 1.0279 & 0.94979 & 0.98173 & 1.01366\tabularnewline
PaBa & JE.MCD1\% & 1.0277 & 1.0289 & 1.0302 & 0.95018 & 0.98689 & 1.02360\tabularnewline
PaBa & JE.Clas1\% & 1.0260 & 1.0269 & 1.0278 & 0.95683 & 0.98838 & 1.01994\tabularnewline
MMDem & CI5\% & 1.0395 & 1.0417 & 1.0442 & 0.87007 & 0.91093 & 0.95179\tabularnewline
MMDem & JE.SDe1\% & 1.0260 & 1.0270 & 1.0280 & 0.93190 & 0.96580 & 0.99970\tabularnewline
MMDem & JE.MCD1\% & 1.0270 & 1.0280 & 1.0291 & 0.93656 & 0.96971 & 1.00287\tabularnewline
MMDem & JE.Clas1\% & 1.0269 & 1.0277 & 1.0287 & 0.95687 & 0.98877 & 1.02066\tabularnewline
WDem & CI5\% & 1.0154 & 1.0160 & 1.0165 & 0.87868 & 0.90425 & 0.92982\tabularnewline
WDem & JE.SDe1\% & 1.0198 & 1.0204 & 1.0211 & 0.81292 & 0.84038 & 0.86784\tabularnewline
WDem & JE.MCD1\% & 1.0217 & 1.0224 & 1.0232 & 0.80975 & 0.83875 & 0.86775\tabularnewline
WDem & JE.Clas1\% & 1.0195 & 1.0200 & 1.0205 & 0.95512 & 0.98171 & 1.00829\tabularnewline
\bottomrule
\end{longtable}

The tabled data confirm that the JE method used with the PaBa regression gives no improvement in the heteroscedastic case (1.0278 for CI vs 1.02886 for the JE.MCD). On the opposite the JE method permit a power gain for the Deming (1.03061 for CI vs 1.021 for the JE.MCD) the MDem (1.02916 for CI vs 1.02254 for the JE.MCD) and also on the less powerful MMDem (1.04172 for CI vs 1.028for the JE.MCD).

For the mixed error the findings are the same: a no significant improvement for the PaBa regression (1.0262 for CI vs 1.02886 for the JE.MCD) and a power gain for the Dem (1.0261 for CI vs 1.02034 for the JE.MCD) the MDem (1.02685 for CI vs 1.0225 for the JE.MCD) and also the MMDem (1.0338 for CI vs 1.028 for the JE.MCD).

The Tab. \ref{tab:powerloss} summarizes these findings

\begin{longtable}[]{@{}ccccccc@{}}
\caption{\label{tab:powerloss}Power summary table for the 100 samples experiments.}\tabularnewline
\toprule
Method & Situation & Dem & MDem & Paba & MMDem & WDem\tabularnewline
\midrule
\endfirsthead
\toprule
Method & Situation & Dem & MDem & Paba & MMDem & WDem\tabularnewline
\midrule
\endhead
CI5\% & Homoscedastic & 1.0238 & 1.0259 & 1.0251 & 1.0300 & 1.0914\tabularnewline
CI5\% & Mixed error & 1.0261 & 1.0269 & 1.0262 & 1.0338 & 1.0598\tabularnewline
CI5\% & Heteroscedastic & 1.0306 & 1.0292 & 1.0278 & 1.0417 & 1.0160\tabularnewline
JE.MCD1\% & Homoscedastic & 1.0166 & 1.0182 & 1.0204 & 1.0208 & 1.0681\tabularnewline
JE.MCD1\% & Mixed error & 1.0203 & 1.0225 & 1.0255 & 1.0280 & 1.0418\tabularnewline
JE.MCD1\% & Heteroscedastic & 1.0210 & 1.0225 & 1.0289 & 1.0280 & 1.0224\tabularnewline
\bottomrule
\end{longtable}

Some conclusions can be drawn about the heteroscedasticity. The first one is that WDem, although attractive at the first glance, is not a reliable solution. The second one is that Deming and MDem JE methods lose less power and retains their advantages compared to the classical CI method.

Thus, the robust MDem regression with the JE method is the robust method of choice also in presence heteroscedasticity. Of course the JE Deming regression performs a little better but in presence of heteroscedasticity the detection of outliers gets even more difficult than in an homoscedastic case and therefore the use of a non-robust method should be carefully evaluated.

It is also worth noting that none of the proposed methods can solve properly this specific problem and further investigations are urgently needed. In particular the role of data transformations should be studied.

\clearpage

\hypertarget{RoleOfTies}{%
\subsection{Ties: the (badly ignored) role of the precision of the methods on the validation}\label{RoleOfTies}}

Very frequently an additional problem arises in method comparisons. All the previous observations are valid with samples that do not contain repeated values. But in reality most instruments have a precision limit and the values are rounded to the best method capability. For several analytes this rounding has a sufficient number of significant digits. But in many other cases the available precision is low and the results are offered with only 3 or even only 2 significant digits. This is particularly true for the so-called ``point-of-care'' medical devices: lightweight, simplified devices that are used by caregivers and are placed next to the patient or carried from patient to patient.

The role of the data precision is investigated with all the available and previously discussed methods for different levels of precision. The lowest level is the 2 digits vs 2 digits precision. Then the precision get raised for one single method to 3 and 4 digits and finally a 3 vs 3 and a 4 vs 4 experiment is also performed.

\hypertarget{the-worst-case-scenario-2-digits-vs-2-digits-precision}{%
\subsubsection{The worst case scenario: 2 digits vs 2 digits precision}\label{the-worst-case-scenario-2-digits-vs-2-digits-precision}}

\begin{figure}

{\centering \includegraphics{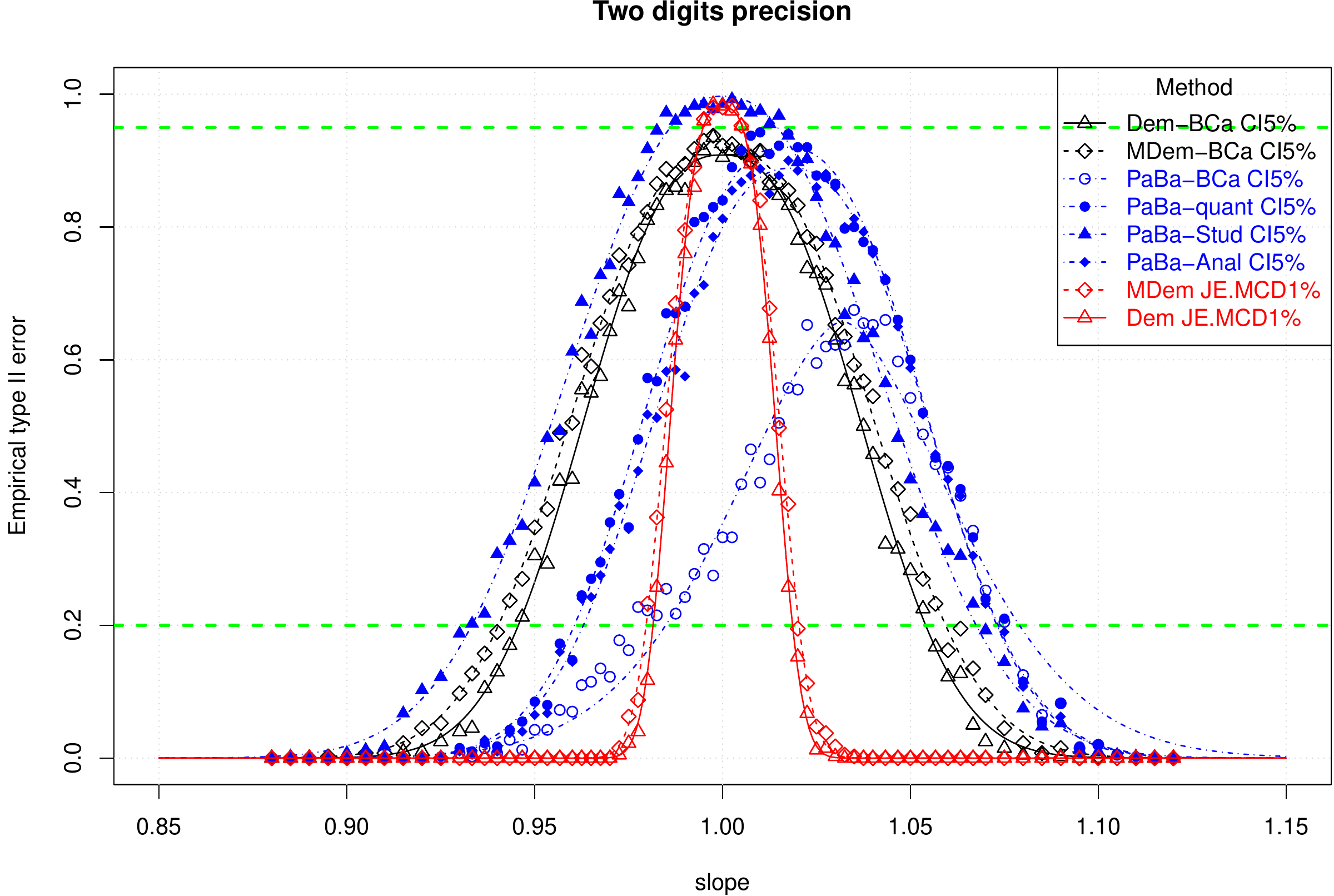} 

}

\caption{Two digit precision, 100 data points and short range.}\label{fig:2vs2}
\end{figure}

The plot in Fig. \ref{fig:2vs2} shows the results of Deming, MDem and PaBa regressions for the short range 100 sample simulations. The JE method is only available for the Deming and the MDem regressions. With the 2 digit precision data the robust covariance determination for the bootstrapped PaBa regressions (of any type) fails because of the very bad quality of the bootstrapped pairs. The computation get stuck repeatedly by singularities. Some convergence issues for the MCD and SDe covariance matrix determination with the 2 digits precision data sets are also registered for the MDem regression if the sample contains only 40 data points. Also in this case the JE power analysis is unfortunately not available. With higher sample sizes the problem disappears for the MDem regression and the JE method behaves as usual.

All the possible mechanisms for the PaBa CI calculation have been recorded. For the bootstrapped family, not only the BCa but also studentized and quantile CI are plotted. Moreover, the non bootstrapped original analytical CI has been determined, too.

The usual empirical type II error plot shows that for all the PaBa classical CI methods a moderate to strong bias is present. Depending upon the type of bootstrap CI determination method, the peak in the empirical type II error plot is not anymore found at the expected slope value of 1. There is an enhanced likelihood to accept slopes with a value higher than 1.

The less biased PaBa classical CI version is the one with the bootstrapped studentized CI where the distortion is visible but still moderate. With the quantile and the BCa CI methods the situation is terrible. Even the original analytical CI calculation (without bootstrap) is strongly biased.

The MDem and Deming regressions do not show any bias either with the classical CI as also with the JE method that is regularly available at sample size of 100 data pairs.

\begin{figure}

{\centering \includegraphics{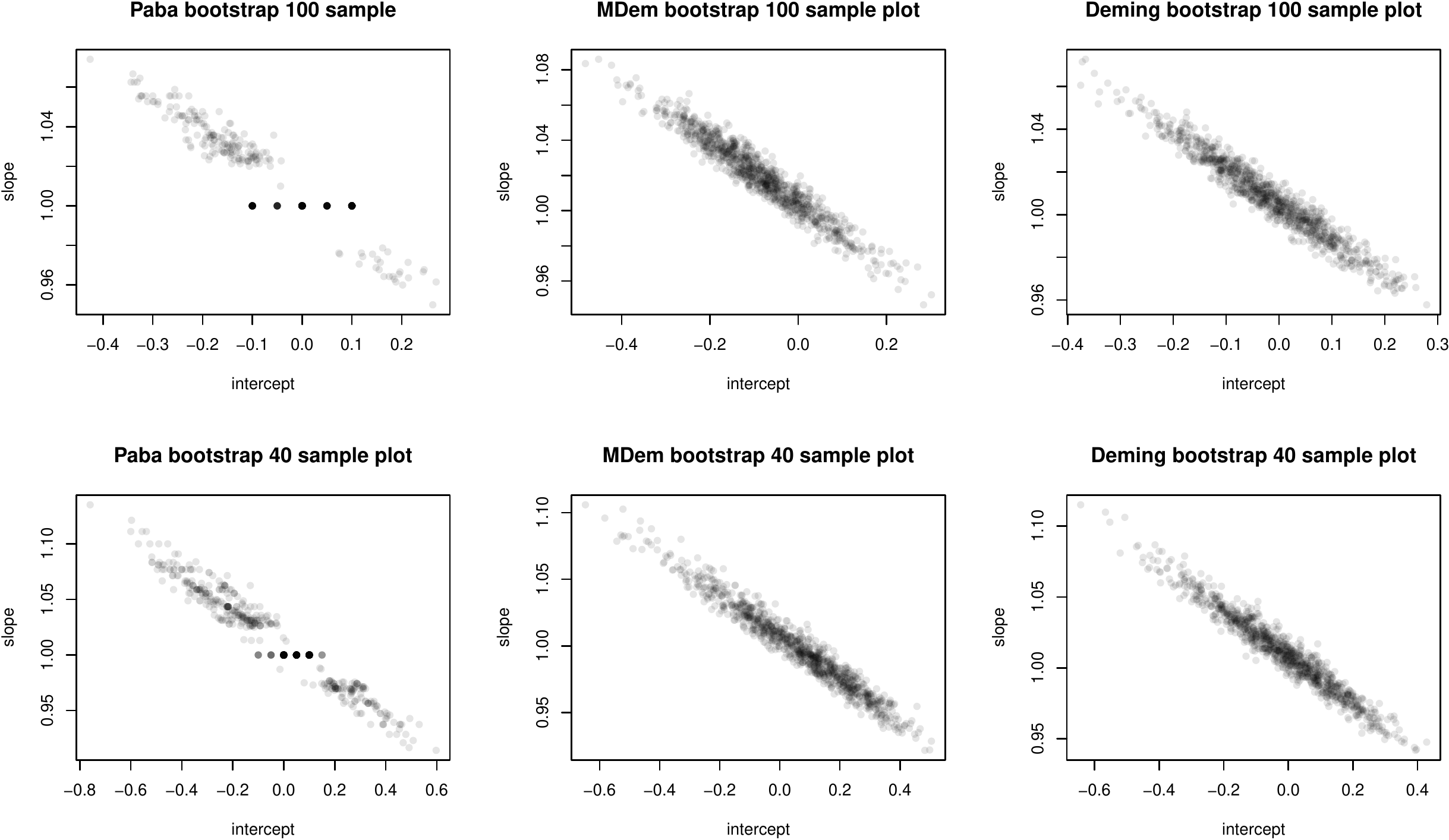} 

}

\caption{Bootstrap pairs plot for the 2 vs 2 digits precision.}\label{fig:2vs2bootplot}
\end{figure}

Typical 2D bootstrapped pair plots are drawn in Fig. \ref{fig:2vs2bootplot}.

Accumulation points caused by the presence of ties in the slope calculation are very manifest in the PaBa regressions in both sample sizes. Some very slight accumulation is also visible in the MDem but only with the smaller sample size. It is also pretty clear that in this kind of analysis the use of a robust covariance matrix should be considered mandatory. The multivariate distributions tend to be quite irregular in their shape.

With the 3 digits precision on the reference instrument and 2 digits on the checked one the obtained bootstrapped values for the PaBa regression are still analyzable with robust covariance matrix methods. This situation could be considered as a simulation for a comparison between a point of care device with a laboratory reference device.

\begin{figure}

{\centering \includegraphics{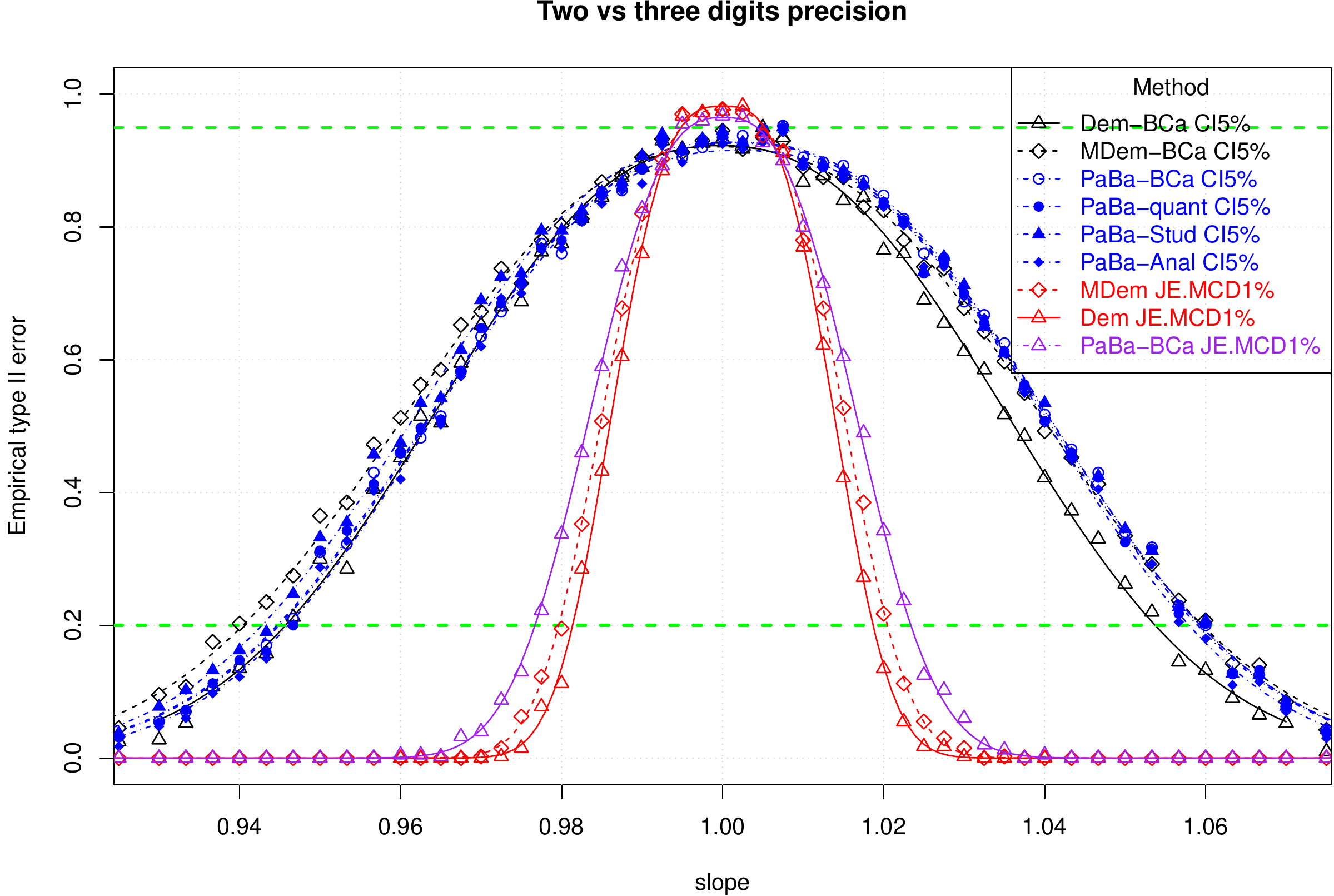} 

}

\caption{2 vs 3 digit precision, 100 data points and short range.}\label{fig:2vs3}
\end{figure}

In the 2 vs 3 digit precision experiment the bias for the CI method is weaker but still well visible if comparing the right and the left side of the PaBa fitted curves (see Fig. \ref{fig:2vs3}); the unbiased Deming regression helps as a reference for the observation.

\begin{figure}

{\centering \includegraphics{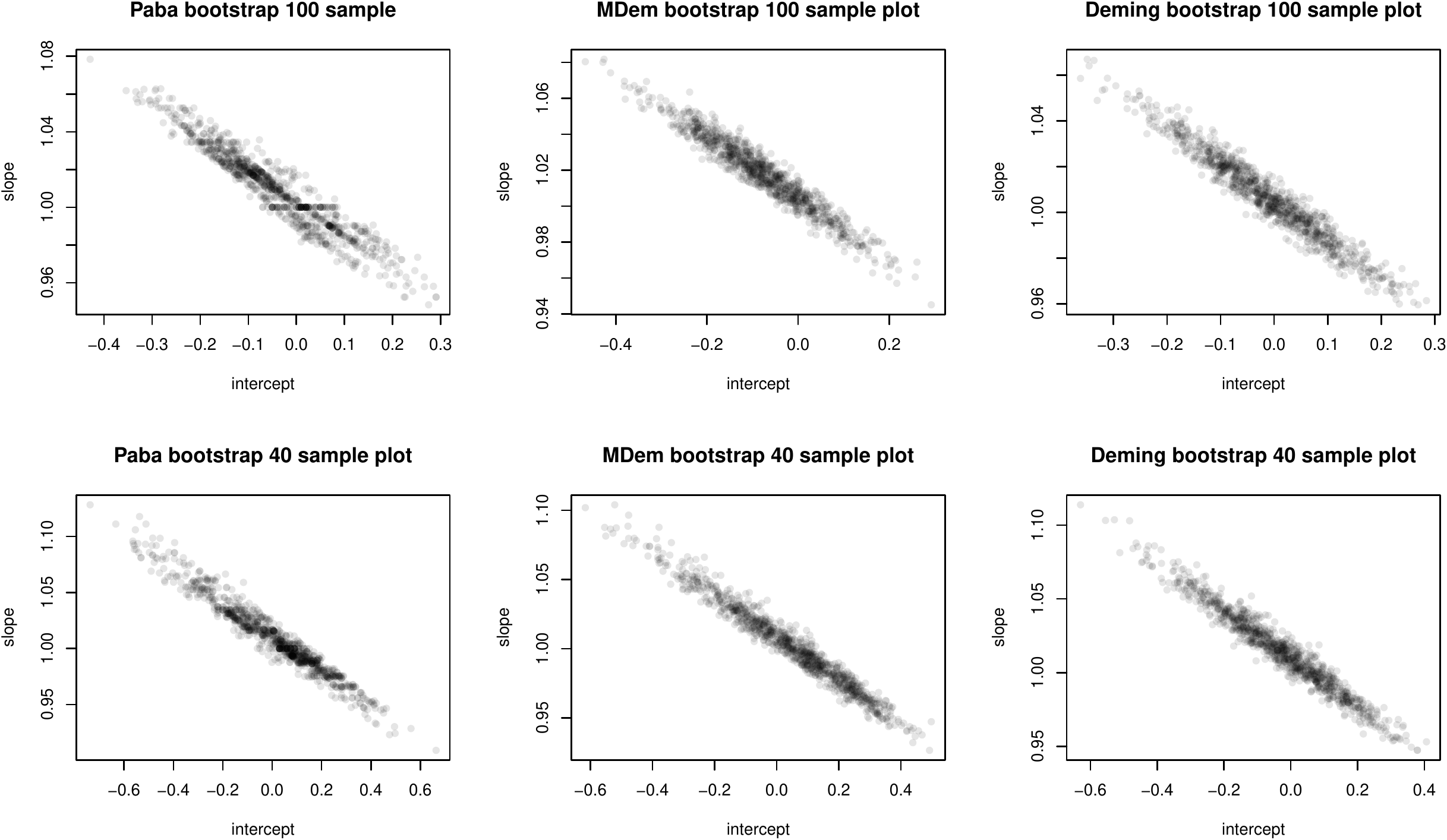} 

}

\caption{Bootstrap pairs plot for the 2 vs 3 digits precision.}\label{fig:3vs2bootplot}
\end{figure}

It is worth noting that all the ellipses are symmetrical and unbiased, even the one obtained from the PaBa BCa bootstrap pairs. The accumulation points in the bootstrap pairs are reduced but not removed as Fig. \ref{fig:3vs2bootplot} shows.

\begin{figure}

{\centering \includegraphics{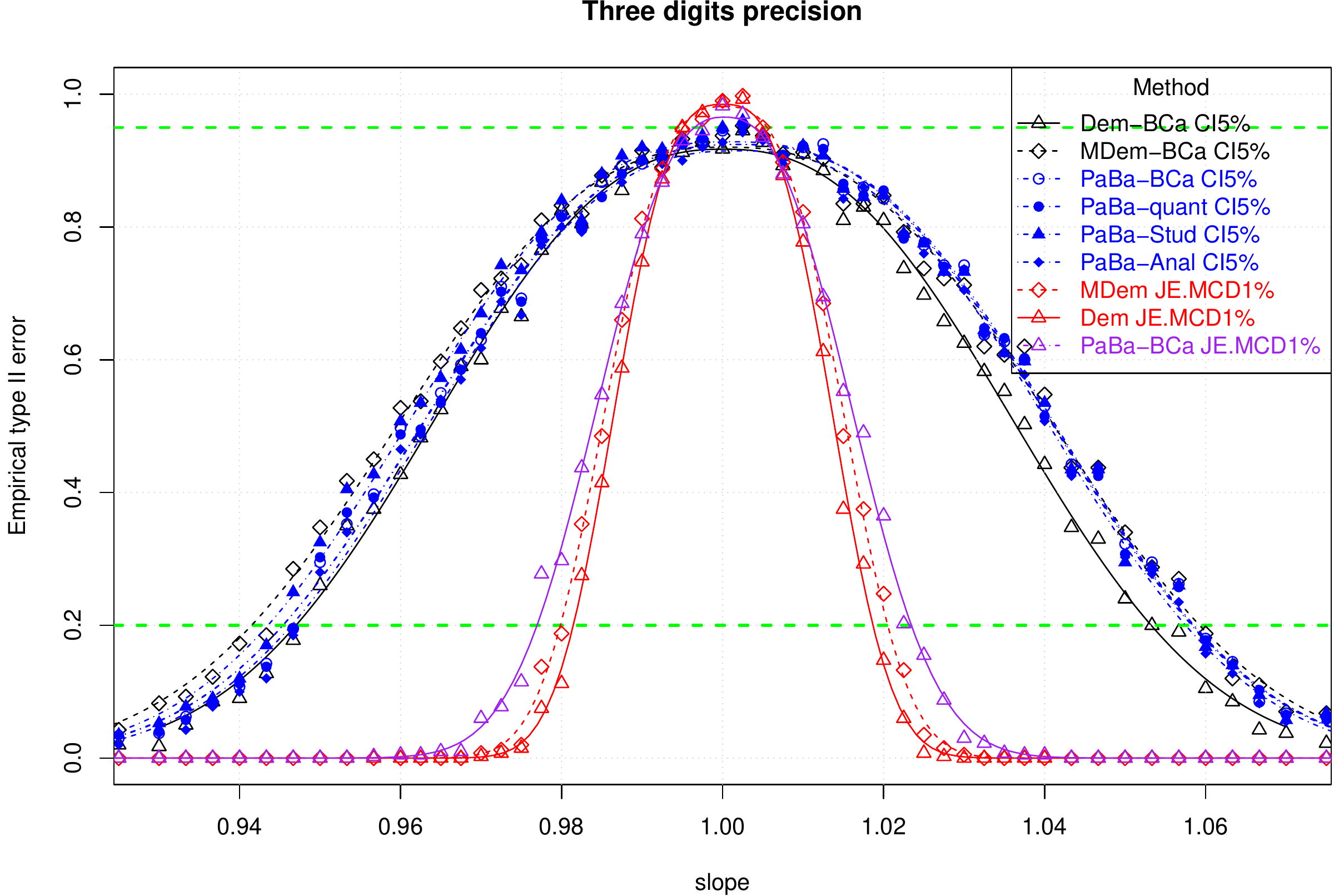} 

}

\caption{3 vs 3 digit precision, 100 data points and short range.}\label{fig:3vs3}
\end{figure}

\begin{figure}

{\centering \includegraphics{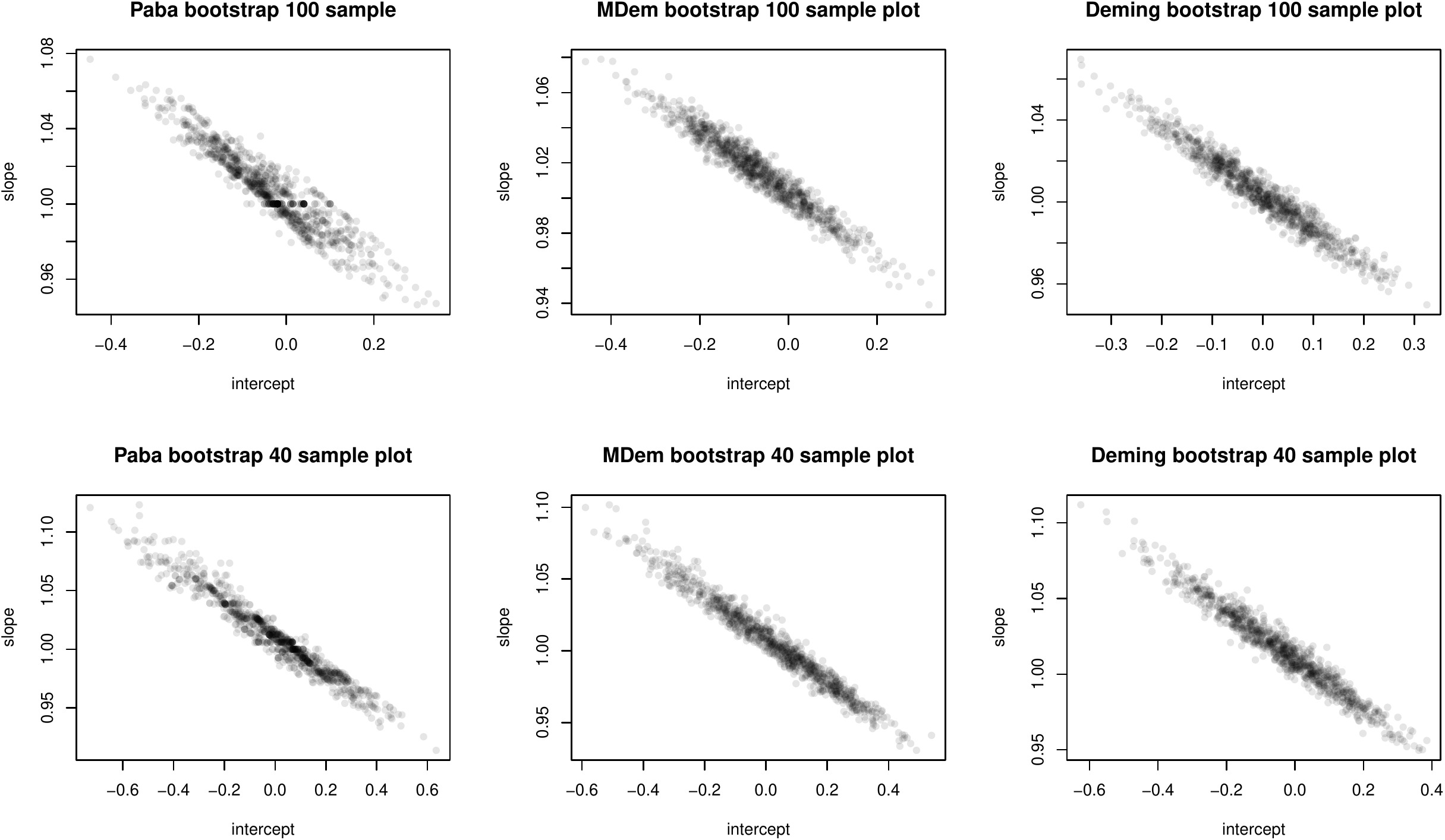} 

}

\caption{Bootstrap pairs plot for the 3 vs 3 digits precision.}\label{fig:3vs3bootplot}
\end{figure}

With a three digit precision for both instruments the bias for the PaBa regression methods is still visible (see Fig. \ref{fig:3vs3}). The related bootstrap pairs plot are reported in Fig. \ref{fig:3vs3bootplot} and are still not ideal.

\begin{figure}

{\centering \includegraphics{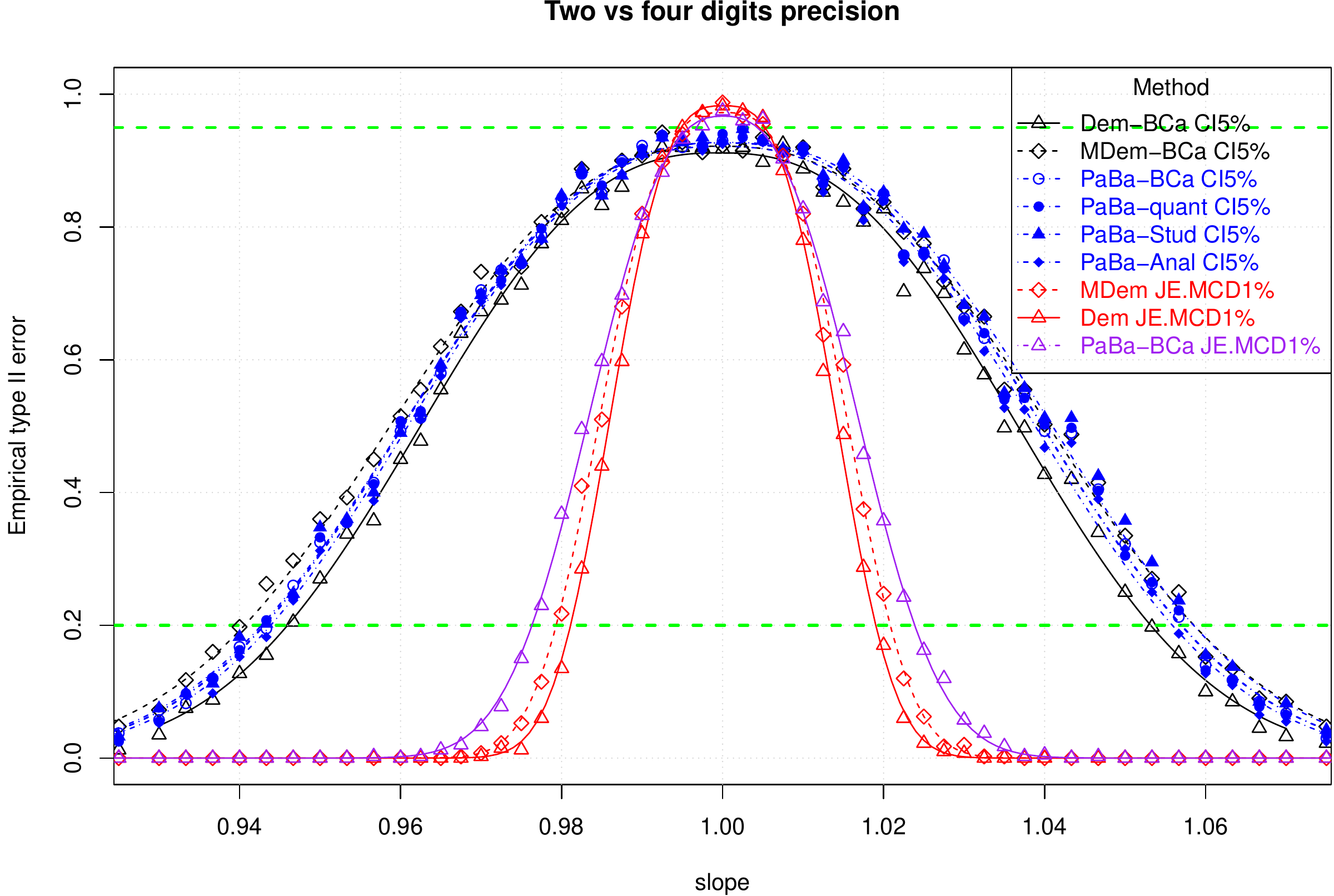} 

}

\caption{2 vs 4 digit precision, 100 data points and short range.}\label{fig:2vs4}
\end{figure}

Apparently a 4 digit precision on at least one instrument is necessary to properly use the PaBa regression. In the 2 vs 4 precision a slight bias is still visible (see Fig. \ref{fig:2vs4}).

\begin{figure}

{\centering \includegraphics{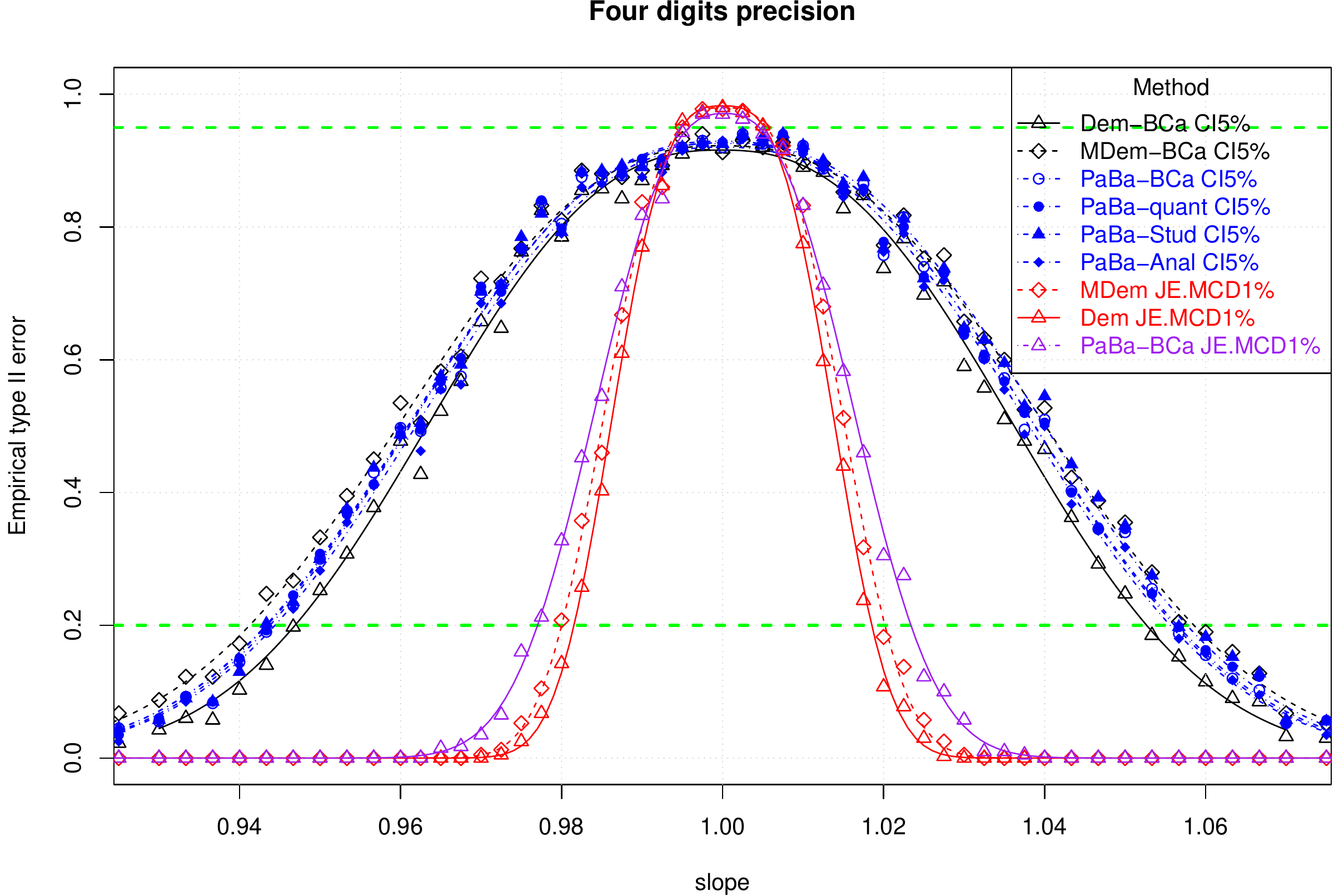} 

}

\caption{4 vs 4 digit precision, 100 data points and short range.}\label{fig:4vs4}
\end{figure}

\begin{figure}

{\centering \includegraphics{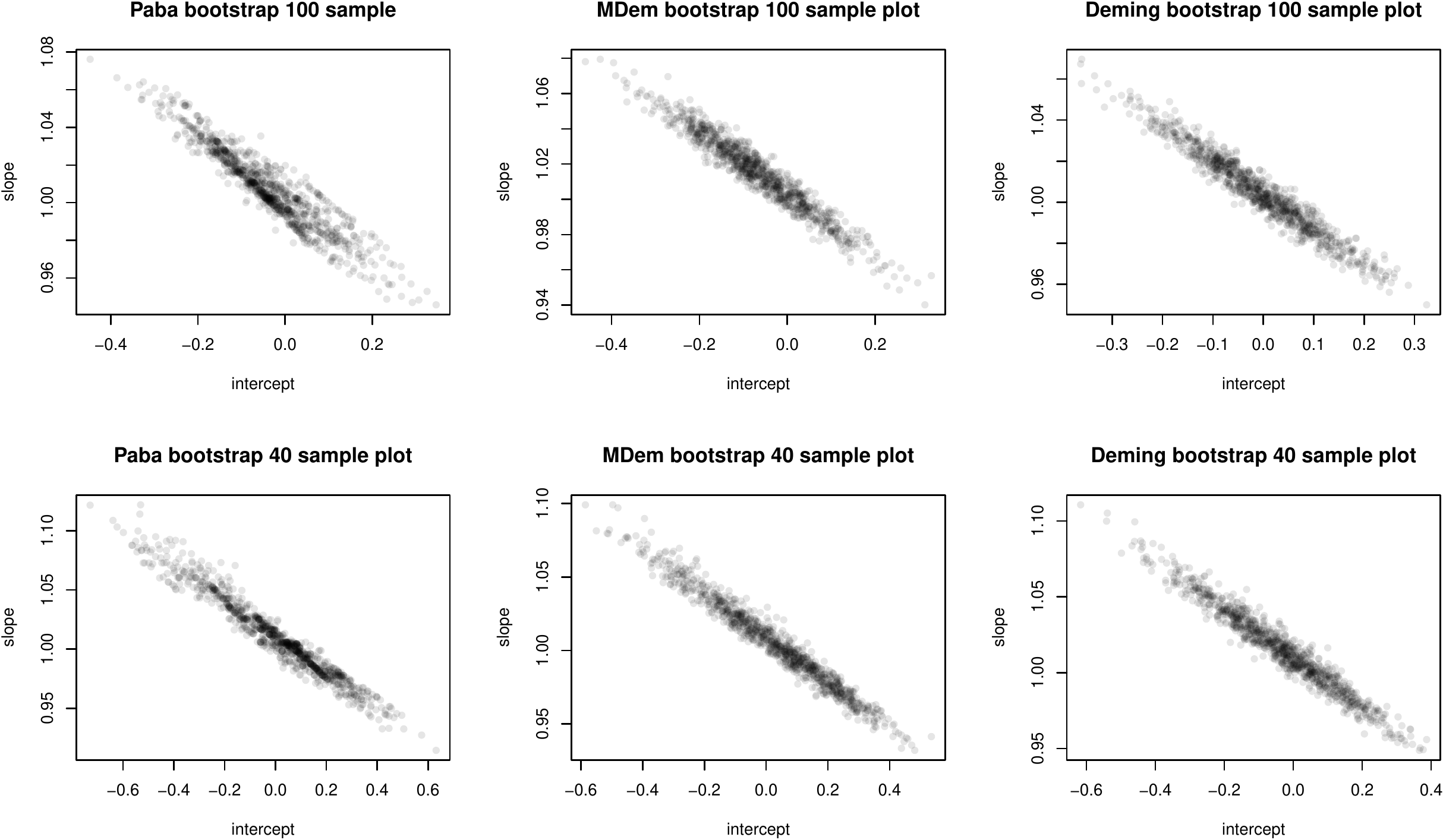} 

}

\caption{Bootstrap pairs plot for the 4 vs 4 digits precision.}\label{fig:4vs4bootplot}
\end{figure}

In the 4 vs 4 digit precision the bias disappears (almost) completely like in Fig. \ref{fig:4vs4}. The 2D bootstrap pairs plot of this last case is reported in Fig. \ref{fig:4vs4bootplot}.

The observed biases are quite astonishing and highly concerning. The reason could be hidden into the mechanism of the slope calculation according to the PaBa method, where an offset is applied in the determination of the median of the pairwise slopes \citep{Passing1983}. It could be argued that in presence of ties, even if the ties are very moderate, the slope determination fails.

\clearpage

\hypertarget{conclusions-about-the-methods}{%
\subsection{Conclusions about the methods}\label{conclusions-about-the-methods}}

A minimum of 4 digits precision on at least one of the instruments, but better on both, is required to use the PaBa regression reliably with the classical CI method. This is a high requirement in the instrument precision that is not frequently met by standard analytical methods.

The MDem regression do not suffer from any bias effect and provides better power especially in presence of moderate heteroscedasticity, as previously discussed.

The MDem is also much better in case of very large samples data sets. The execution time for MDem grows definitely slower in function of the \(n\) number of samples compared to the PaBa method. There is no need to develop an analog PaBaLarge algorithm.

For all these reasons the PaBa regression should be abandoned and replaced by the MDem regression.

The classical Deming regression retains the best results but is not a robust method and should be used with care.

The WDem should be also abandoned because of the very bad instability created by outliers in the low values range. The WDem regression is only an illusory solution for the validation in heteroscedastic conditions.

The power of the JE method largely outperforms the power of the classical CI method. Much smaller sample sizes are needed to attain the same power, although the samples should not be too small to avoid convergence hiccups. This fact can be considered as a break through for all these validation methods that intrinsically have very low range data sets and require very large sample sizes for a full and honest validation. It is still not possible to exactly quantify the reduction ratio that can be attained, but in the observed example a 1/5 ratio could be suggested. Further Monte Carlo simulation should be performed to better assess this ratio.

\clearpage

\hypertarget{application-examples}{%
\section{Application examples}\label{application-examples}}

\hypertarget{creatinine-r-data-set-from-the-mcr-package}{%
\subsection{Creatinine R data set from the \{mcr\} package}\label{creatinine-r-data-set-from-the-mcr-package}}

\begin{figure}

{\centering \includegraphics{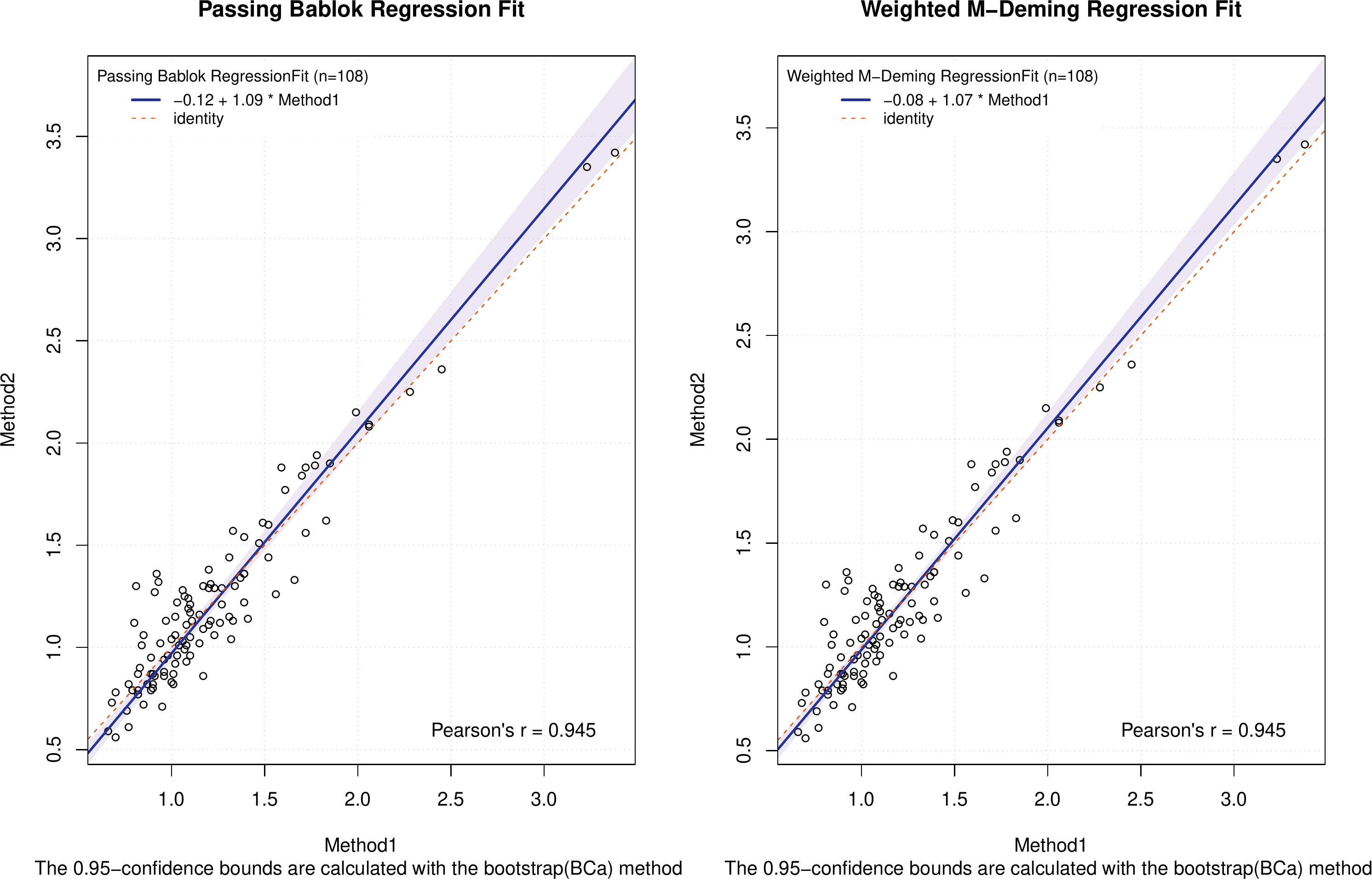} 

}

\caption{Output of {mcr} PaBa and MDem creatinine data set regressions.}\label{fig:mcrcrea}
\end{figure}

The creatinine data set included in the package \{mcr\} is analyzed applying the bootstrap procedure with the PaBa and the MDem regressions (see Fig. \ref{fig:mcrcrea}).

\begin{longtable}[]{@{}ccccc@{}}
\caption{\label{tab:creadata}Cratinine data set validation table.}\tabularnewline
\toprule
Parameter & EST & SE & LCI & UCI\tabularnewline
\midrule
\endfirsthead
\toprule
Parameter & EST & SE & LCI & UCI\tabularnewline
\midrule
\endhead
PaBa - Intercept & -0.11717 & NA & -0.20475 & -0.03427\tabularnewline
PaBa - Slope & 1.08801 & NA & 1.02188 & 1.16854\tabularnewline
MDem - Intercept & -0.08291 & NA & -0.16930 & -0.01615\tabularnewline
MDem - Slope & 1.06891 & NA & 1.02538 & 1.15440\tabularnewline
\bottomrule
\end{longtable}

The classical CI approach leads to a rejection of the \(H_{0}\) (for both the slope and the intercept) whether the test is performed with the PaBa or the Deming regression as reported in Tab. \ref{tab:creadata}.

\begin{figure}

{\centering \includegraphics{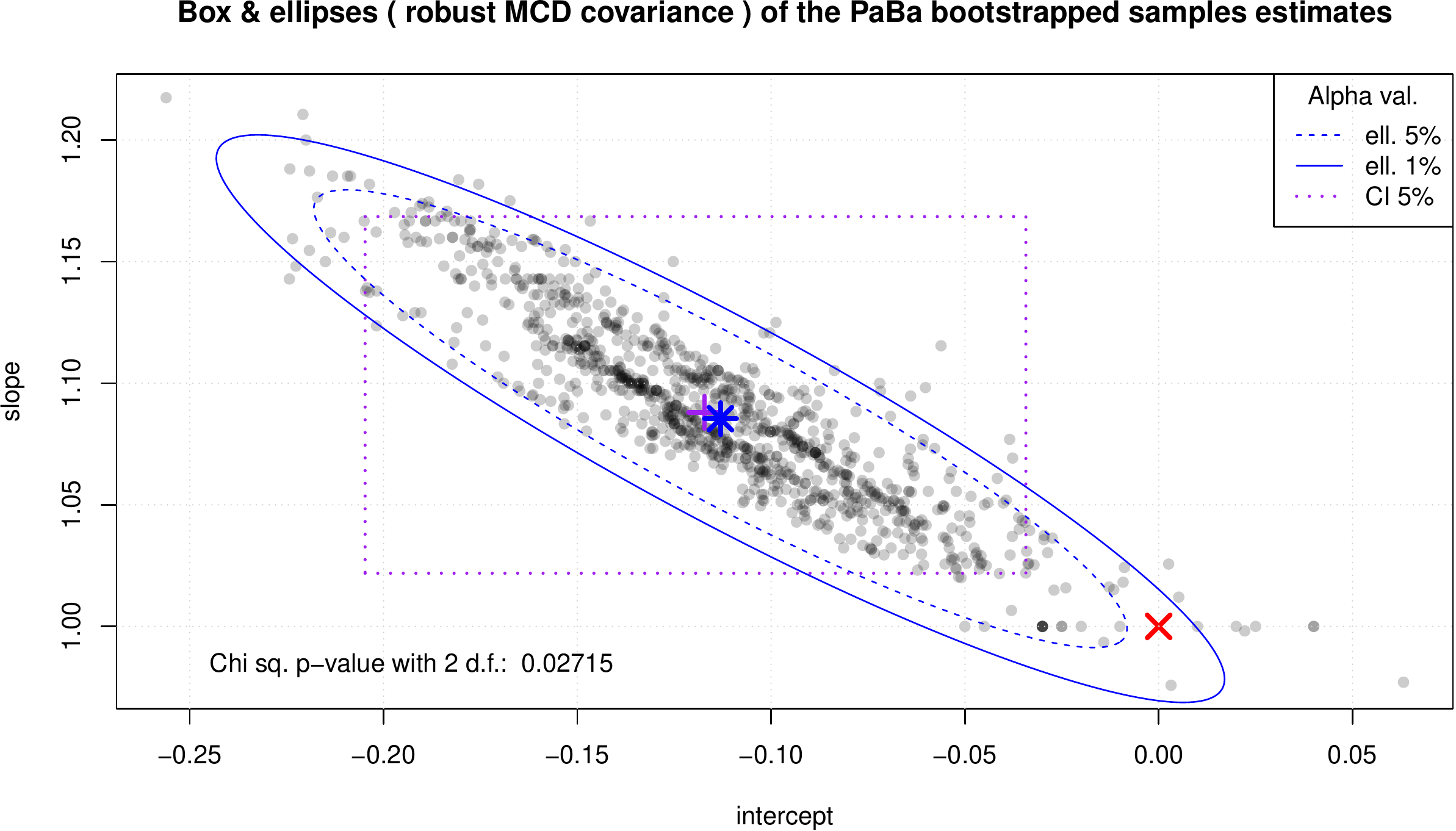} 

}

\caption{Box ellipses plot for the creatinine data set.}\label{fig:creaboxMCD}
\end{figure}

\begin{figure}

{\centering \includegraphics{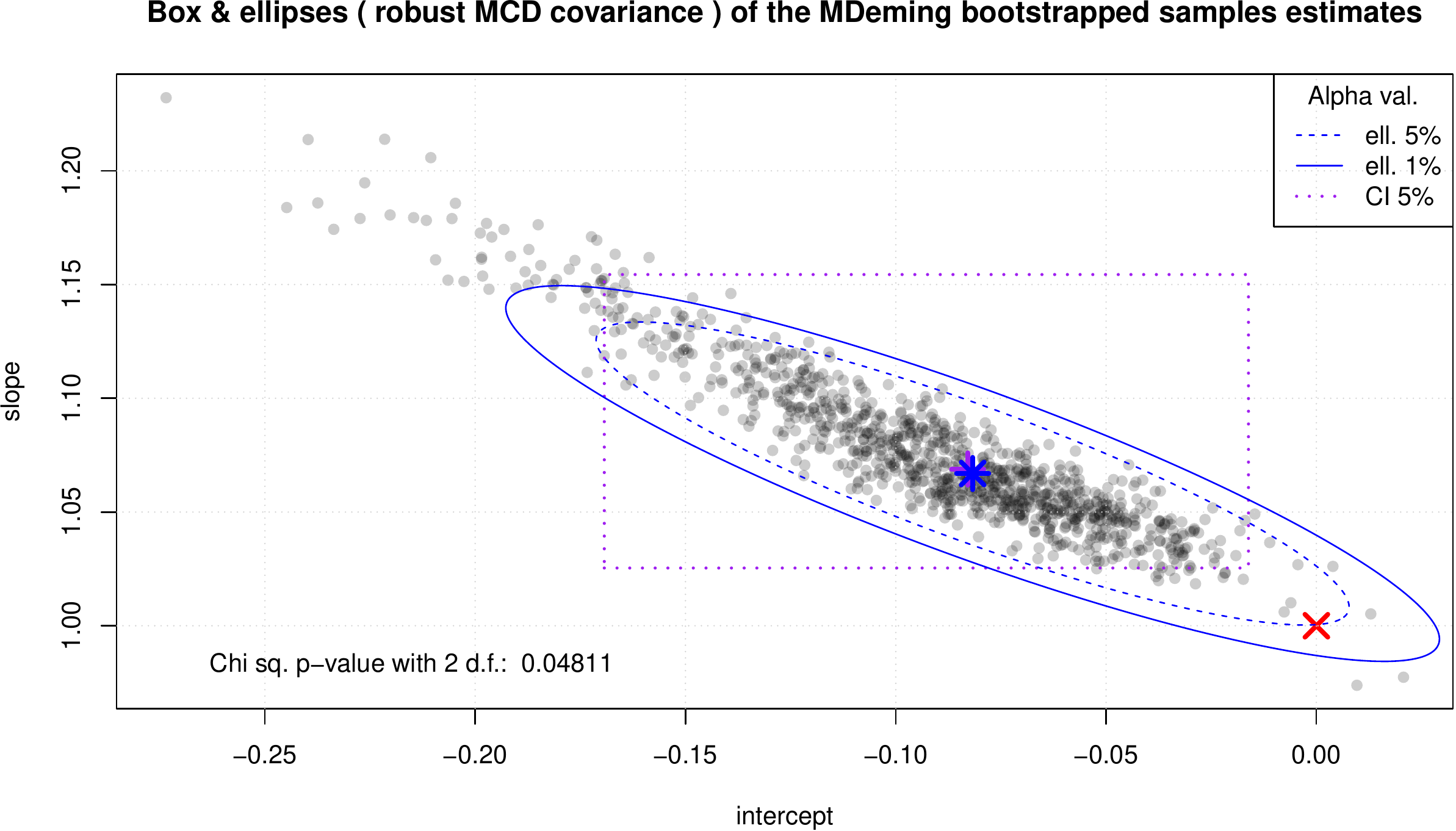} 

}

\caption{Box ellipses plot for the creatinine data set.}\label{fig:creaboxclass}
\end{figure}

The bootstrapped \(\beta_{0}\) and \(\beta_{1}\) can combine into a multivariate distribution through the robust covariance matrix determination. The box ellipses plots for the JE PaBa MCD method (Fig. \ref{fig:creaboxMCD} and the JE MDem one \ref{fig:creaboxclass} show a different picture than the CI table.

Indeed, the JE plots and tests (with Paba and with MDem regression), whatever covariance matrix determination is used (here in the pictures only the MCD method is used) delivers a totally different picture at \(\alpha=0.01\). This one is a much more ``borderline'' case than what it would be barely read from the classical CI table (Tab. \ref{tab:creadata}). Through a visual analysis of the plots and the related \(\chi{2}\) probabilities a disciplinary interpretation leads to a validation.

In Fig. \ref{fig:creaboxMCD} for the PaBa regression are also visible accumulation points in the multivariate distribution that are not present in the analogous plot for the MDem bootstrap pairs (see Fig. \ref{fig:creaboxclass}). For this reason, according to the discussion in Paragraph \ref{RoleOfTies}, the classic PaBa CI method should be considered as suspicious. In fact, the creatinine data set has a 3 digit precision only.

The covariance ellipse and the rectangular box of the classic CI are, for obvious geometrical reasons, badly overlapping. The approach of testing the hypothesis in a step wise fashion appears naif and prone to gross errors. This is especially true if the \(H_{0}\) falls near the bottom left or the upper right corners of the CI rectangle where a clear rejection with the JE method may not take place with the classical CI method, or alternatively near the upper left and bottom right corner where a potential validated case with the JE method may fail with the traditional approach (like in this case).

\pagebreak

\hypertarget{glycated-hemoglobinemodat-extreem-example}{%
\subsection[Glycated hemoglobin (extreem) example]{\texorpdfstring{Glycated hemoglobin\footnote{The data set is owned by the Scuola Superiore Medico Tecnica (SSMT) in Locarno - Switzerland} (extreem) example}{Glycated hemoglobin (extreem) example}}\label{glycated-hemoglobinemodat-extreem-example}}

This is a very controversial historic validation case at the SSMT. For costs reasons the measurements of the glycated hemoglobin were limited to 20 data pairs. The two digits precision would immediately speak against the use of the PaBa regression. But at the time of the validation the bias effect was not known and the calculation were performed as ISO 15189 prescribes with the PaBa regression, with a very questionable sample size of 20 data points\footnote{The minimal sample size of 40 data pairs as prescribed by the ISO 15189 norm is violated.}.

\begin{table}

\caption{\label{tab:emodata}Glycate hemoglobin data set.}
\centering
\resizebox{\linewidth}{!}{
\begin{tabular}[t]{lcccccccccccccccccccc}
\toprule
D10 & 6.4 & 5.6 & 11.1 & 4.9 & 9.0 & 4.7 & 5.7 & 8.4 & 5.4 & 5.8 & 5.6 & 5.9 & 7.5 & 5.7 & 5.6 & 5.8 & 5.4 & 5.2 & 4.9 & 5.9\\
Cobas & 6.0 & 5.9 & 10.6 & 4.7 & 8.2 & 4.8 & 5.5 & 7.6 & 4.9 & 5.3 & 5.3 & 5.6 & 7.1 & 5.2 & 5.3 & 5.5 & 5.2 & 4.8 & 4.7 & 5.7\\
\bottomrule
\end{tabular}}
\end{table}

\begin{figure}

{\centering \includegraphics{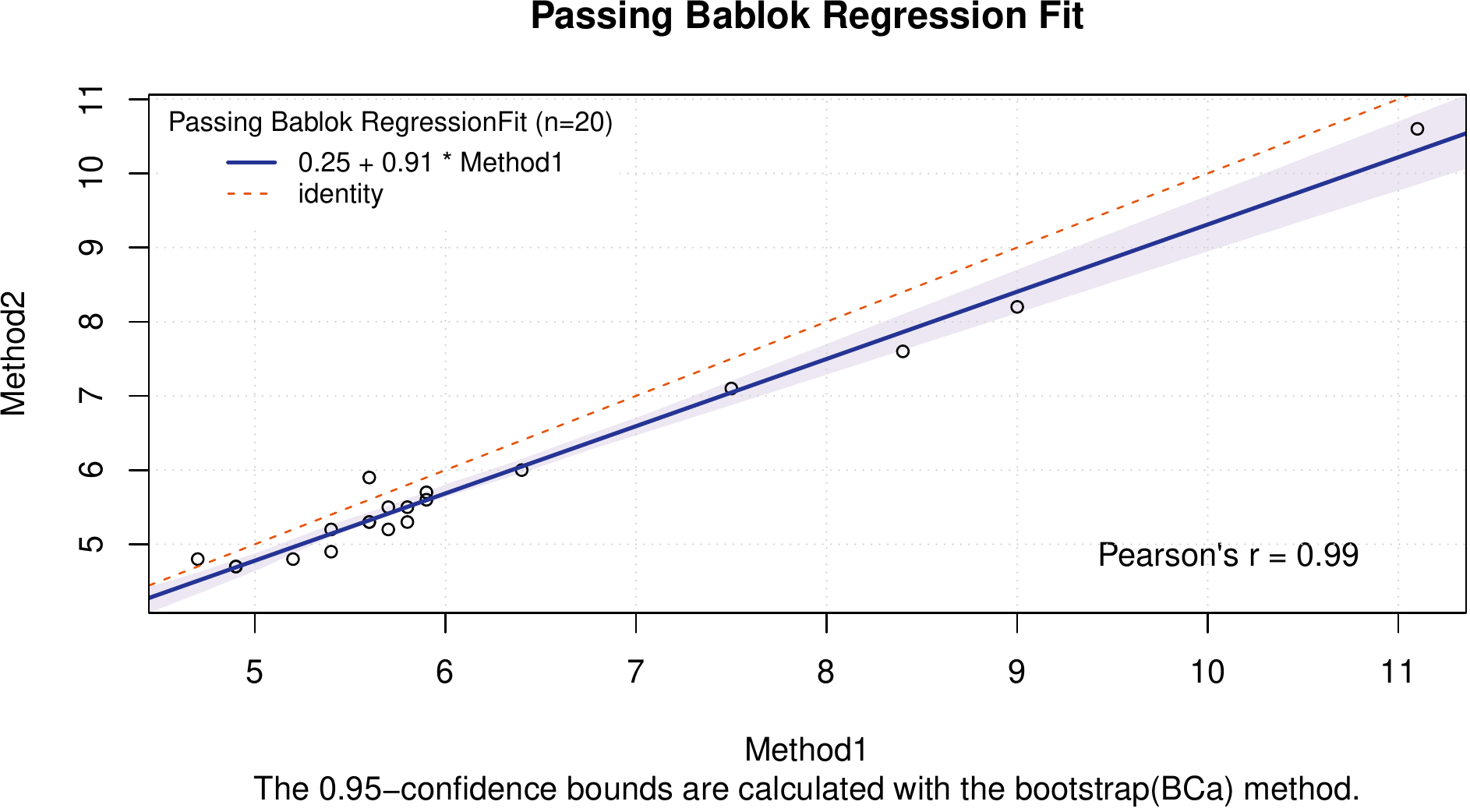} 

}

\caption{Glycated hemoglobin PaBa regression plot.}\label{fig:emopaba}
\end{figure}

At the first glance the plot in Fig. \ref{fig:emopaba} would intuitively suggest a rejection of the validation. The identity line is not running in the grayed CI area.

\begin{longtable}[]{@{}lrrrr@{}}
\caption{\label{tab:emopabatable}Glycate hemoglobin PaBa regression validation table.}\tabularnewline
\toprule
& EST & SE & LCI & UCI\tabularnewline
\midrule
\endfirsthead
\toprule
& EST & SE & LCI & UCI\tabularnewline
\midrule
\endhead
Intercept & 0.24844 & NA & -0.35000 & 0.77235\tabularnewline
Slope & 0.90625 & NA & 0.82008 & 1.00000\tabularnewline
\bottomrule
\end{longtable}

Surprisingly the bootstrapped classical CI coefficients, which constituted the decisive element of the judgment, spoke in the opposite direction. The classical CI testing based on the bootstrap procedure generated a situation in which the UCI boundary of the slope was exactly on the null hypothesis value (see Tab. \ref{tab:emopabatable}). Hence, the limited sample size (and the low precision of the data set) led to a success in the validation.

\pagebreak

The residual analysis showed only a moderate right skewed distribution. No real evidence for heteroscedasticity was visible, but neither could it be excluded because of the reduced sample size.

\begin{figure}

{\centering \includegraphics{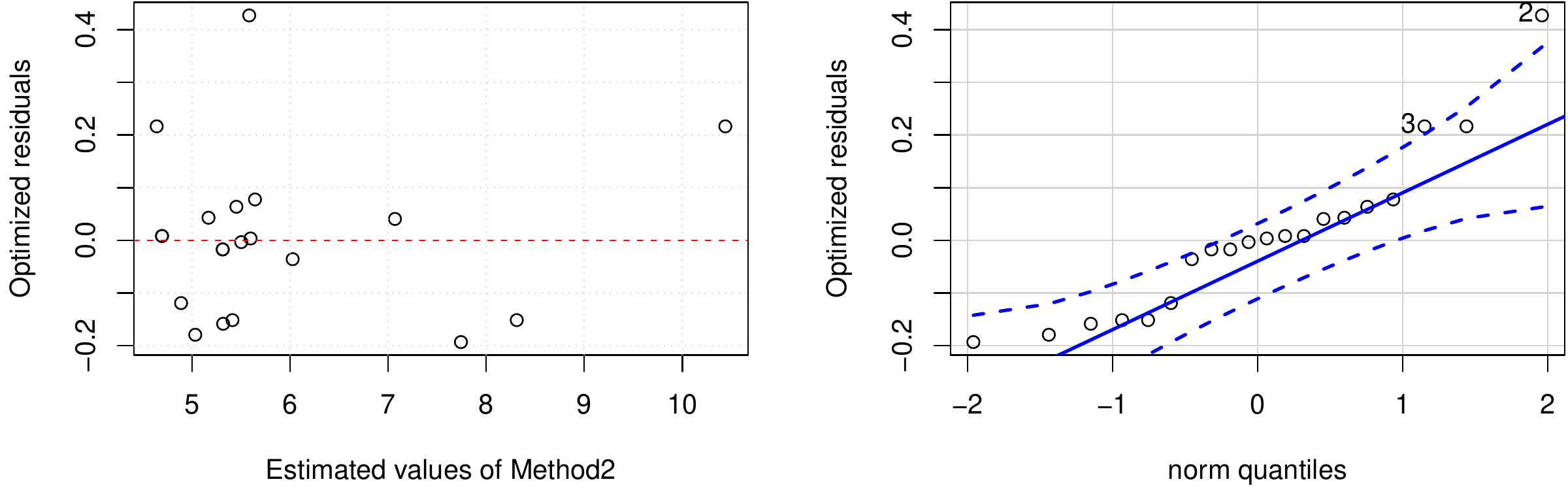} 

}

\caption{Glycated hemoglobin residuals of the PaBa regression.}\label{fig:emopabaresid}
\end{figure}

Moving toward a 2D box ellipses plot representation of the bootstrap estimate pairs could have been decisive.

\begin{figure}

{\centering \includegraphics{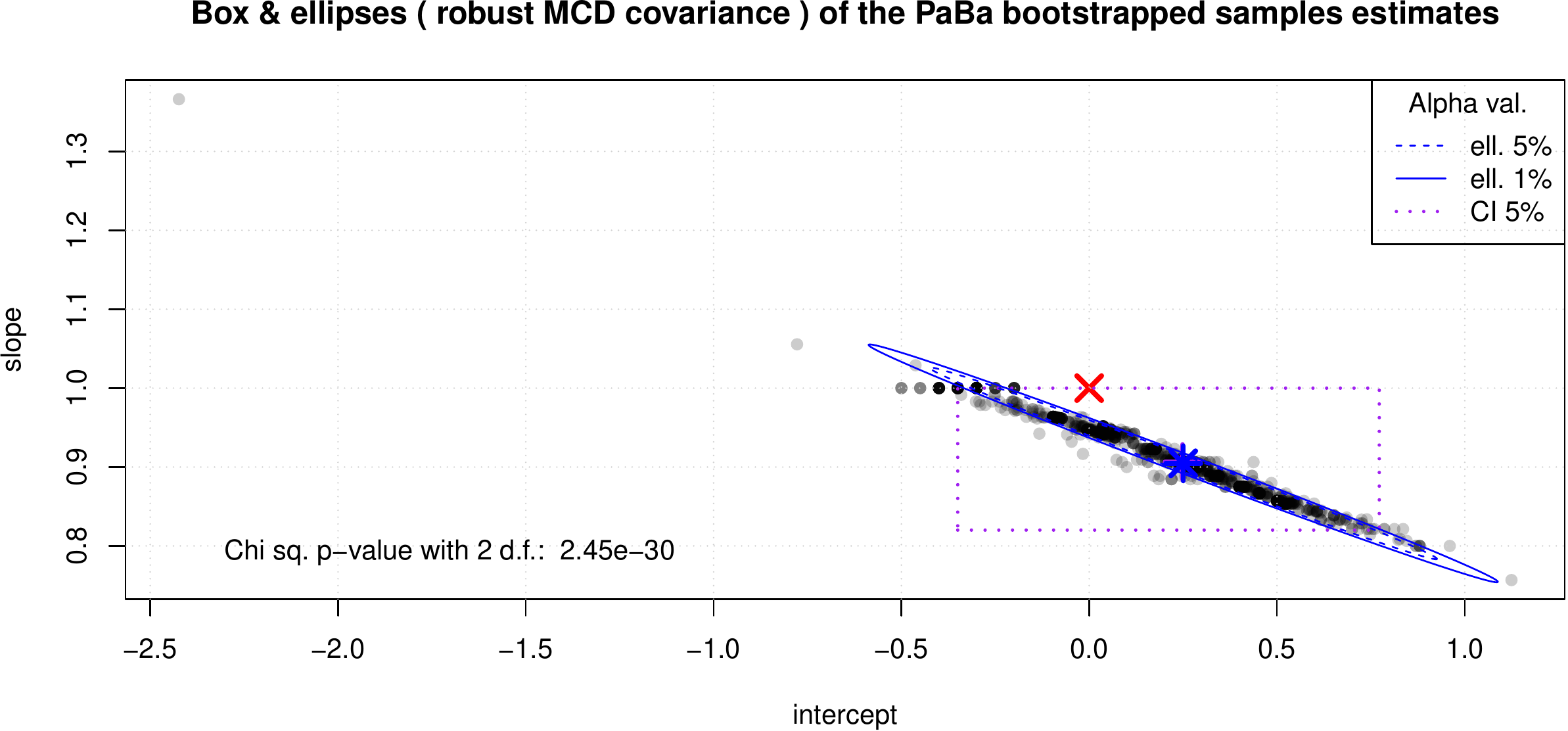} 

}

\caption{Box ellipses plot for the glycated hemoglobine PaBa regression.}\label{fig:emoboxpaba}
\end{figure}

The plot in Fig. \ref{fig:emoboxpaba} clearly shows that the null hypothesis is far away from the ellipse of the bootstrapped pairs. Once again a clear decision can only be taken after a visual inspection of this type of graphical representation. The observation of the 2D plot is according to the overall \(\chi^2\) test result that shows a very low p-value.

The limited number of data points combined with the ties generated by the bootstrap resampling process of low precision data led to a surprisingly high population of slope \(=1\) (visible in upper left corner of the CI box in Fig. \ref{fig:emoboxpaba}) values and to the counterintuitive positive validation result with the CI table. A robust covariance matrix based algorithm can easily remove the distortion and correctly reject the validation.

\begin{figure}

{\centering \includegraphics{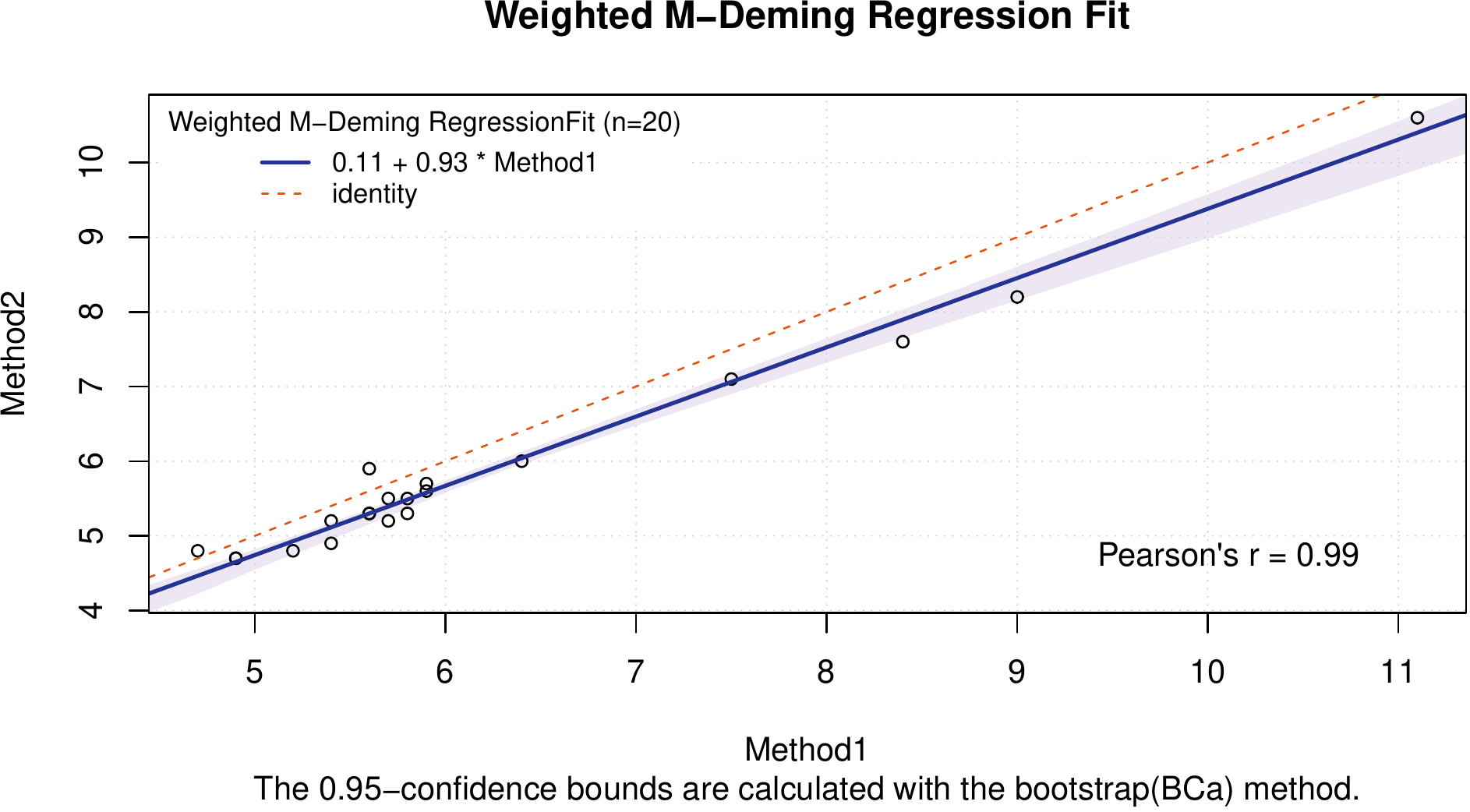} 

}

\caption{Gylcate hemoglobin validation MDem regression plot.}\label{fig:emomdem}
\end{figure}

It is worth noting that the same case with the MDem approach would have provided the following regression plot shown in Fig. \ref{fig:emomdem}.

\begin{longtable}[]{@{}lrrrr@{}}
\caption{\label{tab:emomdemtable}Glycate emoglobin MDem regression validation table.}\tabularnewline
\toprule
& EST & SE & LCI & UCI\tabularnewline
\midrule
\endfirsthead
\toprule
& EST & SE & LCI & UCI\tabularnewline
\midrule
\endhead
Intercept & 0.10586 & NA & -0.33371 & 0.65456\tabularnewline
Slope & 0.92743 & NA & 0.83342 & 0.98283\tabularnewline
\bottomrule
\end{longtable}

The MDem regression do not suffer from the low precision of the data set. The classical CI are slightly narrower (Tab. \ref{tab:emomdemtable}) and with the classical CI approach the validation would have been safely rejected.

\pagebreak

\begin{figure}

{\centering \includegraphics{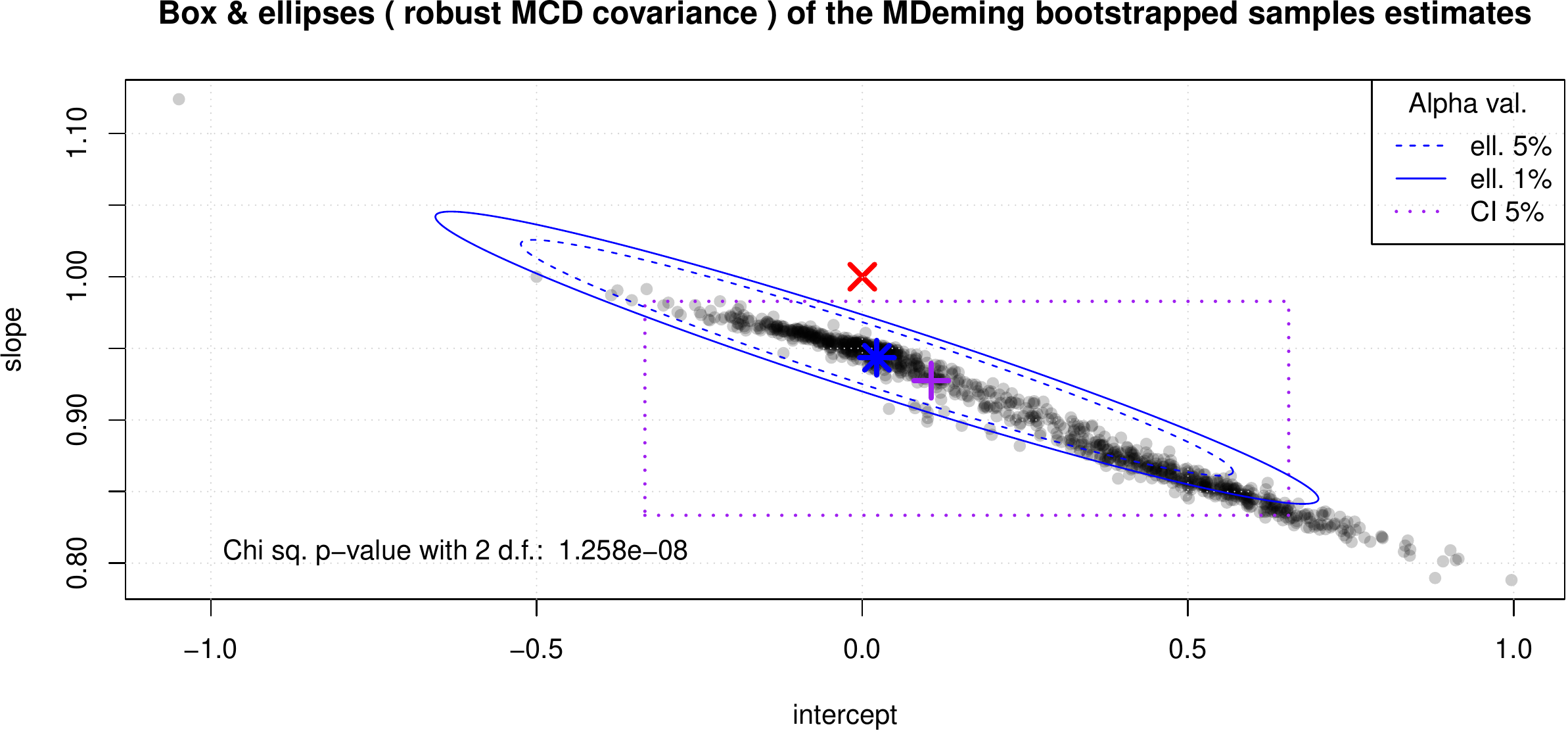} 

}

\caption{Box ellipses plot for the glycated hemoglobine MDem regression.}\label{fig:emoboxmdeming}
\end{figure}

The box ellipses plot (see Fig. \ref{fig:emoboxmdeming}) shows a coherent situation of a clear rejection.

\clearpage

\hypertarget{conclusions}{%
\section{Conclusions}\label{conclusions}}

A new visual type of approach for the interpretation of the results of method validation tests is presented.

The 2D visual inspection of the bootstrap pairs takes a central role. A single multivariate test based on the use of a robust covariance estimation substitutes the separated test for slope and intercept and helps for the visual interpretation. For sake of completeness also the box corresponding to the classical CI is drawn on the same plot.

The FWER is managed properly with the JE method, without the risk of an enhanced rejection caused by a repeated 95\% testing. Moreover, the robustness of the covariance determination offers the warranty that the role of distortions and outliers in the multivariate bootstrap data pair distribution is reduced.

In modern statistics many \(H_{0}\) confirming test like the normality tests are less and less used and a visual approach (like the use of Q-Q plots) is more and more preferred. The ``absence of evidence is not evidence of absence'' principle gets every day more attention. This 2D plot technique, combined with a single \(\chi^2\) helper test could become a central tool for the correct interpretation of the data.

\hypertarget{aknowledgments}{%
\section{Aknowledgments}\label{aknowledgments}}

This work has been performed as a Diploma work for the DAS course in applied statistics at the ETH-Zürich. Special thanks go to Prof.~Andreas Ruckstuhl who supervised the writing of this thesis and contributed to the scientific discussion. In addition, special thanks go also to Dr Lukas Meier, the DAS course director, and to the computing cluster teams of the ETH (leonhard.open.ethz.ch and euler.ethz.ch) for the support in the computational intensive Monte Carlo simulations. Finally thanks are due to the Scuola Superiore Medico Tecnica (Locarno-Switzerland) for making available the data set used as an example.

\clearpage

\hypertarget{appendix}{%
\section{Appendix}\label{appendix}}

\hypertarget{DataGenFunc}{%
\subsection{Appendix A: the data generation function}\label{DataGenFunc}}

\hypertarget{homoscedastic-data-set-generator}{%
\subsubsection{Homoscedastic data set generator}\label{homoscedastic-data-set-generator}}

\begin{verbatim}
## function (xmin, xmax, n = 40, slope = 1, intercept = 0, sigmax = 1, 
##     sigmay = 1) 
## {
##     xr <- runif(n, xmin, xmax)
##     yr <- slope * xr + intercept
##     xe <- rnorm(1:length(xr), mean = 0, sd = sigmax)
##     ye <- rnorm(1:length(yr), mean = 0, sd = sigmay)
##     detmin <- 0.1
##     X <- ifelse((xr + xe) > detmin, xr + xe, detmin/2)
##     Y <- ifelse((yr + ye) > detmin, yr + ye, detmin/2)
##     return(data.frame(cbind(X = X, Y = Y)))
## }
## <bytecode: 0x55ae3301a9f0>
\end{verbatim}

\hypertarget{mixed-heteroscedastic-data-set-generator}{%
\subsubsection{Mixed heteroscedastic data set generator}\label{mixed-heteroscedastic-data-set-generator}}

\begin{verbatim}
## function (xmin, xmax, n = 40, slope = 1, intercept = 0, sigmax = 1, 
##     sigmay = 1) 
## {
##     xr <- runif(n, xmin, xmax)
##     yr <- slope * xr + intercept
##     xe <- xr/mean(xr) * rnorm(1:length(xr), mean = 0, sd = sigmax)
##     ye <- yr/mean(yr) * rnorm(1:length(yr), mean = 0, sd = sigmay)
##     detmin <- 0.1
##     X <- ifelse((xr + xe) > detmin, xr + xe, detmin/2)
##     Y <- ifelse((yr + ye) > detmin, yr + ye, detmin/2)
##     return(data.frame(cbind(X = X, Y = Y)))
## }
## <bytecode: 0x55ae30fcb7b8>
\end{verbatim}

\cleardoublepage

\hypertarget{pure-heteroscedastic-data-set-generator}{%
\subsubsection{Pure heteroscedastic data set generator}\label{pure-heteroscedastic-data-set-generator}}

\begin{verbatim}
## function (xmin, xmax, n = 40, slope = 1, intercept = 0, sigmax = 1, 
##     sigmay = 1) 
## {
##     xr <- runif(n, xmin, xmax)
##     yr <- slope * xr + intercept
##     ranx <- rnorm(1:length(xr), mean = 0, sd = sigmax)
##     rany <- rnorm(1:length(yr), mean = 0, sd = sigmay)
##     xe <- xr/mean(xr) * ranx/2 + ranx/2
##     ye <- yr/mean(yr) * rany/2 + rany/2
##     detmin <- 0.1
##     X <- ifelse((xr + xe) > detmin, xr + xe, detmin/2)
##     Y <- ifelse((yr + ye) > detmin, yr + ye, detmin/2)
##     return(data.frame(cbind(X = X, Y = Y)))
## }
## <bytecode: 0x55ae31a5d690>
\end{verbatim}

\hypertarget{homoscedastic-data-set-generator-with-limited-precision}{%
\subsubsection{Homoscedastic data set generator with limited precision}\label{homoscedastic-data-set-generator-with-limited-precision}}

\begin{verbatim}
## function(xmin,xmax,n=40,slope=1,intercept=0,sigmax=1,sigmay=1){
##   xr<-runif(n,xmin,xmax)
##   yr<-slope*xr+intercept
##   xe<-rnorm(1:length(xr),mean=0,sd=sigmax)
##   ye<-rnorm(1:length(yr),mean=0,sd=sigmay)
##   detmin<-0.1
##   X<-ifelse((xr+xe) > detmin, signif(xr+xe,4),detmin/2)
##   Y<-ifelse((yr+ye) > detmin, signif(yr+ye,4),detmin/2)
##   return(data.frame(cbind(X=X,Y=Y)))
## }
## <bytecode: 0x55ae32be8ba0>
\end{verbatim}

\nopagebreak

\hypertarget{RejectionLong}{%
\subsection{Appendix B: empirical rejection plot for the long range experiments}\label{RejectionLong}}

\begin{figure}[H]

{\centering \includegraphics{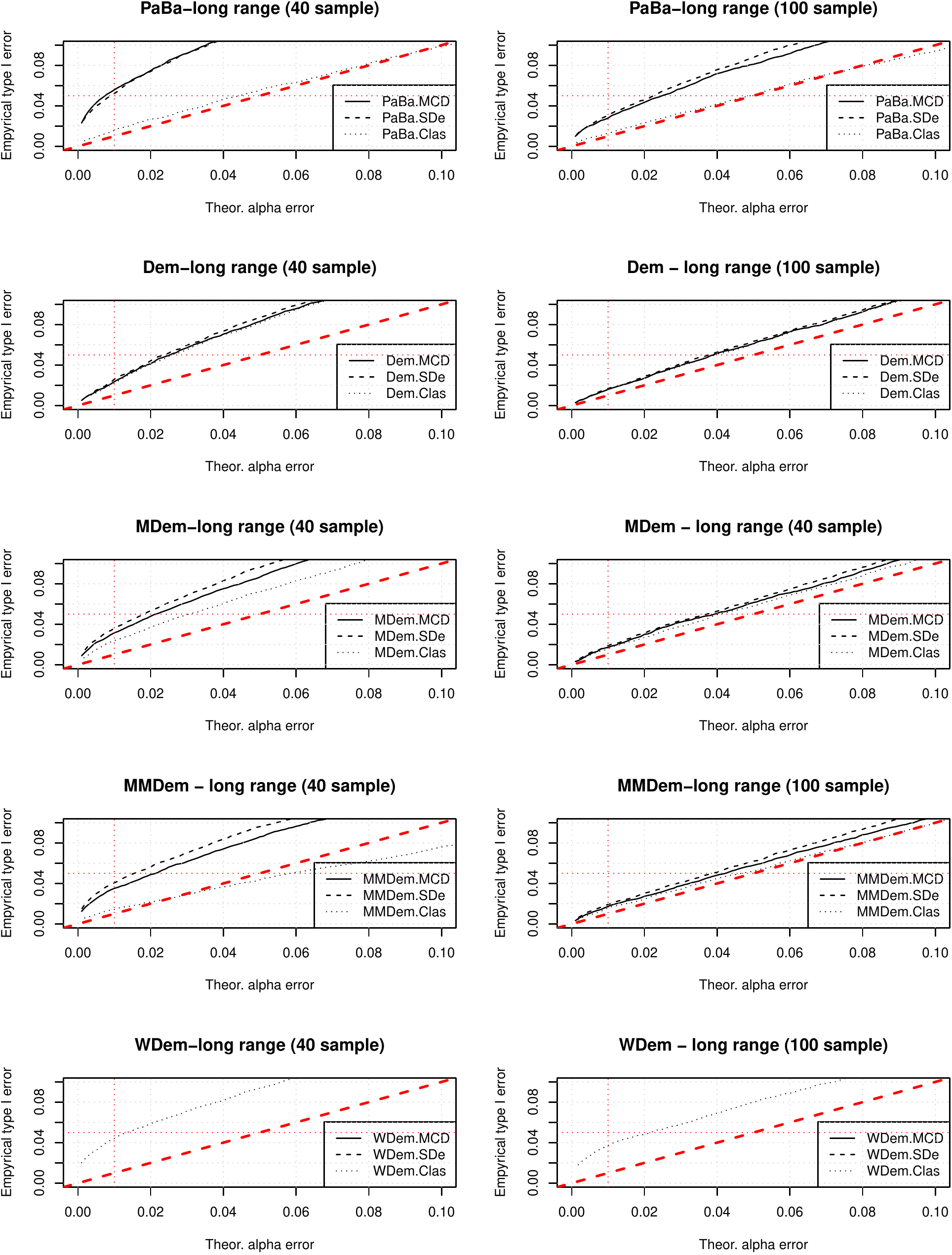} 

}

\caption{P-P plot with long range data comparing 40 and 100 samples data set results.}\label{fig:rejectionlong}
\end{figure}
\clearpage

\hypertarget{SampleRange}{%
\subsection{Appendix C: the plots for Dem, MMDem, Paba and WDem}\label{SampleRange}}

\begin{figure}

{\centering \includegraphics{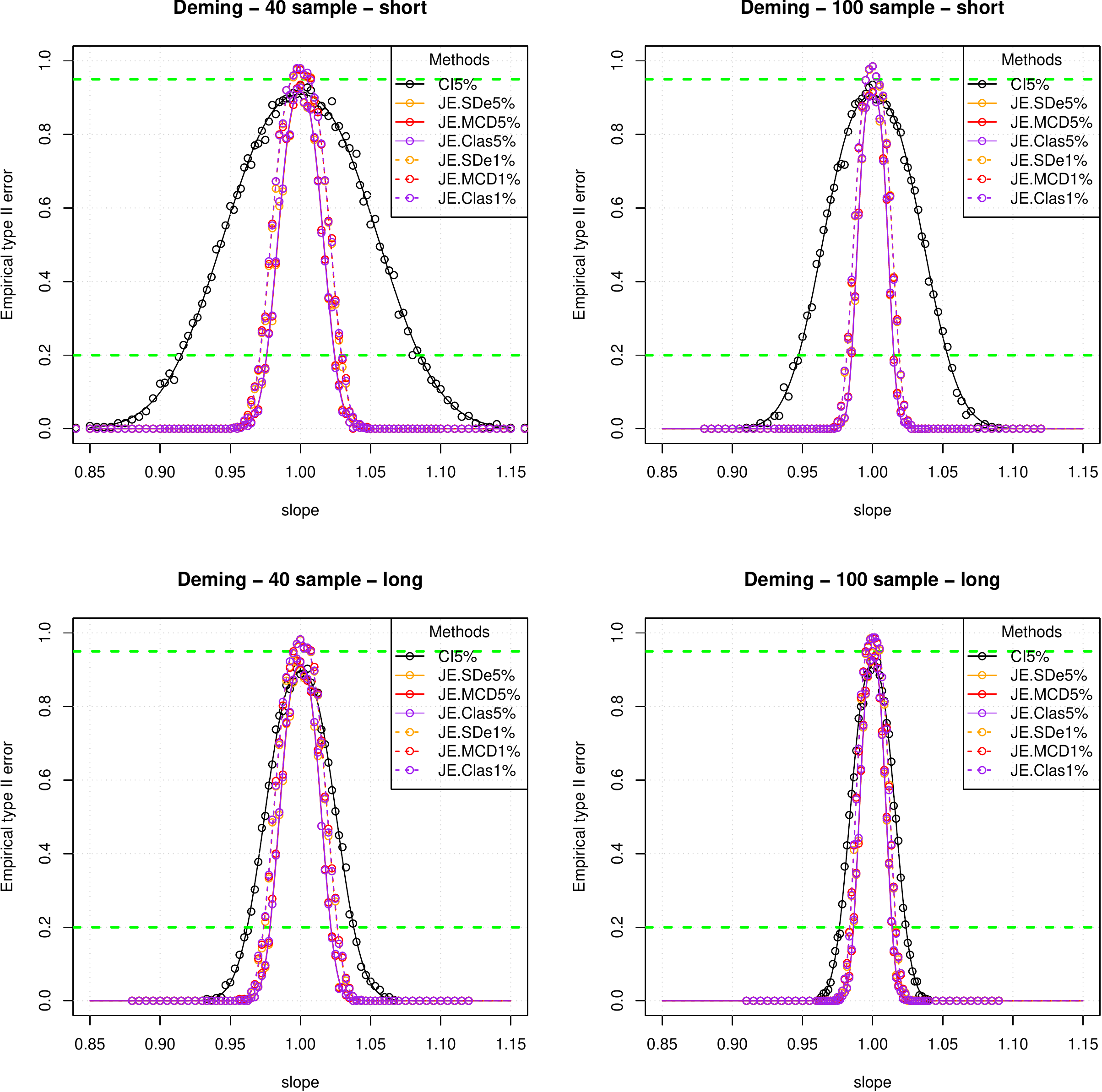} 

}

\caption{Classic CI vs JE type II errors for Passing Bablok regression.}\label{fig:dcompare}
\end{figure}

\pagebreak

\begin{figure}

{\centering \includegraphics{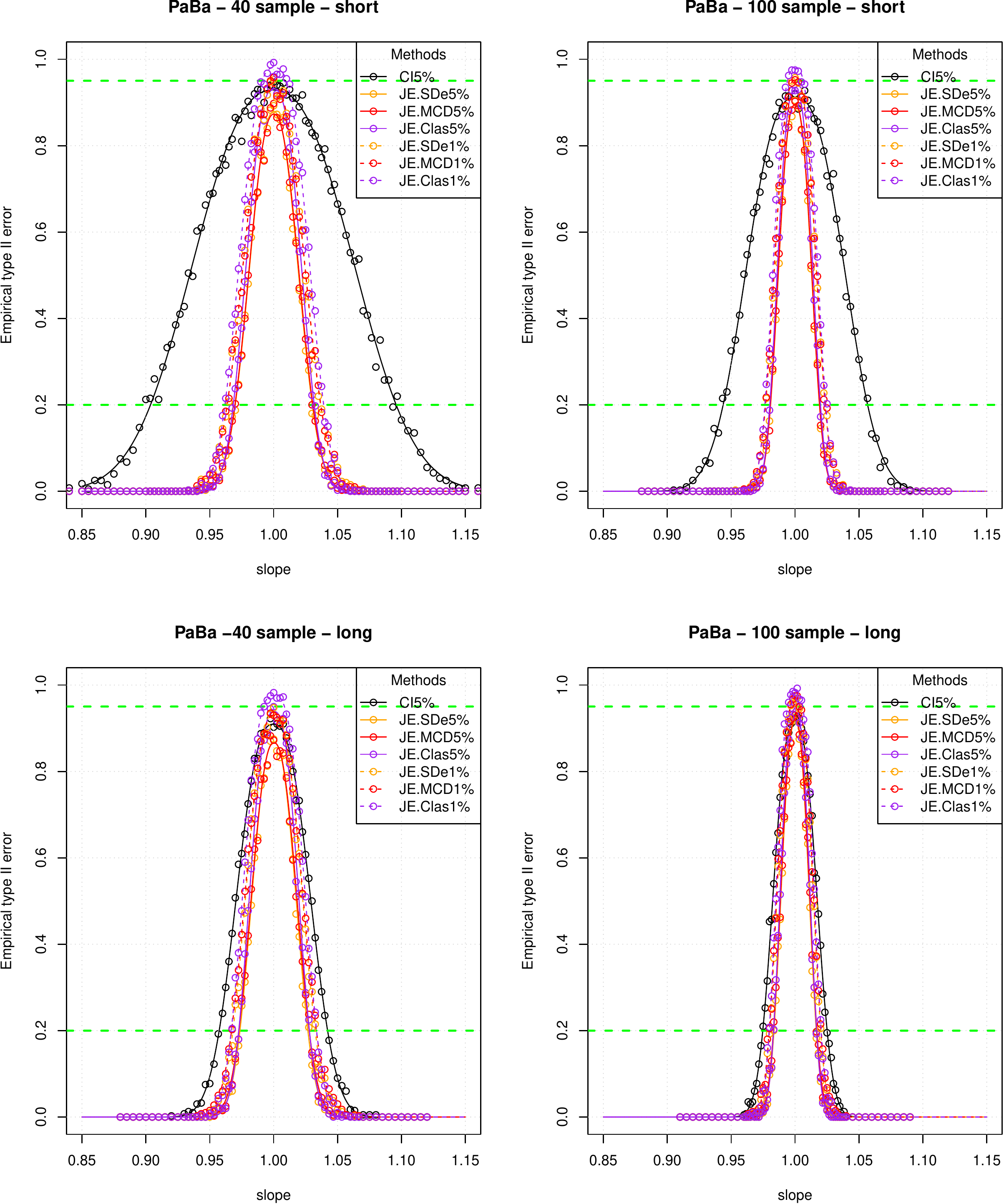} 

}

\caption{Classic CI vs JE type II errors for Passing Bablok regression.}\label{fig:pcompare}
\end{figure}

\pagebreak

\begin{figure}

{\centering \includegraphics{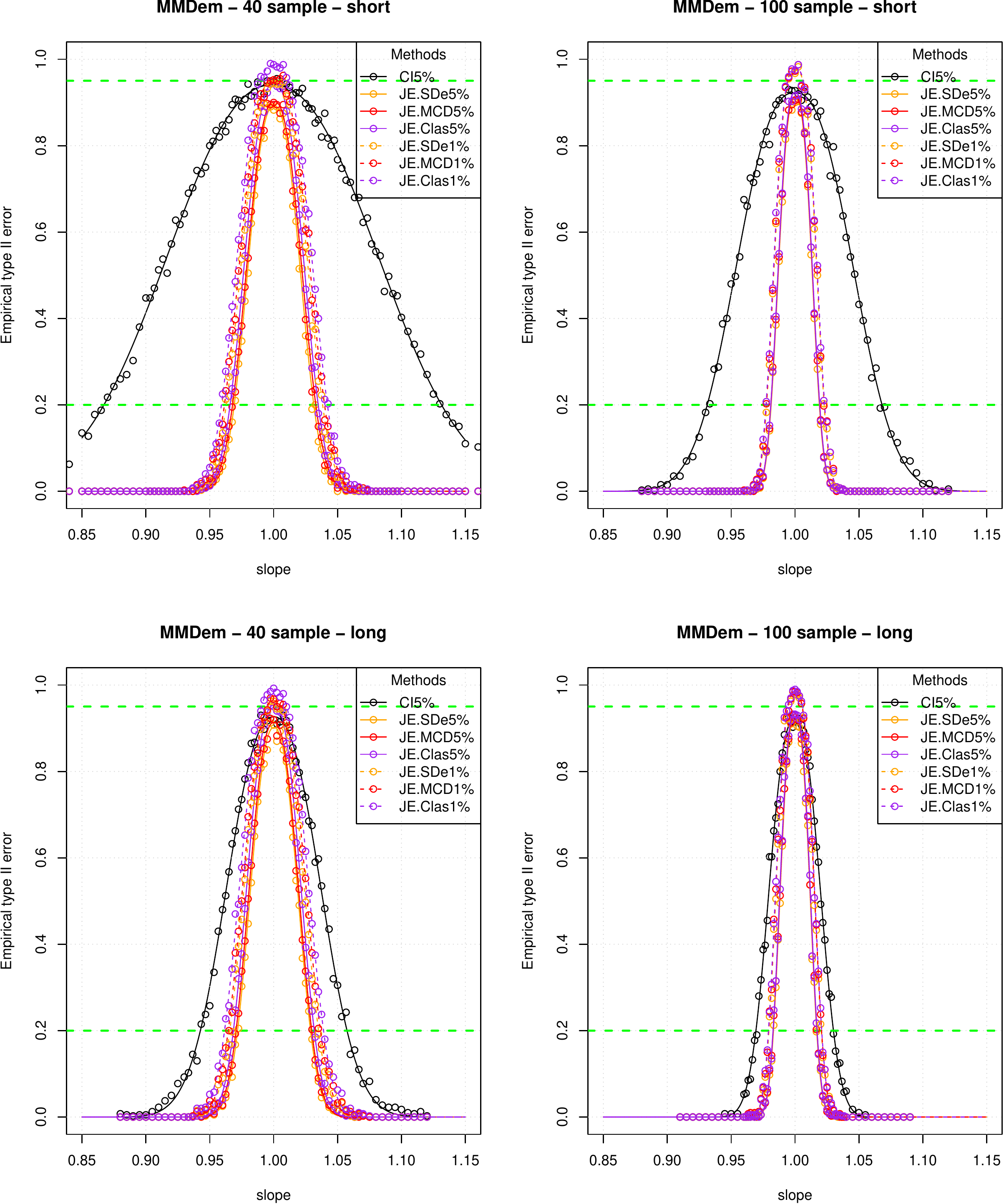} 

}

\caption{Classic CI vs JE type II errors for Passing Bablok regression.}\label{fig:mmcompare}
\end{figure}

\clearpage

\hypertarget{NonLinOpt}{%
\subsection{Appendix D: Nonlinear exponential power regression.}\label{NonLinOpt}}

\begin{figure}

{\centering \includegraphics{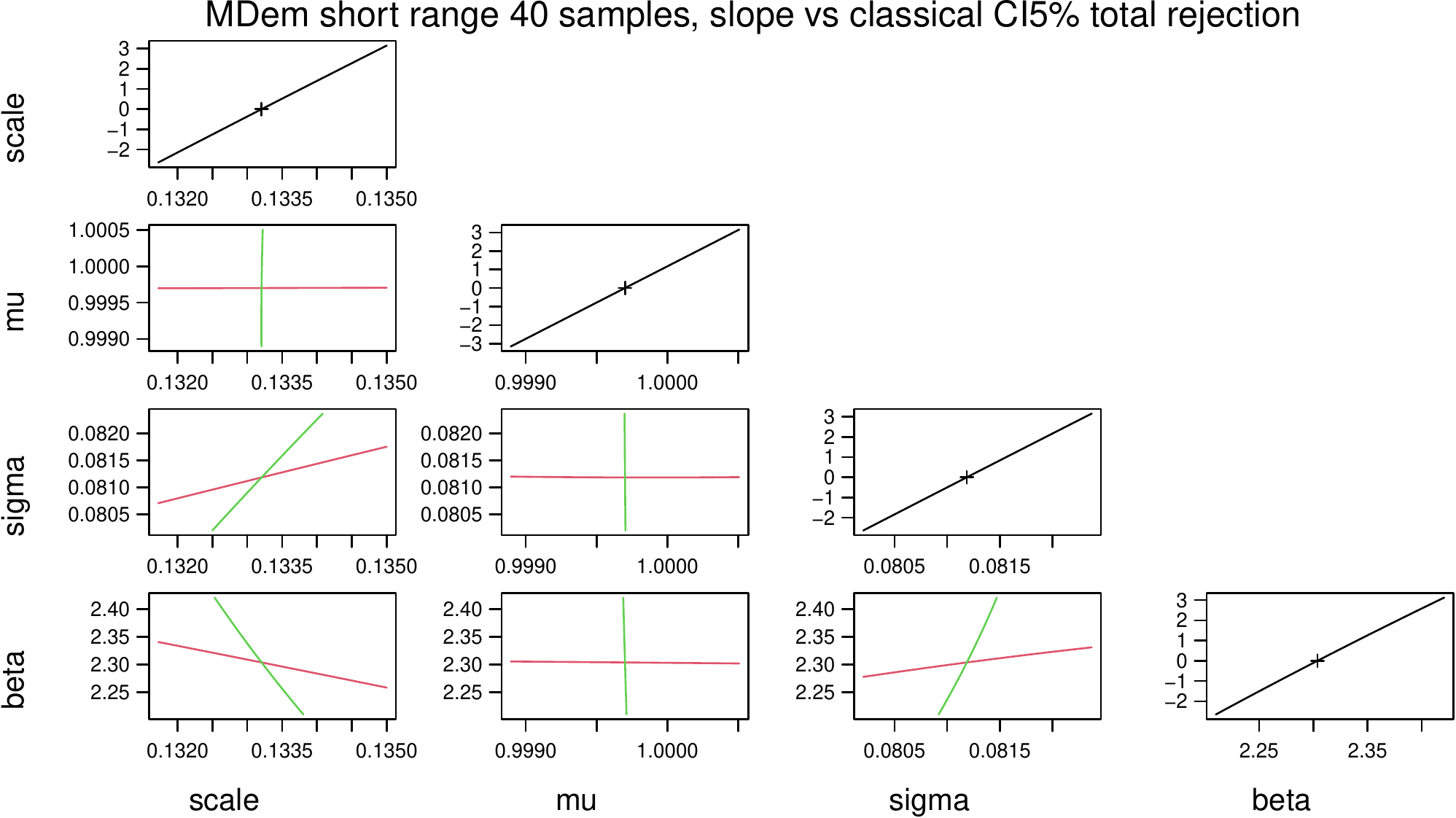} 

}

\caption{Profile plot example of the MDem regression for the classical CI type II error rate vs the slope in the 40 samples short range experiment.}\label{fig:profileplot}
\end{figure}

\pagebreak

\hypertarget{ComparisonTables}{%
\subsection{Appendix E: comparison tables}\label{ComparisonTables}}

\begin{longtable}[]{@{}cccccccc@{}}
\caption{\label{tab:100shorttable}Power and type I error comparison for the slope with the short range 100 data points simulations of the slopes.}\tabularnewline
\toprule
Regression & Method & p80\% LCI & p80\% est & p80\% UCI & T-I LCI & T-I est & T-I UCI\tabularnewline
\midrule
\endfirsthead
\toprule
Regression & Method & p80\% LCI & p80\% est & p80\% UCI & T-I LCI & T-I est & T-I UCI\tabularnewline
\midrule
\endhead
Deming & CI5\% & 1.0510 & 1.0534 & 1.0560 & 0.86984 & 0.90535 & 0.94086\tabularnewline
Deming & JE.SDe1\% & 1.0185 & 1.0189 & 1.0192 & 0.96055 & 0.97637 & 0.99219\tabularnewline
Deming & JE.MCD1\% & 1.0186 & 1.0189 & 1.0192 & 0.96240 & 0.97909 & 0.99578\tabularnewline
Deming & JE.Clas1\% & 1.0185 & 1.0188 & 1.0192 & 0.96180 & 0.97854 & 0.99528\tabularnewline
MDem & CI5\% & 1.0560 & 1.0588 & 1.0618 & 0.87659 & 0.91411 & 0.95163\tabularnewline
MDem & JE.SDe1\% & 1.0200 & 1.0203 & 1.0207 & 0.95780 & 0.97465 & 0.99150\tabularnewline
MDem & JE.MCD1\% & 1.0200 & 1.0204 & 1.0208 & 0.95745 & 0.97586 & 0.99427\tabularnewline
MDem & JE.Clas1\% & 1.0200 & 1.0204 & 1.0207 & 0.96287 & 0.97910 & 0.99533\tabularnewline
PaBa & CI5\% & 1.0542 & 1.0568 & 1.0596 & 0.87476 & 0.91186 & 0.94895\tabularnewline
PaBa & JE.SDe1\% & 1.0231 & 1.0235 & 1.0240 & 0.93682 & 0.95357 & 0.97031\tabularnewline
PaBa & JE.MCD1\% & 1.0231 & 1.0236 & 1.0242 & 0.93124 & 0.95168 & 0.97213\tabularnewline
PaBa & JE.Clas1\% & 1.0235 & 1.0241 & 1.0247 & 0.95695 & 0.97886 & 1.00077\tabularnewline
MMDem & CI5\% & 1.0646 & 1.0681 & 1.0720 & 0.88215 & 0.92250 & 0.96285\tabularnewline
MMDem & JE.SDe1\% & 1.0223 & 1.0227 & 1.0232 & 0.95487 & 0.97495 & 0.99503\tabularnewline
MMDem & JE.MCD1\% & 1.0224 & 1.0230 & 1.0236 & 0.95321 & 0.97758 & 1.00194\tabularnewline
MMDem & JE.Clas1\% & 1.0227 & 1.0231 & 1.0237 & 0.95750 & 0.97829 & 0.99907\tabularnewline
WDem & CI5\% & 1.0575 & 1.0605 & 1.0638 & 0.85934 & 0.89592 & 0.93249\tabularnewline
WDem & JE.SDe1\% & 1.0189 & 1.0192 & 1.0196 & 0.96270 & 0.97853 & 0.99436\tabularnewline
WDem & JE.MCD1\% & 1.0190 & 1.0193 & 1.0196 & 0.96720 & 0.98114 & 0.99508\tabularnewline
WDem & JE.Clas1\% & 1.0190 & 1.0192 & 1.0196 & 0.96407 & 0.97852 & 0.99297\tabularnewline
\bottomrule
\end{longtable}

\begin{longtable}[]{@{}cccccccc@{}}
\caption{\label{tab:40longtable}Power and type I error comparison for the slope with the long range 40 data points simulations.}\tabularnewline
\toprule
Regression & Method & p80\% LCI & p80\% est & p80\% UCI & T-I LCI & T-I est & T-I UCI\tabularnewline
\midrule
\endfirsthead
\toprule
Regression & Method & p80\% LCI & p80\% est & p80\% UCI & T-I LCI & T-I est & T-I UCI\tabularnewline
\midrule
\endhead
Deming & CI5\% & 1.0367 & 1.0380 & 1.0395 & 0.86575 & 0.89359 & 0.92143\tabularnewline
Deming & JE.SDe1\% & 1.0256 & 1.0262 & 1.0269 & 0.95016 & 0.97292 & 0.99567\tabularnewline
Deming & JE.MCD1\% & 1.0259 & 1.0265 & 1.0272 & 0.95181 & 0.97451 & 0.99720\tabularnewline
Deming & JE.Clas1\% & 1.0257 & 1.0264 & 1.0271 & 0.95335 & 0.97611 & 0.99887\tabularnewline
MDem & CI5\% & 1.0416 & 1.0429 & 1.0444 & 0.88799 & 0.91237 & 0.93676\tabularnewline
MDem & JE.SDe1\% & 1.0285 & 1.0293 & 1.0302 & 0.93556 & 0.96050 & 0.98543\tabularnewline
MDem & JE.MCD1\% & 1.0289 & 1.0297 & 1.0306 & 0.94262 & 0.96676 & 0.99090\tabularnewline
MDem & JE.Clas1\% & 1.0291 & 1.0298 & 1.0305 & 0.95086 & 0.97261 & 0.99436\tabularnewline
PaBa & CI5\% & 1.0410 & 1.0422 & 1.0435 & 0.88692 & 0.90945 & 0.93197\tabularnewline
PaBa & JE.SDe1\% & 1.0315 & 1.0325 & 1.0335 & 0.90982 & 0.93505 & 0.96028\tabularnewline
PaBa & JE.MCD1\% & 1.0324 & 1.0334 & 1.0345 & 0.90540 & 0.92989 & 0.95438\tabularnewline
PaBa & JE.Clas1\% & 1.0329 & 1.0336 & 1.0343 & 0.95682 & 0.97612 & 0.99542\tabularnewline
MMDem & CI5\% & 1.0546 & 1.0574 & 1.0606 & 0.88966 & 0.92660 & 0.96355\tabularnewline
MMDem & JE.SDe1\% & 1.0340 & 1.0351 & 1.0363 & 0.93463 & 0.96076 & 0.98690\tabularnewline
MMDem & JE.MCD1\% & 1.0355 & 1.0366 & 1.0377 & 0.93807 & 0.96368 & 0.98930\tabularnewline
MMDem & JE.Clas1\% & 1.0379 & 1.0390 & 1.0402 & 0.96434 & 0.99055 & 1.01675\tabularnewline
WDem & CI5\% & 1.1007 & 1.1106 & 1.1221 & 0.74405 & 0.78543 & 0.82681\tabularnewline
WDem & JE.SDe1\% & 1.0526 & 1.0620 & 1.0736 & 0.55452 & 0.61620 & 0.67787\tabularnewline
WDem & JE.MCD1\% & 1.0552 & 1.0657 & 1.0785 & 0.51943 & 0.57930 & 0.63916\tabularnewline
WDem & JE.Clas1\% & 1.0708 & 1.0785 & 1.0881 & 0.92709 & 0.98993 & 1.05276\tabularnewline
\bottomrule
\end{longtable}

\pagebreak

\hypertarget{Comparison250Samples}{%
\subsection{Appendix F: 250 samples CI vs 40 samples ellipse power comparison}\label{Comparison250Samples}}

\begin{longtable}[]{@{}cccccccc@{}}
\caption{\label{tab:dem250shorttable}Deming regression: classical CI method with 250 samples vs JE method with 40 samples, power and Type I error comparison.}\tabularnewline
\toprule
Samples & Method & p80\% LCI & p80\% est & p80\% UCI & T-I LCI & T-I est & T-I UCI\tabularnewline
\midrule
\endfirsthead
\toprule
Samples & Method & p80\% LCI & p80\% est & p80\% UCI & T-I LCI & T-I est & T-I UCI\tabularnewline
\midrule
\endhead
250 & CI5\% & 1.0322 & 1.0338 & 1.0356 & 0.87738 & 0.91632 & 0.95526\tabularnewline
250 & JE.SDe1\% & 1.0118 & 1.0120 & 1.0122 & 0.96597 & 0.98329 & 1.00060\tabularnewline
250 & JE.MCD1\% & 1.0118 & 1.0119 & 1.0121 & 0.96591 & 0.98187 & 0.99783\tabularnewline
250 & JE.Clas1\% & 1.0118 & 1.0120 & 1.0122 & 0.96507 & 0.98289 & 1.00072\tabularnewline
40 & CI5\% & 1.0811 & 1.0850 & 1.0894 & 0.87433 & 0.90834 & 0.94236\tabularnewline
40 & JE.SDe1\% & 1.0284 & 1.0290 & 1.0297 & 0.94656 & 0.96790 & 0.98925\tabularnewline
40 & JE.MCD1\% & 1.0287 & 1.0294 & 1.0301 & 0.94639 & 0.96875 & 0.99112\tabularnewline
40 & JE.Clas1\% & 1.0286 & 1.0293 & 1.0300 & 0.95034 & 0.97118 & 0.99202\tabularnewline
\bottomrule
\end{longtable}

\begin{longtable}[]{@{}cccccccc@{}}
\caption{\label{tab:mm250shorttable}MMDem regression: classical CI method with 250 samples vs JE method with 40 samples, power and Type I error comparison.}\tabularnewline
\toprule
Samples & Method & p80\% LCI & p80\% est & p80\% UCI & T-I LCI & T-I est & T-I UCI\tabularnewline
\midrule
\endfirsthead
\toprule
Samples & Method & p80\% LCI & p80\% est & p80\% UCI & T-I LCI & T-I est & T-I UCI\tabularnewline
\midrule
\endhead
250 & CI5\% & 1.0388 & 1.0407 & 1.0429 & 0.88429 & 0.92309 & 0.96189\tabularnewline
250 & JE.SDe1\% & 1.0139 & 1.0141 & 1.0143 & 0.97662 & 0.99078 & 1.00495\tabularnewline
250 & JE.MCD1\% & 1.0140 & 1.0142 & 1.0144 & 0.97785 & 0.99222 & 1.00659\tabularnewline
250 & JE.Clas1\% & 1.0139 & 1.0141 & 1.0144 & 0.97749 & 0.99292 & 1.00835\tabularnewline
40 & CI5\% & 1.1253 & 1.1313 & 1.1381 & 0.90772 & 0.94090 & 0.97407\tabularnewline
40 & JE.SDe1\% & 1.0372 & 1.0383 & 1.0394 & 0.92305 & 0.94758 & 0.97211\tabularnewline
40 & JE.MCD1\% & 1.0388 & 1.0398 & 1.0409 & 0.93230 & 0.95544 & 0.97859\tabularnewline
40 & JE.Clas1\% & 1.0411 & 1.0420 & 1.0430 & 0.96489 & 0.98605 & 1.00721\tabularnewline
\bottomrule
\end{longtable}

\begin{longtable}[]{@{}cccccccc@{}}
\caption{\label{tab:w250shorttable}WDem regression: classical CI method with 250 samples vs JE method with 40 samples, power and Type I error comparison.}\tabularnewline
\toprule
Samples & Method & p80\% LCI & p80\% est & p80\% UCI & T-I LCI & T-I est & T-I UCI\tabularnewline
\midrule
\endfirsthead
\toprule
Samples & Method & p80\% LCI & p80\% est & p80\% UCI & T-I LCI & T-I est & T-I UCI\tabularnewline
\midrule
\endhead
250 & CI5\% & 1.0388 & 1.0407 & 1.0429 & 0.88429 & 0.92309 & 0.96189\tabularnewline
250 & JE.SDe1\% & 1.0139 & 1.0141 & 1.0143 & 0.97662 & 0.99078 & 1.00495\tabularnewline
250 & JE.MCD1\% & 1.0140 & 1.0142 & 1.0144 & 0.97785 & 0.99222 & 1.00659\tabularnewline
250 & JE.Clas1\% & 1.0139 & 1.0141 & 1.0144 & 0.97749 & 0.99292 & 1.00835\tabularnewline
40 & CI5\% & 1.0900 & 1.0953 & 1.1013 & 0.85356 & 0.89079 & 0.92803\tabularnewline
40 & JE.SDe1\% & 1.0295 & 1.0300 & 1.0307 & 0.93664 & 0.95482 & 0.97300\tabularnewline
40 & JE.MCD1\% & 1.0297 & 1.0303 & 1.0309 & 0.93734 & 0.95569 & 0.97405\tabularnewline
40 & JE.Clas1\% & 1.0295 & 1.0301 & 1.0308 & 0.93739 & 0.95793 & 0.97848\tabularnewline
\bottomrule
\end{longtable}

\pagebreak

\hypertarget{InterceptTables}{%
\subsection{Appendix G. tables for the intercept experiments}\label{InterceptTables}}

\begin{longtable}[]{@{}cccccccc@{}}
\caption{\label{tab:int100shorttable}Power and type I error comparison for the intercept with short range 100 samples simulations.}\tabularnewline
\toprule
Regression & Method & p80\% LCI & p80\% est & p80\% UCI & T-I LCI & T-I est. & T-I UCI\tabularnewline
\midrule
\endfirsthead
\toprule
Regression & Method & p80\% LCI & p80\% est & p80\% UCI & T-I LCI & T-I est. & T-I UCI\tabularnewline
\midrule
\endhead
Deming & CI5\% & 0.29641 & 0.31082 & 0.32690 & 0.87824 & 0.91420 & 0.95016\tabularnewline
Deming & JE.SDe1\% & 0.10149 & 0.10380 & 0.10625 & 0.95368 & 0.97597 & 0.99826\tabularnewline
Deming & JE.MCD1\% & 0.10149 & 0.10380 & 0.10625 & 0.95368 & 0.97597 & 0.99826\tabularnewline
Deming & JE.Clas1\% & 0.10149 & 0.10380 & 0.10625 & 0.95368 & 0.97597 & 0.99826\tabularnewline
MDem & CI5\% & 0.32314 & 0.33885 & 0.35640 & 0.88341 & 0.91991 & 0.95641\tabularnewline
MDem & JE.SDe1\% & 0.11046 & 0.11361 & 0.11700 & 0.94668 & 0.97385 & 1.00103\tabularnewline
MDem & JE.MCD1\% & 0.11046 & 0.11361 & 0.11700 & 0.94668 & 0.97385 & 1.00103\tabularnewline
MDem & JE.Clas1\% & 0.11046 & 0.11361 & 0.11700 & 0.94668 & 0.97385 & 1.00103\tabularnewline
PaBa & CI5\% & 0.31532 & 0.32938 & 0.34497 & 0.88245 & 0.91682 & 0.95118\tabularnewline
PaBa & JE.SDe1\% & 0.12919 & 0.13381 & 0.13884 & 0.93402 & 0.96281 & 0.99159\tabularnewline
PaBa & JE.MCD1\% & 0.12919 & 0.13381 & 0.13884 & 0.93402 & 0.96281 & 0.99159\tabularnewline
PaBa & JE.Clas1\% & 0.12919 & 0.13381 & 0.13884 & 0.93402 & 0.96281 & 0.99159\tabularnewline
MMDem & CI5\% & 0.37337 & 0.39538 & 0.42058 & 0.87905 & 0.92365 & 0.96825\tabularnewline
MMDem & JE.SDe1\% & 0.12508 & 0.12900 & 0.13328 & 0.93708 & 0.96708 & 0.99708\tabularnewline
MMDem & JE.MCD1\% & 0.12508 & 0.12900 & 0.13328 & 0.93708 & 0.96708 & 0.99708\tabularnewline
MMDem & JE.Clas1\% & 0.12508 & 0.12900 & 0.13328 & 0.93708 & 0.96708 & 0.99708\tabularnewline
WDem & CI5\% & 0.32028 & 0.33998 & 0.36239 & 0.86193 & 0.90362 & 0.94532\tabularnewline
WDem & JE.SDe1\% & 0.10022 & 0.10278 & 0.10552 & 0.94498 & 0.97023 & 0.99547\tabularnewline
WDem & JE.MCD1\% & 0.10022 & 0.10278 & 0.10552 & 0.94498 & 0.97023 & 0.99547\tabularnewline
WDem & JE.Clas1\% & 0.10022 & 0.10278 & 0.10552 & 0.94498 & 0.97023 & 0.99547\tabularnewline
\bottomrule
\end{longtable}

\begin{longtable}[]{@{}cccccccc@{}}
\caption{\label{tab:int100longtable}Power and type I error comparison for the intercept with long range 100 samples simulations.}\tabularnewline
\toprule
Regression & Method & p80\% LCI & p80\% est & p80\% UCI & T-I LCI & T-I est. & T-I UCI\tabularnewline
\midrule
\endfirsthead
\toprule
Regression & Method & p80\% LCI & p80\% est & p80\% UCI & T-I LCI & T-I est. & T-I UCI\tabularnewline
\midrule
\endhead
Deming & CI5\% & 1.5152 & 1.5529 & 1.5927 & 0.89232 & 0.91216 & 0.93199\tabularnewline
Deming & JE.SDe1\% & 1.0206 & 1.0399 & 1.0602 & 0.95971 & 0.97844 & 0.99717\tabularnewline
Deming & JE.MCD1\% & 1.0206 & 1.0399 & 1.0602 & 0.95971 & 0.97844 & 0.99717\tabularnewline
Deming & JE.Clas1\% & 1.0206 & 1.0399 & 1.0602 & 0.95971 & 0.97844 & 0.99717\tabularnewline
MDem & CI5\% & 1.6450 & 1.6904 & 1.7387 & 0.90670 & 0.92845 & 0.95019\tabularnewline
MDem & JE.SDe1\% & 1.1147 & 1.1295 & 1.1448 & 0.96505 & 0.97774 & 0.99043\tabularnewline
MDem & JE.MCD1\% & 1.1147 & 1.1295 & 1.1448 & 0.96505 & 0.97774 & 0.99043\tabularnewline
MDem & JE.Clas1\% & 1.1147 & 1.1295 & 1.1448 & 0.96505 & 0.97774 & 0.99043\tabularnewline
PaBa & CI5\% & 1.6849 & 1.7423 & 1.8041 & 0.89677 & 0.92280 & 0.94882\tabularnewline
PaBa & JE.SDe1\% & 1.2855 & 1.3177 & 1.3518 & 0.94987 & 0.97039 & 0.99092\tabularnewline
PaBa & JE.MCD1\% & 1.2855 & 1.3177 & 1.3518 & 0.94987 & 0.97039 & 0.99092\tabularnewline
PaBa & JE.Clas1\% & 1.2855 & 1.3177 & 1.3518 & 0.94987 & 0.97039 & 0.99092\tabularnewline
MMDem & CI5\% & 1.8842 & 1.9557 & 2.0334 & 0.91144 & 0.93923 & 0.96701\tabularnewline
MMDem & JE.SDe1\% & 1.2698 & 1.2866 & 1.3038 & 0.96594 & 0.97834 & 0.99074\tabularnewline
MMDem & JE.MCD1\% & 1.2698 & 1.2866 & 1.3038 & 0.96594 & 0.97834 & 0.99074\tabularnewline
MMDem & JE.Clas1\% & 1.2698 & 1.2866 & 1.3038 & 0.96594 & 0.97834 & 0.99074\tabularnewline
WDem & CI5\% & 2.6324 & 2.9391 & 3.3184 & 0.78550 & 0.85055 & 0.91559\tabularnewline
WDem & JE.SDe1\% & 1.5045 & 1.6531 & 1.8226 & 0.62565 & 0.67968 & 0.73370\tabularnewline
WDem & JE.MCD1\% & 1.5045 & 1.6531 & 1.8226 & 0.62565 & 0.67968 & 0.73370\tabularnewline
WDem & JE.Clas1\% & 1.5045 & 1.6531 & 1.8226 & 0.62565 & 0.67968 & 0.73370\tabularnewline
\bottomrule
\end{longtable}

\begin{longtable}[]{@{}cccccccc@{}}
\caption{\label{tab:int40longtable}Power and type I error comparison for the intercept with the long range 40 samples simulations.}\tabularnewline
\toprule
Regression & Method & p80\% LCI & p80\% est & p80\% UCI & T-I LCI & T-I est. & T-I UCI\tabularnewline
\midrule
\endfirsthead
\toprule
Regression & Method & p80\% LCI & p80\% est & p80\% UCI & T-I LCI & T-I est. & T-I UCI\tabularnewline
\midrule
\endhead
Deming & CI5\% & 2.3724 & 2.4822 & 2.6036 & 0.86041 & 0.89496 & 0.92952\tabularnewline
Deming & JE.SDe1\% & 1.6034 & 1.6477 & 1.6953 & 0.93488 & 0.96079 & 0.98671\tabularnewline
Deming & JE.MCD1\% & 1.6034 & 1.6477 & 1.6953 & 0.93488 & 0.96079 & 0.98671\tabularnewline
Deming & JE.Clas1\% & 1.6034 & 1.6477 & 1.6953 & 0.93488 & 0.96079 & 0.98671\tabularnewline
MDem & CI5\% & 2.6351 & 2.7799 & 2.9434 & 0.86504 & 0.90649 & 0.94794\tabularnewline
MDem & JE.SDe1\% & 1.7653 & 1.8364 & 1.9148 & 0.92647 & 0.96031 & 0.99414\tabularnewline
MDem & JE.MCD1\% & 1.7653 & 1.8364 & 1.9148 & 0.92647 & 0.96031 & 0.99414\tabularnewline
MDem & JE.Clas1\% & 1.7653 & 1.8364 & 1.9148 & 0.92647 & 0.96031 & 0.99414\tabularnewline
PaBa & CI5\% & 2.7810 & 2.9184 & 3.0718 & 0.87280 & 0.91053 & 0.94826\tabularnewline
PaBa & JE.SDe1\% & 2.0217 & 2.0927 & 2.1694 & 0.90510 & 0.93156 & 0.95803\tabularnewline
PaBa & JE.MCD1\% & 2.0217 & 2.0927 & 2.1694 & 0.90510 & 0.93156 & 0.95803\tabularnewline
PaBa & JE.Clas1\% & 2.0217 & 2.0927 & 2.1694 & 0.90510 & 0.93156 & 0.95803\tabularnewline
MMDem & CI5\% & 3.4970 & 3.7127 & 3.9595 & 0.88120 & 0.92442 & 0.96763\tabularnewline
MMDem & JE.SDe1\% & 2.0528 & 2.1183 & 2.1892 & 0.92958 & 0.95679 & 0.98399\tabularnewline
MMDem & JE.MCD1\% & 2.0528 & 2.1183 & 2.1892 & 0.92958 & 0.95679 & 0.98399\tabularnewline
MMDem & JE.Clas1\% & 2.0528 & 2.1183 & 2.1892 & 0.92958 & 0.95679 & 0.98399\tabularnewline
WDem & CI5\% & 3.2593 & 3.5729 & 3.9418 & 0.76193 & 0.81126 & 0.86059\tabularnewline
WDem & JE.SDe1\% & 1.7598 & 2.1936 & 2.7496 & 0.52564 & 0.61507 & 0.70451\tabularnewline
WDem & JE.MCD1\% & 1.7598 & 2.1936 & 2.7496 & 0.52564 & 0.61507 & 0.70451\tabularnewline
WDem & JE.Clas1\% & 1.7598 & 2.1936 & 2.7496 & 0.52564 & 0.61507 & 0.70451\tabularnewline
\bottomrule
\end{longtable}

\clearpage

\hypertarget{HetCompPlot}{%
\subsection{Appendix H: heteroscedastic additional comparative plots}\label{HetCompPlot}}

\begin{figure}[H]

{\centering \includegraphics{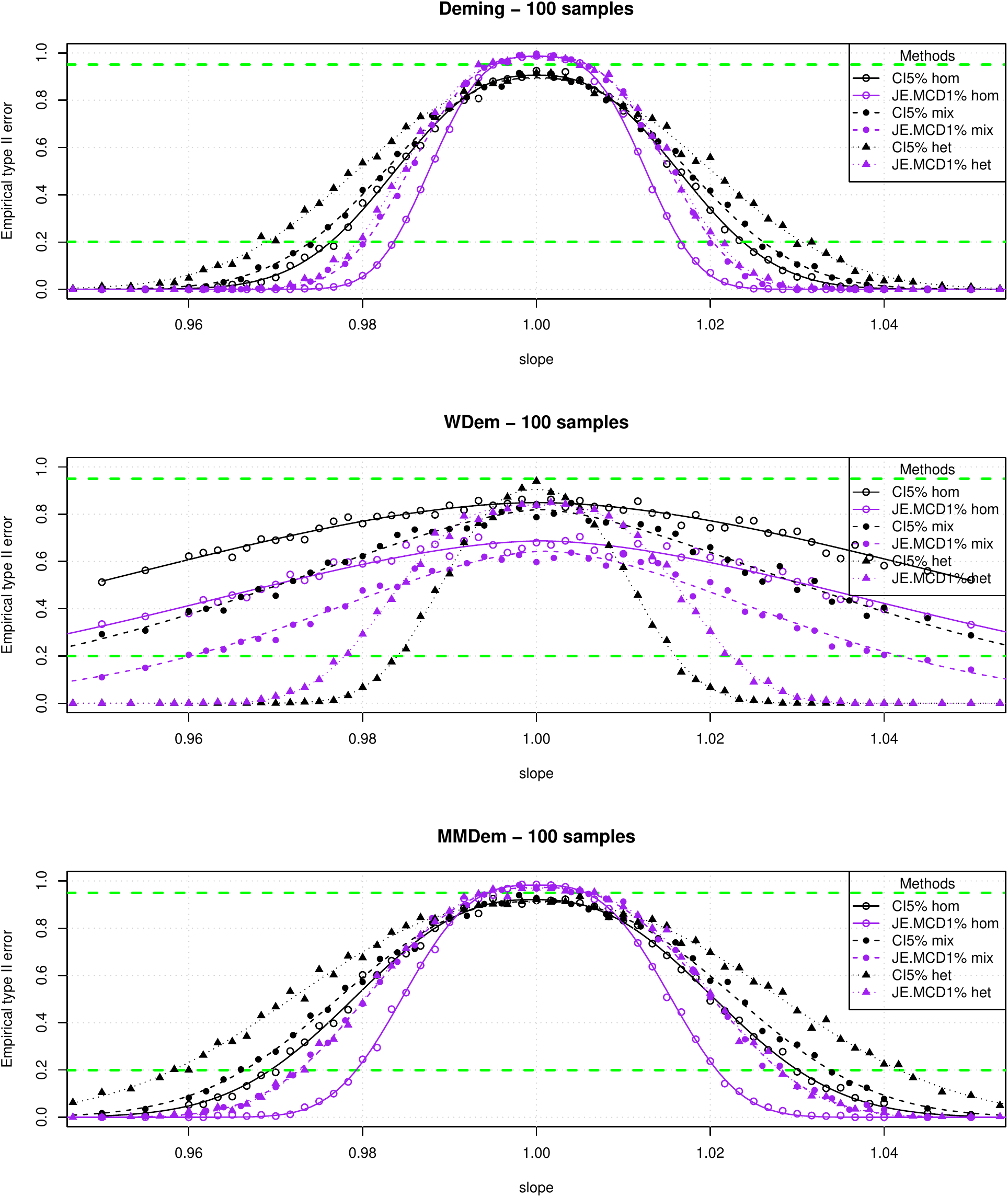} 

}

\caption{Role of the heteroscedasticity: one plot per regression method with the 40 samples data sets.}\label{fig:homohetmixcompdmmw}
\end{figure}

\clearpage

  \bibliography{thesis.bib}

\end{document}